\documentclass[12pt,preprint]{emulateapj}

\begin{document}
\title{$UV$-Continuum Slopes of $>$4000 $z\sim4$-8 Galaxies from the
  HUDF/XDF, HUDF09, ERS, CANDELS-South, and CANDELS-North
  Fields\altaffilmark{1}}

\author{R. J. Bouwens\altaffilmark{2,3}, G. D. Illingworth\altaffilmark{3}, P. A. Oesch\altaffilmark{3,\dag}, I. Labb{\'e}\altaffilmark{2}, P.G. van Dokkum\altaffilmark{4}, M. Trenti\altaffilmark{5}, M. Franx\altaffilmark{2}, R. Smit\altaffilmark{2}, V. Gonzalez\altaffilmark{3,6}, D. Magee\altaffilmark{3}
}
\altaffiltext{2}{Leiden Observatory, Leiden University, NL-2300 RA Leiden, Netherlands}
\altaffiltext{3}{UCO/Lick Observatory, University of California, Santa Cruz, CA 95064}
\altaffiltext{4}{Department of Astronomy, Yale University, New Haven, CT 06520}
\altaffiltext{5}{Institute of Astronomy, University of Cambridge, Madingley Road, Cambridge CB3 0HA, UK}
\altaffiltext{6}{University of California, Riverside, CA 92521, USA}
\altaffiltext{7}{Institute for Astronomy, ETH Zurich, 8092 Zurich, Switzerland}
\altaffiltext{\dag}{Hubble Fellow}
\begin{abstract}
We measure the UV-continuum slope $\beta$ for over 4000 high-redshift
galaxies over a wide range of redshifts $z\sim4$-8 and luminosities
from the HST HUDF/XDF, HUDF09-1, HUDF09-2, ERS, CANDELS-N, and
CANDELS-S data sets.  Our new $\beta$ results reach very faint levels
at $z\sim4$ ($-$15.5 mag: $0.006 L_{z=3}^{*}$), $z\sim5$ ($-$16.5 mag:
$0.014L_{z=3}^{*}$), and $z\sim 6$ and $z\sim 7$ ($-$17 mag:
$0.025L_{z=3}^{*}$).  Inconsistencies between previous studies led us
to conduct a comprehensive review of systematic errors and develop a
new technique for measuring $\beta$ that is robust against biases that
arise from the impact of noise.  We demonstrate, by object-by-object
comparisons, that all previous studies, including our own and those
done on the latest HUDF12 dataset, suffered from small systematic
errors in $\beta$.  We find that after correcting for the systematic
errors (typically $\Delta\beta\sim0.1$-0.2) all $\beta$ results at
$z\sim7$ from different groups are in excellent agreement.  The mean
$\beta$ we measure for faint ($-$18 mag: $0.1L_{z=3}^{*}$) $z\sim4$,
$z\sim5$, $z\sim6$, and $z\sim7$ galaxies is $-$2.03$\pm$0.03$\pm$0.06
(random and systematic errors), $-$2.14$\pm$0.06$\pm$0.06,
$-$2.24$\pm$0.11$\pm$0.08, and $-$2.30$\pm$0.18$\pm$0.13,
respectively.  Our new $\beta$ values are redder than we have reported
in the past, but bluer than other recent results.  Our previously
reported trend of bluer $\beta$'s at lower luminosities is confirmed,
as is the evolution to bluer $\beta$'s at high redshifts.  $\beta$
appears to show only a mild luminosity dependence faintward of
$M_{UV,AB}\sim-19$ mag, suggesting that the mean $\beta$ asymptotes to
$\sim-$2.2 to $-$2.4 for faint $z\geq4$ galaxies.  At $z\sim7$, the
observed $\beta$'s suggest non-zero, but low dust extinction, and they
agree well with values predicted in cosmological hydrodynamical
simulations.
\end{abstract}

\section{Introduction}

An important frontier in the study of very high redshift galaxies
remains the study of their stellar populations.  Galaxies within a few
hundred million years of the Big Bang are expected to be quite
different from galaxies at lower redshift, with significantly younger
ages and lower metallicities.  For very young and chemically immature
systems, changes in the stellar population could include a transition
to a more top-heavy IMF (e.g., Bromm \& Larson 2004), evolution in the
dust composition (e.g., due to changes in the dust production
mechanism: Maiolino et al.\ 2004), as well as a much lower dust
extinction overall (e.g., Bouwens et al.\ 2009; Finlator et al.\ 2011;
Dayal \& Ferrara 2012).

Over the last few years, considerable progress has been made in
characterizing the changes in stellar populations of galaxies back to
the earliest times.  We have constraints on both the ages and emission
line strengths of galaxies at $z\gtrsim4$ (e.g., Stark et al.\ 2009;
Labb{\'e} et al.\ 2010; Gonz{\'a}lez et al.\ 2011; Gonz{\'a}lez et
al.\ 2012; Stark et al.\ 2013; Labb{\'e} et al.\ 2013; Lee et
al.\ 2012; Oesch et al.\ 2013a), and we have measurements of the
$UV$-continuum slopes $\beta$ ($f_{\lambda}\propto \lambda^{\beta}$:
e.g., Meurer et al.\ 1999) of high-redshift galaxies (e.g., Bouwens et
al.\ 2010; Bunker et al.\ 2010).  Quantifying the dependence of
$\beta$ on both cosmic time and as a function of other quantities like
luminosity or stellar mass has been very revealing.  The
$UV$-continuum slope $\beta$ is particularly useful due to its
sensitivity to the metallicity, age, and especially the dust content
within a galaxy.  Bouwens et al.\ (2012) demonstrated that the mean
$UV$-continuum slope $\beta$ of galaxies shows a dependence on $UV$
luminosity at $z\sim4$, $z\sim5$, and $z\sim6$, with almost an
identical slope independent of redshift (see also Bouwens et al.\ 2009
for an earlier similar, but more limited, demonstration).  A gradual
reddening of the $UV$-continuum slope $\beta$ with cosmic time is also
observed (see also work by Stanway et al.\ 2005, Wilkins et al.\ 2011,
Finkelstein et al.\ 2012, and Castellano et al.\ 2012).  These studies
suggest a general trend of decreasing dust content of galaxies to
earlier cosmic times, to lower luminosities, and to lower masses
(though the observed trends may be enhanced by changes in the ages or
metallicities).  Similar trends are found as a function of the
rest-frame optical luminosity of galaxies, as seen with Spitzer/IRAC
(Oesch et al.\ 2013a; see also Papovich et al.\ 2004) and also in large
cosmological hydrodynamical simulations (e.g., Finlator et al.\ 2011:
see Bouwens et al.\ 2012; Finkelstein et al.\ 2012).

\begin{deluxetable*}{ccccccccccc}
\tablewidth{0cm}
\tabletypesize{\footnotesize}
\tablecaption{Observational Data Considered in Establishing the $UV$-continuum Slope Distribution for $z\sim4$-7 Galaxies.\label{tab:obsdata}}
\tablehead{\colhead{} & \colhead{Area} & \multicolumn{8}{c}{$5\sigma$ Depth\tablenotemark{a}}\\
\colhead{Field}  & \colhead{(arcmin$^2$)} & \colhead{$B_{435}$} & \colhead{$V_{606}$} & \colhead{$i_{775}$} & \colhead{$I_{814}$} & \colhead{$z_{850}$} & \colhead{$Y_{098}/Y_{105}$} & \colhead{$J_{125}$} & \colhead{$JH_{140}$} & \colhead{$H_{160}$}}
\startdata
XDF\tablenotemark{b}  & 4 & 29.8\tablenotemark{c} & 30.3\tablenotemark{c} & 30.3\tablenotemark{c} & 29.1 & 29.4\tablenotemark{c} & 30.1 & 29.8 & 29.8 & 29.8 \\
HUDF09-1       & 4 & --- & 29.0 & 29.0 & --- & 29.0 & 29.0 & 29.3 & ---\tablenotemark{d} & 29.1 \\
HUDF09-2       & 4 & 28.8 & 29.9 & 29.3 & 29.0 & 29.2 & 29.2 & 29.5 & ---\tablenotemark{d} & 29.3 \\
CANDELS-S/Deep & 66 & 28.2 & 28.5 & 28.0 & 28.8 & 28.0 & 28.5 & 28.8 & ---\tablenotemark{d} & 28.5 \\
ERS & 39 & 28.2 & 28.5 & 28.0 & 28.0 & 28.0 & 27.9 & 28.4 & ---\tablenotemark{d} & 28.1 \\
CANDELS-N/Deep & 60 & 28.2 & 28.5 & 28.0 & 28.8 & 28.0 & 28.5 & 28.8 & ---\tablenotemark{d} & 28.5 \\
\\
\multicolumn{11}{c}{Also Used in Establishing the $\beta$ Distribution for $z\sim4$-6 Galaxies}\\
CANDELS-S/Wide & 40 & 28.2 & 28.5 & 28.0 & 28.1 & 28.0 & 28.0 & 27.8 & ---\tablenotemark{d} & 27.8 \\
CANDELS-N/Wide & 75 & 28.2 & 28.5 & 28.0 & 28.0 & 28.0 & 28.0 & 27.8 & ---\tablenotemark{d} & 27.8 
\enddata
\tablenotetext{a}{The $5\sigma$ depths are based on the light
  within a $0.35''$-diameter aperture.  No correction is made for the
  light outside this aperture.  This is in contrast to many other
  studies where the quoted depths are corrected for the missing light
  (which can result in a $\sim$0.3 mag and $\sim$0.5 mag correction to
  the quoted depths for the ACS and WFC3/IR data, respectively, but
  depend upon the profile assumed).}
\tablenotetext{b}{The XDF refers to the 4.7 arcmin$^2$ region over the
  HUDF with ultra-deep near-IR observations from the HUDF09 and HUDF12
  programs (Illingworth et al.\ 2013).  It includes all ACS and
  WFC3/IR observations acquired over this region for the 10-year
  period 2002 to 2012.}
\tablenotetext{c}{The present XDF reduction (Illingworth et al.\ 2013)
  is typically $\sim$0.2 mag deeper than the original reduction of the
  HUDF ACS data provided by Beckwith et al.\ (2006).}
\tablenotetext{d}{Our approach for deriving the mean $\beta$ for
  $z\sim7$ galaxies is free of systematic biases with just three
  filters (see \S4.2), and so we do not require deep $JH_{140}$
  observations.  This enables us to make full use of the large number
  of datasets to measure $\beta$ at $z\sim7$ for which no deep
  $JH_{140}$ data are available.}
\end{deluxetable*}

The study of the $UV$-continuum slopes $\beta$ at $z\sim7$, while more
uncertain, has been improving due to ever more substantial data sets
in the near-IR with WFC3/IR.  While there was rapid consensus that
$\beta$ is moderately blue for the most luminous galaxies
($\beta\sim-1.8$ to $\sim-2.1$: Bouwens et al.\ 2010, 2012;
Finkelstein et al.\ 2012; Rogers et al.\ 2013), the $\beta$ values for
lower luminosity systems has become the subject of a prolonged debate.
There is good reason for all the attention given to these lower
luminosity galaxies: it is likely that the lowest-luminosity $z\sim7$
galaxies may be the least chemically-enriched galaxies accessible to
us and could potentially tell us something important about the
spectral properties or dust extinction of such young systems.  In this
regard, the possible discovery of galaxies with $\beta$'s as blue as
$-3$ in the first-year WFC3/IR observations from the HUDF09 program
over the Hubble Ultra-Deep Field (Beckwith et al.\ 2006) was therefore
potentially exciting (Bouwens et al.\ 2010; Finkelstein et al.\ 2010).
However, subsequent work has consistently yielded somewhat redder
values.  Values ranging from $-2.1$ to $-2.7$ were reported at low
$UV$ luminosities by Wilkins et al.\ (2011), Bouwens et al.\ (2012),
Finkelstein et al.\ (2012), and Dunlop et al.\ (2013).

While there has been considerable speculation as to why the subsequent
measurements of the mean $\beta$'s were redder than the initial
estimates, the most important question going forwards regards the
actual value for the faintest galaxies at $z\sim7$.  Despite
significant scatter in the measured $\beta$'s for the faintest sources
based upon the full HUDF09 dataset (e.g., Bouwens et al.\ 2012;
Finkelstein et al.\ 2012; Rogers et al.\ 2013), the availability of
even deeper observations over the HUDF from the HUDF12 program raised
the prospect that these differences could be resolved.  In a first
analysis of their HUDF12 data set, Dunlop et al.\ al.\ (2013) find a
mean $\beta$ of $-2.1\pm0.2$ at $z\sim7$.  Dunlop et al.\ (2013) also
arrive at determinations of the mean $\beta$ for faint galaxies at
$z\sim8$ and $z\sim9$, reporting $-1.9\pm0.3$ and $-1.8\pm0.6$,
respectively, suggesting that $\beta$ may be not be especially bluer
than $\beta\sim-2$ somewhat contrary to earlier results.

Given the noteworthy contrast of the Dunlop et al.\ (2013) $\beta$
results with previous results, it seems clear that an independent
analysis of the HUDF12 and other deep field observations is required
to further clarify the situation.  Fortunately, there is now a
considerable amount of additional information we can utilize, beyond
that already considered in previous work, to obtain the best possible
constraints on the mean value of $\beta$ for faint $z\sim7$-8
galaxies.  For example, while the HUDF provides the highest quality
information on $\beta$ for faint $z\sim7$-8 galaxies, the numbers are
still small and there is also high-quality information on the $\beta$
distribution for the faintest $z\sim7$ galaxies from the two parallel
fields to the HUDF, HUDF09-1 and HUDF09-2 (hereinafter, referred to as
HUDF09-Ps), that have thus far not been exploited.  While both of
these fields have very deep $J_{125}$ and $H_{160}$-band observations,
Dunlop et al.\ (2013) did not use the faintest sources from these
fields to avoid concerns about potential systematic biases which they
suggest could be present without the addition of deep imaging in a
fourth WFC3/IR band F140W.  Fortunately, as we will demonstrate
(\S4.2), there are techniques to avoid such systematic biases.

In addition, the availability of new ultra-deep WFC3/IR data, and
deeper reductions of the ACS data from the eXtreme Deep Field (XDF)
effort (Illingworth et al.\ 2013), over the Hubble Ultra-Deep Field
(hereinafter we refer to this data set as the XDF) make it possible to
obtain even more precise determinations of the $UV$-continuum slopes
$\beta$ for galaxies at $z\sim4$, $z\sim5$, and $z\sim6$.  Together
with the much greater wavelength leverage and sample sizes available
for galaxies at these redshifts and significant gains in depth (e.g.,
$2\times$ deeper observations in the $Y_{105}$ band: Ellis et
al.\ 2013), the increasingly accurate measures of the mean $\beta$'s
at $z\sim4$-6 allow for very precise constraints on how the mean
$\beta$ evolves towards early times.  Finally, given the existence of
several significant compilations of $\beta$ measurements in the
literature (Bouwens et al.\ 2012; Finkelstein et al.\ 2012; Dunlop et
al.\ 2013), we are able to perform extensive intercomparisons with
previous measurements of $\beta$ to obtain the broadest possible
perspective on where systematic biases may have affected $\beta$
measurements in the past and as cross checks on our own results.  This
is the first time such extensive intercomparisons have been made.  As
we will demonstrate in \S5, they are essential for resolving the
current debate on the $UV$-continuum slopes at high redshift.

The purpose of this paper is to utilize the ultra-deep observations
from the XDF and other programs to provide an independent assessment
of the mean $\beta$ for faint galaxies at $z\sim7$-9 and also at
$z\sim4$-6.  To ensure that our improved constraints on $\beta$ at
$z\sim7$ are as robust as possible, we introduce a new method for
fully leveraging the faintest galaxies in existing legacy data sets
while remaining robust against systematic biases resulting from noise.
We motivate and develop this new approach in \S4.1-\S4.2.  Our
analysis includes sources from the XDF (Illingworth et al. 2013
combining the HUDF09 [Bouwens et al.\ 2011] and HUDF12 [Ellis et
  al.\ 2013] datasets), the HUDF09-1 and HUDF09-2 (Bouwens et
al.\ 2011), the CANDELS-South and CANDELS-North (Grogin et al.\ 2011;
Koekemoer et al.\ 2011), and the ERS (Windhorst et al.\ 2011) field.
This is the most comprehensive compilation of data sets thus far
utilized to study $\beta$.  The ultimate goal of this study is to use
this information to obtain insight into the likely physical properties
of star-forming galaxies in the early universe.

A brief plan for this paper follows.  In \S2, we describe the
observational data sets we will use to study the $UV$-continuum slopes
at $z\sim4$-9 and our procedure for performing photometry.  In \S3, we
quantify the mean $\beta$'s for faint $z\sim4$-6 galaxies and then
briefly explore the implications of these $\beta$ determinations for
faint galaxies at $z\sim7$-9.  In \S4, we describe our basic
methodology for selecting $z\sim7$ galaxies and measuring their
$UV$-continuum slopes and then present our basic results.  In \S5, we
compare our results with those previously obtained in the literature,
to better understand current and past differences.  In \S6, we examine
the results for a small sample of $z\sim 8$-8.5 galaxies from the
HUDF.  In \S7, we discuss the physical implications of our results,
and finally in \S8, we conclude with a summary and prospective.  In
the appendices, we describe in detail our derivation of our
PSF-matching procedure, extensive intercomparisons of the $\beta$
measurements from this work with previous work, detailed simulations
to validate our new algorithm to measure $\beta$ at $z\sim7$, and an
alternate procedure to derive $\beta$ at $z\sim8$.  Throughout this
work, we find it convenient to quote results in terms of the
luminosity $L_{z=3}^{*}$ Steidel et al.\ (1999) derived at $z\sim3$,
i.e., $M_{1700,AB}=-21.07$, for consistency with previous studies.  We
refer to the HST F435W, F606W, F775W, F814W, F850LP, F098M, F105W,
F125W, F140W, and F160W bands as $B_{435}$, $V_{606}$, $i_{775}$,
$I_{814}$, $z_{850}$, $Y_{098}$, $Y_{105}$, $J_{125}$, $JH_{140}$, and
$H_{160}$, respectively, for simplicity.  Where necessary, we assume
$\Omega_0 = 0.3$, $\Omega_{\Lambda} = 0.7$, $H_0 =
70\,\textrm{km/s/Mpc}$.  All magnitudes are in the AB system (Oke \&
Gunn 1983).

\section{Observational Data and Photometry}

\subsection{Data Sets}

To accurately quantify the $UV$-continuum slope of galaxies as a
function of luminosity, we require observations over a wide range in
depths.  These include both the very deep observations over the XDF
that extend to very faint luminosities in 9 bands from $B_{435}$ to
$H_{160}$ and the deep, wide-area observations over CANDELS that
provide statistically-useful samples of the rarer bright galaxies.
Table~\ref{tab:obsdata} lists the data sets we consider and their
approximate depths, filters, and total area on the sky.

Our deepest data set, the $\sim$4.7 arcmin$^2$ XDF, incorporates all
available WFC3/IR observations from the HUDF09, HUDF12, and CANDELS
programs, including $\sim$100-orbit $Y_{105}$-band, $\sim$40-orbit
$J_{125}$-band, 30-orbit $JH_{140}$-band, and $\sim$85-orbit $H_{160}$
observations.  Importantly, the current XDF data set includes all of
the available observations in the $J_{125}$ and $H_{160}$ bands over
the HUDF, thereby providing the strongest possible constraints on the
$UV$-continuum slopes $\beta$ of star-forming galaxies at $z\sim4$-8.
Reductions of these data use Multidrizzle (Koekemoer et al.\ 2003),
both for the purpose of generating the full stack in the case of the
$Y_{105}$ and $JH_{140}$ band observations and for generating the two
$\sim$50\% stacks in the case of the $J_{125}$ and $H_{160}$-band
observations (see \S4.2).  Full stacks of the $J_{125}$ and
$H_{160}$-band observations are also made to allow for more optimal
determinations of $\beta$ for faint $z\sim4$-6 galaxies.  At optical
wavelengths, we utilize our new XDF reductions which take advantage of
all observations obtained over the HUDF with ACS over the last ten
years (Illingworth et al.\ 2013) and reach $\sim$0.2 mag deeper than
the 2004 release (Beckwith et al.\ 2006).  These deeper optical
observations are important for ensuring that any lower-redshift
interlopers in our samples are kept to a minimum to the faint-end
limit of the XDF observations and also for improving the $\beta$
measurements we make for faint $z\sim4$-5 galaxies.

The other deep data set, reaching to nearly HUDF depths, consists of
the very deep optical and near-IR observations over the two HUDF09
parallel fields HUDF09-1 and HUDF09-2 (Bouwens et al.\ 2011).  These
fields cover $\sim$9 arcmin$^2$ in total area and include observations
in the $V_{606}i_{775}z_{850}Y_{105}J_{125}H_{160}$ and
$B_{435}V_{606}i_{775}I_{814}z_{850}Y_{105}J_{125}H_{160}$ bands,
respectively.  These two fields are described in Bouwens et
al.\ (2011: see also Bouwens et al.\ 2012) and are $\sim$0.4-0.8 mag
shallower in the near-IR and optical than the XDF.

Finally, the shallower, but wide-area data sets, crucial for the
rarer, brighter sources, are the optical/ACS and near-IR WFC3/IR
observations over the CANDELS DEEP region of the GOODS South (all 10
epochs), the CANDELS DEEP region of the GOODS North (all 10 epochs),
and ERS field is made, including the full
$B_{435}V_{606}i_{775}I_{814}z_{850}Y_{098}Y_{105}J_{125}H_{160}$
observations where available (see Bouwens et al.\ 2012 for a
description of the reductions).  For our $z\sim7$ $\beta$
determinations, we utilize only those areas where $\geq2$ orbits of
WFC3/IR observations are available in both the $J_{125}$ and $H_{160}$
bands to ensure a sufficient number of exposures to obtain a good
reduction of the 50\% splits of the $J_{125}$ and $H_{160}$ data.  We
also only utilize those areas where WFC3/IR $Y_{098}/Y_{105}$
observations have been acquired.  The availability of
$Y_{098}/Y_{105}$-band observations over these fields is useful both
for obtaining more accurate constraints on the redshifts of the
sources under study and for making it possible to largely detect and
select sources, without significant reliance on deep data in other
bands like the $J_{125}$ band.

Zeropoints for the ACS and WFC3/IR observations are the latest values
taken from the STScI zeropoint
calculator\footnote{\texttt{http://www.stsci.edu/hst/acs/analysis/zeropoints/zpt.py}}
and from the WFC3/IR data handbook (Dressel et al.\ 2012).
Corrections for foreground dust extinction from our galaxy are
performed based on the Schlafly \& Finkbeiner (2011) maps.

\subsection{Catalog Construction and Photometry}

Our procedure for constructing catalogs and doing photometry for
sources in our fields is similar to much of our previous work (e.g.,
Bouwens et al.\ 2012).  Briefly, we run SExtractor (Bertin \& Arnouts
1996) in dual-image mode, taking the detection images to be the square
root of a $\chi^2$ image (Szalay et al.\ 1997: similar to a coadded
image) and using the PSF-matched images for photometry.  The $\chi^2$
image is constructed from the WFC3/IR bands where we expect sources to
be significantly detected (except in cases where it would bias our
$\beta$ measurements: see \S3.1 and \S4.3).  Colors are measured in
small scalable apertures using a Kron (1980) factor of 1.2.  Typical
effective radii for these small scalable apertures are $0.38''$,
$0.29''$, $0.22''$ for $z\sim7$ sources from the XDF with $\sim$27
mag, $\sim$28.5 mag, and $\sim$29.5 mag, respectively.  Fluxes in
these small scalable apertures are then corrected to total fluxes by
(1) accounting for the additional flux in a larger scalable aperture
(Kron factor 2.5) over that in a smaller scalable aperture (computed
from the square root of $\chi^2$ image) and (2) by correcting for the
light on the wings of the PSF based on the tabulated encircled energy
distribution (Table 7.7 of Dressel et al.\ 2012).

We exercised extra care in PSF-matching our observations given the
short lever arm available in wavelength to derive $UV$-continuum
slopes $\beta$ at $z\sim4$-8 (particularly $z\sim7$-8) and therefore
the sensitivity of these measurements to small systematics in the
derived colors (see \S5.2).  Normally the PSF-matching is
automatically handled with our software with encircled energy
distributions being reproduced to $\lesssim\,$2-3\%.  However, because
of the sensitivity of our results to small errors in the PSF matching,
we took a particularly rigorous approach and further optimized our
algorithm to do the PSF-matching between the $J_{125}$-band
observations and the $H_{160}$-band observations.  We constructed the
PSF in each band using the average two-dimensional profiles of a small
number of relatively high S/N, unsaturated stars from our HUDF
reductions (here 5) to construct the core of the PSF (to radii
$<$0.6$''$) and then using the tabulated encircled energy
distributions to construct the profile of stars at larger radii
$>$0.6$''$.  We verified that the encircled energy distribution we
extracted for the PSF in each band reproduced that from the tabulated
encircled energy distribution (Dressel et al.\ 2012: after accounting
for the non-zero size of the drizzle kernel) to within 1\%.  We also
compared the encircled energy distributions from the PSFs we
constructed with those from those derived by the 3D HST team (van
Dokkum et al.\ 2013; Skelton et al.\ 2014) and found agreement to
within 1\% at a radius of $\sim$0.2$''$.  Finally, we computed the
radii containing 70\% of the light for each of these PSFs and compared
them with the results from the HUDF12 team (e.g., Dunlop et
al.\ 2013).  We found excellent agreement overall (typically
$<$0.005$''$).

In using our derived PSFs to PSF-match the observations, we explicitly
solved for the kernel that when convolved with the $J_{125}$-band PSF
would reproduce the encircled-energy distribution for the
$H_{160}$-band PSF to $\lesssim$1.2\% (see Appendix A).  A similar
procedure was used for PSF-matching the observations in the other
bands to the $H_{160}$-band observations.

\begin{figure}
\epsscale{1.15}
\plotone{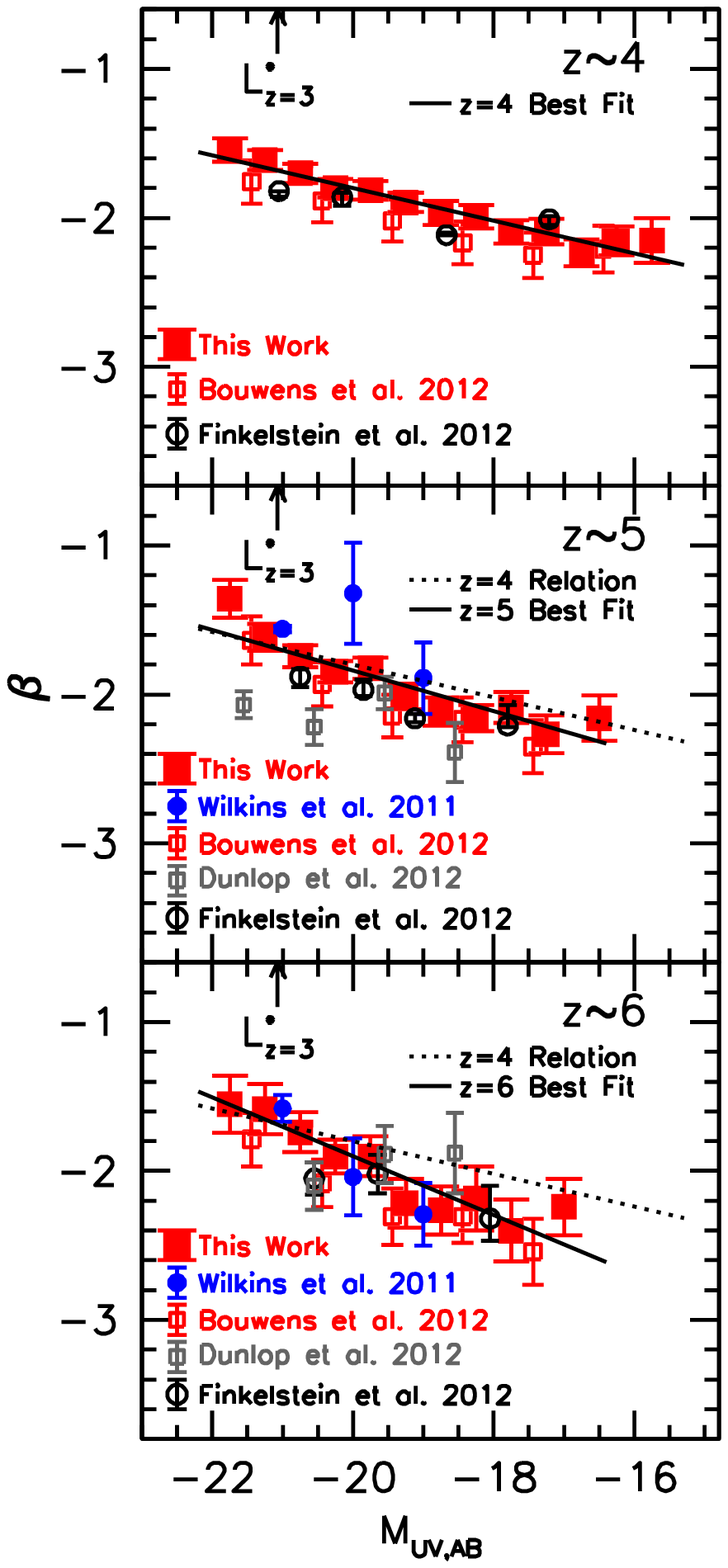}
\caption{Biweight-mean observed $\beta$'s for our $z\sim4$
  (\textit{upper panel}), $z\sim5$ (\textit{middle panel}), and
  $z\sim6$ (\textit{lower panel}) galaxy samples as a function of the
  rest-frame $UV$ luminosity $M_{UV}$ (see also
  Table~\ref{tab:medianbeta} for the binned measurements).  These
  samples take advantage of the deeper ACS+WFC3/IR observations over
  the XDF and also the wider-area CANDELS-North observations, in
  addition to the CANDELS-South, ERS, and HUDF-Ps observations that
  Bouwens et al.\ (2012) utilized.  For context, the $\beta$
  vs. $M_{UV}$ relations derived by Bouwens et al.\ (2012), Wilkins et
  al.\ (2011), Dunlop et al.\ (2012), and Finkelstein et al.\ (2012)
  are also shown.  The median $\beta$'s we quote for Finkelstein et
  al.\ (2012) for the lowest luminosity $z\sim4$ and $z\sim5$ galaxies
  rely exclusively on their $\beta$ measurements from the HUDF.  While
  Finkelstein et al.\ (2012) do not explicitly quote these median
  $\beta$'s in their paper, these median $\beta$'s can be extracted
  from the information provided and were previously presented in
  Figure 11 of Bouwens et al.\ (2012).  The solid lines give the
  best-fit linear relationship to the mean $\beta$ vs. $M_{UV}$
  relationship.  The present determination of this relationship is in
  excellent agreement with those previously derived by Bouwens et
  al.\ (2012), albeit with a slight offset to the intercept consistent
  with the estimated systematic errors (\S3.3).\label{fig:colmag456}}
\end{figure}

\section{Lower Redshift $\beta$ Results as a Baseline for Interpreting $z>6$ Results}

\subsection{$z\sim4$-6 Samples: Source Selection and $\beta$ Measurements}

Previously, Bouwens et al.\ (2012) made use of the substantial
quantity of observations over the HUDF09, ERS, and CANDELS-South
fields to conduct a thorough characterization of the $\beta$
distribution over a wide luminosity range for star-forming galaxies at
$z\sim4$, $z\sim5$, and $z\sim6$.  The availability of even deeper
near-IR observations over the HUDF region (provided by the XDF data
set) and observations over the CANDELS-North field allow us to further
refine this characterization.  An improved measurement of the $\beta$
distribution for faint galaxies at $z\sim4$-6 will also be useful for
establishing the approximate evolution in $\beta$ versus redshift and
therefore interpreting the $\beta$ results for faint galaxies at
$z\sim7$-8 (\S3.5).

We select our $z\sim4$-6 samples slightly differently for the deeper
and shallower datasets.  First, we consider the $z\sim4$-6 sample
selection for the shallower, wide-area datasets, i.e., the
CANDELS-North, CANDELS-South, or ERS data sets.  The selection
criteria we utilize are identical to what we previously utilized in
Bouwens et al.\ (2012).  These criteria are
\begin{eqnarray*}
(B_{435}-V_{606} > 1.1) \wedge (B_{435}-V_{606} > (V_{606}-z_{850})+1.1) \\
\wedge (V_{606}-z_{850}<1.6)
\end{eqnarray*}
for our $z\sim4$ sample,
\begin{eqnarray*}
[(V_{606}-i_{775} > 0.9(i_{775}-z_{850})+1.5) \vee \\
(V_{606}-i_{775} > 2))] \wedge \\ 
(V_{606}-i_{775}>1.2) \wedge (i_{775}-z_{850}<0.8)
\end{eqnarray*}
for our $z\sim5$ selection, and
\begin{displaymath}
(i_{775}-z_{850}>1.3)\wedge(z_{850}-J_{125}<0.9)
\end{displaymath}
where $\wedge$ and $\vee$ represent the logical \textbf{AND} and
\textbf{OR} symbols, respectively.  Sources in our $z\sim5$ selections
must be detected at less than $2\sigma$ in the $B_{435}$-band
observations.  Similarly, we require that sources in our $z\sim6$
selections be undetected at $2\sigma$ in the $B_{435}$-band
observations and that the sources be either undetected in the
$V_{606}$-band observations or have a $V_{606}-z_{850}$ color redder
than 2.8 (identical to the criteria utilized by Bouwens et al.\ 2006
in selecting $z\sim6$ sources).  Sources are required to be detected
at $>$5.5$\sigma$ in the $\chi^2$ image we generated from the
$Y_{105}J_{125}H_{160}$ imaging observations (or
$Y_{098}J_{125}H_{160}$ imaging observations for our ERS samples).

\begin{deluxetable*}{ccccc}
\tablewidth{16.7cm}
\tabletypesize{\footnotesize}
\tablecaption{The biweight mean, median, and inverse-variance-weighted mean $\beta$ for $z\sim4$-8 galaxies, as a function of $UV$ luminosity $M_{UV}$\label{tab:medianbeta}}
\tablehead{
\colhead{} & \colhead{} & \colhead{} & \colhead{Inverse-Variance} \\
\colhead{$<M_{UV,AB}>$\tablenotemark{a}} & \colhead{Biweight Mean $\beta$\tablenotemark{b,c}} & \colhead{Median $\beta$\tablenotemark{c}} & \colhead{Weighted Mean $\beta$\tablenotemark{c}} & \colhead{\# of Sources\tablenotemark{d}}}
\startdata
\multicolumn{5}{c}{\textbf{z$\,\sim\,$4 (\S3.2)}} \\
$-$21.75 & $-$1.54$\pm$0.07$\pm$0.06 & $-$1.49$_{-0.10}^{+0.08}$$\pm$0.06  & $-$1.42$\pm$0.05$\pm$0.06 & 54\\
$-$21.25 & $-$1.61$\pm$0.04$\pm$0.06 & $-$1.61$_{-0.05}^{+0.04}$$\pm$0.06  & $-$1.52$\pm$0.03$\pm$0.06 & 141\\
$-$20.75 & $-$1.70$\pm$0.03$\pm$0.06 & $-$1.70$_{-0.02}^{+0.07}$$\pm$0.06  & $-$1.57$\pm$0.02$\pm$0.06 & 285\\
$-$20.25 & $-$1.80$\pm$0.02$\pm$0.06 & $-$1.81$_{-0.03}^{+0.04}$$\pm$0.06  & $-$1.71$\pm$0.02$\pm$0.06 & 457\\
$-$19.75 & $-$1.81$\pm$0.03$\pm$0.06 & $-$1.81$_{-0.03}^{+0.04}$$\pm$0.06  & $-$1.74$\pm$0.02$\pm$0.06 & 552\\
$-$19.25 & $-$1.90$\pm$0.02$\pm$0.06 & $-$1.88$_{-0.02}^{+0.03}$$\pm$0.06  & $-$1.85$\pm$0.02$\pm$0.06 & 586\\
$-$18.75 & $-$1.97$\pm$0.06$\pm$0.06 & $-$1.96$_{-0.06}^{+0.05}$$\pm$0.06  & $-$1.90$\pm$0.05$\pm$0.06 & 57\tablenotemark{e}\\
$-$18.25 & $-$1.99$\pm$0.06$\pm$0.06 & $-$1.98$_{-0.09}^{+0.07}$$\pm$0.06  & $-$1.97$\pm$0.05$\pm$0.06 & 70\tablenotemark{e}\\
$-$17.75 & $-$2.09$\pm$0.08$\pm$0.06 & $-$2.00$_{-0.12}^{+0.06}$$\pm$0.06  & $-$1.92$\pm$0.05$\pm$0.06 & 94\\
$-$17.25 & $-$2.09$\pm$0.07$\pm$0.06 & $-$2.07$_{-0.08}^{+0.03}$$\pm$0.06  & $-$2.04$\pm$0.06$\pm$0.06 & 69\\
$-$16.75 & $-$2.23$\pm$0.10$\pm$0.06 & $-$2.25$_{-0.11}^{+0.07}$$\pm$0.06  & $-$2.10$\pm$0.07$\pm$0.06 & 86\\
$-$16.25 & $-$2.15$\pm$0.12$\pm$0.06 & $-$2.19$_{-0.10}^{+0.12}$$\pm$0.06  & $-$1.94$\pm$0.08$\pm$0.06 & 96\\
$-$15.75 & $-$2.15$\pm$0.12$\pm$0.06 & $-$2.16$_{-0.07}^{+0.12}$$\pm$0.06  & $-$1.95$\pm$0.14$\pm$0.06 & 53\\
\multicolumn{5}{c}{\textbf{z$\,\sim\,$5 (\S3.2)}} \\
$-$21.75 & $-$1.36$\pm$0.48$\pm$0.06 & $-$1.14$_{-0.28}^{+0.61}$$\pm$0.06  & $-$0.72$\pm$0.11$\pm$0.06 & 12\\
$-$21.25 & $-$1.62$\pm$0.11$\pm$0.06 & $-$1.55$_{-0.11}^{+0.13}$$\pm$0.06  & $-$1.44$\pm$0.07$\pm$0.06 & 35\\
$-$20.75 & $-$1.74$\pm$0.05$\pm$0.06 & $-$1.77$_{-0.03}^{+0.07}$$\pm$0.06  & $-$1.61$\pm$0.04$\pm$0.06 & 83\\
$-$20.25 & $-$1.85$\pm$0.05$\pm$0.06 & $-$1.88$_{-0.02}^{+0.06}$$\pm$0.06  & $-$1.75$\pm$0.04$\pm$0.06 & 134\\
$-$19.75 & $-$1.82$\pm$0.04$\pm$0.06 & $-$1.82$_{-0.05}^{+0.05}$$\pm$0.06  & $-$1.82$\pm$0.04$\pm$0.06 & 150\\
$-$19.25 & $-$2.01$\pm$0.07$\pm$0.06 & $-$2.00$_{-0.04}^{+0.10}$$\pm$0.06  & $-$1.99$\pm$0.06$\pm$0.06 & 72\\
$-$18.75 & $-$2.12$\pm$0.10$\pm$0.06 & $-$2.15$_{-0.05}^{+0.12}$$\pm$0.06  & $-$2.08$\pm$0.07$\pm$0.06 & 38\tablenotemark{e}\\
$-$18.25 & $-$2.16$\pm$0.09$\pm$0.06 & $-$2.19$_{-0.10}^{+0.12}$$\pm$0.06  & $-$2.08$\pm$0.07$\pm$0.06 & 58\tablenotemark{e}\\
$-$17.75 & $-$2.09$\pm$0.10$\pm$0.06 & $-$2.02$_{-0.14}^{+0.07}$$\pm$0.06  & $-$2.02$\pm$0.09$\pm$0.06 & 38\\
$-$17.25 & $-$2.27$\pm$0.14$\pm$0.06 & $-$2.26$_{-0.12}^{+0.09}$$\pm$0.06  & $-$2.33$\pm$0.11$\pm$0.06 & 31\\
$-$16.50 & $-$2.16$\pm$0.17$\pm$0.06 & $-$2.20$_{-0.15}^{+0.12}$$\pm$0.06  & $-$2.34$\pm$0.14$\pm$0.06 & 26\\
\multicolumn{5}{c}{\textbf{z$\,\sim\,$6 (\S3.2)}} \\
$-$21.75 & $-$1.55$\pm$0.17$\pm$0.08 & $-$1.53$_{-0.25}^{+0.18}$$\pm$0.08  & $-$1.59$\pm$0.17$\pm$0.08 & 6\\
$-$21.25 & $-$1.58$\pm$0.10$\pm$0.08 & $-$1.55$_{-0.15}^{+0.09}$$\pm$0.08  & $-$1.60$\pm$0.15$\pm$0.08 & 10\\
$-$20.75 & $-$1.74$\pm$0.10$\pm$0.08 & $-$1.73$_{-0.12}^{+0.06}$$\pm$0.08  & $-$1.71$\pm$0.11$\pm$0.08 & 23\\
$-$20.25 & $-$1.90$\pm$0.09$\pm$0.08 & $-$1.88$_{-0.07}^{+0.09}$$\pm$0.08  & $-$1.81$\pm$0.08$\pm$0.08 & 53\\
$-$19.75 & $-$1.90$\pm$0.13$\pm$0.08 & $-$1.93$_{-0.06}^{+0.14}$$\pm$0.08  & $-$1.91$\pm$0.11$\pm$0.08 & 37\\
$-$19.25 & $-$2.22$\pm$0.18$\pm$0.08 & $-$2.17$_{-0.22}^{+0.15}$$\pm$0.08  & $-$2.24$\pm$0.14$\pm$0.08 & 12\\
$-$18.75 & $-$2.26$\pm$0.14$\pm$0.08 & $-$2.23$_{-0.10}^{+0.10}$$\pm$0.08  & $-$2.19$\pm$0.13$\pm$0.08 & 17\\
$-$18.25 & $-$2.19$\pm$0.22$\pm$0.08 & $-$2.12$_{-0.08}^{+0.44}$$\pm$0.08  & $-$2.07$\pm$0.19$\pm$0.08 & 11\\
$-$17.75 & $-$2.40$\pm$0.30$\pm$0.08 & $-$2.45$_{-0.27}^{+0.55}$$\pm$0.08  & $-$2.28$\pm$0.18$\pm$0.08 & 17\\
$-$17.00 & $-$2.24$\pm$0.20$\pm$0.08 & $-$2.25$_{-0.18}^{+0.17}$$\pm$0.08  & $-$2.19$\pm$0.16$\pm$0.08 & 25\\
\multicolumn{5}{c}{\textbf{z$\,\sim\,$7 (\S4.7)}} \\
$-$21.25 & $-$1.75$\pm$0.18$\pm$0.13 & $-$1.74$_{-0.12}^{+0.14}$$\pm$0.13  & $-$1.67$\pm$0.10$\pm$0.13 & 26\\
$-$19.95 & $-$1.89$\pm$0.13$\pm$0.13 & $-$1.88$_{-0.08}^{+0.07}$$\pm$0.13  & $-$1.86$\pm$0.08$\pm$0.13 & 102\\
$-$18.65 & $-$2.30$\pm$0.18$\pm$0.13 & $-$2.66$_{-0.08}^{+0.20}$$\pm$0.13  & $-$2.39$\pm$0.14$\pm$0.13 & 43\\
$-$17.35 & $-$2.42$\pm$0.28$\pm$0.13 & $-$2.15$_{-0.21}^{+0.17}$$\pm$0.13  & $-$2.39$\pm$0.39$\pm$0.13 & 13\\
\multicolumn{5}{c}{\textbf{z$\,\sim\,$8 (\S6.2)}} \\
$-$19.95 & $-$2.30$\pm$0.01$\pm$0.27\tablenotemark{f} & $-$2.30$_{-0.01}^{+0.01}$$\pm$0.27\tablenotemark{f}  
& $-$2.30$\pm$0.54$\pm$0.27 & 2\\
$-$18.65 & $-$1.41$\pm$0.60$\pm$0.27 & $-$1.51$_{-1.07}^{+0.14}$$\pm$0.27  & $-$1.88$\pm$0.74$\pm$0.27 & 4\\
\multicolumn{5}{c}{\textbf{z$\,\sim\,$8.5 (\S6.3)}} \\
$-$18.50 & $-$2.06$\pm$0.51$\pm$0.27 & $-$1.82$_{-0.73}^{+0.73}$$\pm$0.27  & $-$2.07$\pm$0.92$\pm$0.27 & 2
\enddata
\tablenotetext{a}{Each $UV$ luminosity bin probes a 0.5-mag range for
  our $z\sim4$ selection, our brighter ($M_{UV}<-17$) $z\sim5$ and
  ($M_{UV}<-18$) $z\sim6$ sources, a 1-mag range for our faintest
  $z\sim5$ and $z\sim6$ sources, and a 1.3-mag range for sources in
  our $z\sim7$ and $z\sim8$ samples.}
\tablenotetext{b}{The biweight mean is our preferred method for quoting 
the central value for the $\beta$ distribution.}
\tablenotetext{c}{Both random and systematic errors are quoted
  (presented first and second, respectively).}
\tablenotetext{d}{To better represent the \# of real sources
  incorporated in our $\beta$ measurements (given that we select
  $z\sim7$ sources on each field twice), we divide the number of selected
  $z\sim7$ sources by two.}
\tablenotetext{e}{The biweight mean over this luminosity interval
  $-19<M_{UV,AB}<-18$ can also be derived based on the 476 sources
  with these luminosities over the full data set, and it is
  $-1.90\pm0.03$.  We elected to use the XDF+HUDF09-2 measurement here
  since the systematics will be smaller.}
\tablenotetext{f}{Since the two bright sources in this luminosity subsample have 
essentially the same measured $\beta$, the error we derive on the biweight 
mean/median from bootstrap resampling is clearly too low.}
\end{deluxetable*}

The catalogs we utilize, in applying the above criteria, were derived
purely from the optical/ACS $B_{435}V_{606}i_{775}I_{814}z_{850}$
observations (PSF-matched to the $z_{850}$ band) where possible.  This
ensures that the color measurements we use to select our $z\sim4$-6
samples have very high S/N.

Second, we consider source selection over our deepest data sets, i.e.,
XDF, HUDF09-1, and HUDF09-2.  We have revised the criteria we utilize
from our shallower data sets.  This was necessary to ensure that
source selection is entirely decoupled from the measurement of
$\beta$, so that we can obtain an unbiased measurement of $\beta$ to
the limit of our $z\sim4$-6 XDF+HUDF09-Ps samples.  A coupling of
source selection with the measurement of $\beta$ was an issue with
previous $z\sim7$ samples (Bouwens et al.\ 2010: see Appendix B.2),
and so we must take special care to avoid it here (see Dunlop et
al.\ 2012; Bouwens et al.\ 2012).

The selection criteria we settled upon for our deeper fields are
\begin{eqnarray*}
(B_{435}-V_{606} > 1.1) \wedge (V_{606}-Y_{105}<1.3) \wedge \\
(V_{606}-z_{850}<1.6)
\end{eqnarray*}
for our $z\sim4$ sample,
\begin{displaymath}
(V_{606}-i_{775} > 1.2) \wedge (i_{775}-J_{125}<0.9)
\end{displaymath}
for our $z\sim5$ selection, and
\begin{displaymath}
(i_{775}-z_{850}>1.0)\wedge(z_{850}-J_{125}<0.5)
\end{displaymath}
for our $z\sim6$ selection.  When applying color criteria involving
only ACS observations, smaller-aperture color measurements
(PSF-matched to the $z_{850}$-band data) were utilized on special
0.03$''$-reductions generated for the XDF, HUDF09-1, and HUDF09-2 data
sets to ensure more optimal results.

Source selection and photometry for our $z\sim4$, $z\sim5$, and
$z\sim6$ samples is based on the square root of $\chi^2$ image (Szalay
et al.\ 1999) constructed from the
$V_{606}Y_{105}JH_{140}H_{160}$-band, $J_{125}JH_{140}$-band, and
$J_{125}JH_{140}$-band images, respectively.  The $Y_{105}$ or
$H_{160}$ band images were not used in constructing the $\chi^2$ image
for our $z\sim5$ and $z\sim6$ samples (or the $J_{125}$-band image for
our $z\sim4$ samples), given our use of these images to derive
$\beta$.  Sources are required to be detected at $5\sigma$ in the
$\chi^2$ image to ensure that they are real.

Using the same simulations as provided in Appendix F, we verified that
the above selection criteria resulted in very similar mean redshifts
for our $z\sim4$, $z\sim5$, and $z\sim6$ selections from the deep
XDF+HUDF09-Ps data sets as for our shallower ERS+CANDELS selections
since the respective selection criteria differed only slightly.  The
mean redshifts we found for $z\sim5$ and $z\sim6$ for both selections
were identically equal to 5.0 and 5.9, respectively, while for our
$z\sim4$ selections, the mean redshifts were equal to 3.8 (for our
ERS+CANDELS selections) and 4.0 (for our XDF+HUDF09-Ps selection).

\begin{figure}
\epsscale{1.15}
\plotone{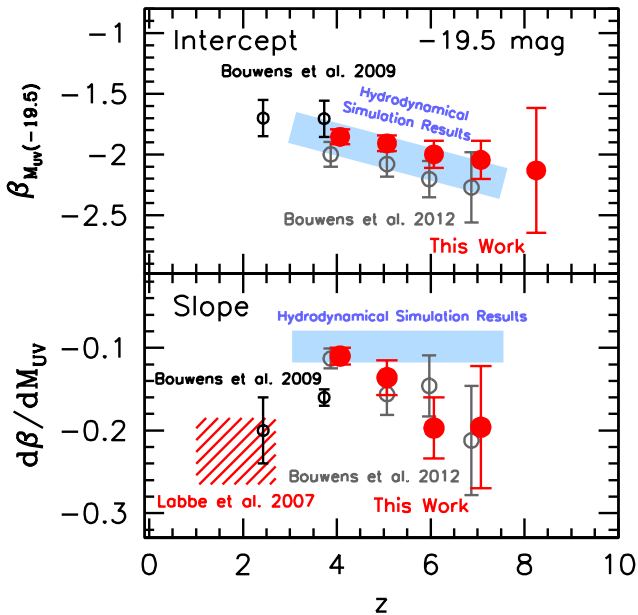}
\caption{Slope and Intercept of the $\beta$-$M_{UV}$ color-luminosity
  relationship for star-forming galaxies at various redshifts (see
  also Table~\ref{tab:slopes}).  (\textit{Upper Panel}) Intercept of
  the $\beta$-$M_{UV}$ color-luminosity relationship at a fixed $UV$
  luminosity ($M_{UV}$) of $-19.5$ AB mag.  The present determination
  is shown with the large red circles, while our earlier
  determinations at $z\sim2.5$-4 and $z\sim4$-6 (Bouwens et al.\ 2009;
  Bouwens et al.\ 2012) are shown with smaller open black circles and
  gray circles.  The mean $\beta$ derived at fixed luminosity
  ($M_{UV,AB}=-19.5$) shows some reddening with cosmic time, but the
  evolution is not large.  The derived trend is in remarkably good
  agreement with the predictions of the hydrodynamical
  simulations from Finlator et al.\ (2011: \textit{light blue shaded
    region}).  (\textit{Lower Panel}) Slope of the $\beta$-$M_{UV}$
  color-luminosity relationship.  This relation is such that brighter
  galaxies exhibit redder $\beta$'s than fainter galaxies.  The
  symbols are as in the upper panel.  The slope to this relation at
  $z\sim1.0$-2.7, as inferred by Labb{\'e} et al.\ 2007, is also shown
  with the red cross-hatched region (see Figure 7 from Bouwens et
  al.\ 2012).  The slope of the $\beta$-$M_{UV}$ relation does not
  show statistically significant evolution with cosmic time (see also
  Bouwens et al.\ 2012).\label{fig:slope}}
\end{figure}

The total number of sources that satisfied our $z\sim4$, $z\sim5$, and
$z\sim6$ criteria were 2925, 670, and 210, respectively.  $\beta$'s
are derived for individual sources in our $z\sim4$, $z\sim5$, and
$z\sim6$ samples using the same power-law fits ($f_{\lambda} \propto
\lambda^{\beta}$) used for our $z\sim7$ samples (see also Bouwens et
al.\ 2012).  For sources in the ERS, CANDELS-North, and CANDELS-South
fields, $\beta$'s are then derived for $z\sim4$, $z\sim5$, and
$z\sim6$ galaxies from power-law fits to the photometry in the
$i_{775}I_{814}z_{850}Y_{105}J_{125}$, $z_{850}Y_{105}J_{125}H_{160}$,
and $Y_{105}J_{125}H_{160}$ bands, respectively.  Where available
(e.g., over the ERS field), use of the $Y_{098}$-band photometry was
made.  All the photometry used for the $\beta$ determinations are
PSF-matched to the $H_{160}$ band.  By selecting sources using
photometry in smaller (and different) apertures than the photometry we
use to derive $\beta$, we are able to ensure that our $\beta$ are much
more immune to noise-driven systematic biases, such as the photometric
error coupling bias discussed in Appendix B.1.2 of Bouwens et
al. (2012).

For sources in the deeper XDF, HUDF09-1, and HUDF09-2 data sets, the
flux measurements are made in the $i_{775}I_{814}z_{850}J_{125}$ bands
for our $z\sim4$ sample, in the $z_{850}Y_{105}H_{160}$ bands for our
$z\sim5$ sample, and in the $Y_{105}H_{160}$ bands for our $z\sim6$
sample.  We restrict our fits to these bands to ensure that there is
no coupling between the selection of sources and the determination of
$\beta$, therefore ensuring that noise-driven biases are identically
zero.  Similar to our procedure on our shallower data sets, all the
photometry used for the $\beta$ determinations are performed from
images PSF-matched to the $H_{160}$ band.

\begin{figure*}
\epsscale{1.15}
\plotone{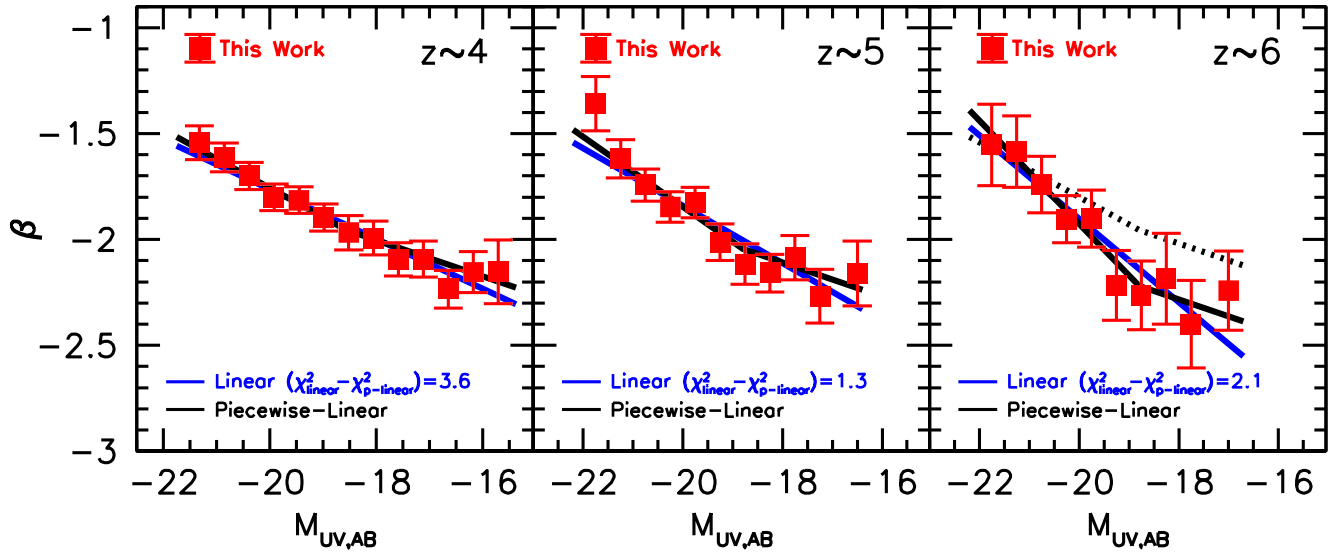}
\caption{Comparison of simple linear (\textit{blue lines}) and
  piecewise-linear fits (\textit{black lines}) to our $z\sim4$
  (\textit{left}), $z\sim5$ (\textit{center}), $z\sim6$ (\textit{right
    panels}) $\beta$ vs. $M_{UV}$ results (see \S3.4).  Our
  piecewise-linear fits are made over two linear segments, with the
  slope of the faint piece fixed to $d\beta/dM_{UV}=-0.08$.  For
  simplicity, we have kept the break luminosity fixed in our
  piecewise-linear models to the average best-fit break luminosity
  found at $z\sim4$, $z\sim5$, and $z\sim6$, which is $-18.8\pm0.2$
  mag.  The slope that we use faintward of the break luminosity
  $d\beta/dM_{UV}$ is taken to be equal the average best-fit slope we
  found for our $\beta$ results at $z\sim4$, $z\sim5$, and $z\sim6$
  (when allowing this slope to vary).  The best-fit parameters we find
  for our piecewise-linear model are given in Table~\ref{tab:bslopes}.
  The reduced $\chi^2$ values we compute with our piecewise-linear
  models provide a significantly better fit to the observed $\beta$
  vs. $M_{UV}$ relationship than simple linear fits (96\%, 75\%, and
  85\% confidence at $z\sim4$, $z\sim5$, and $z\sim6$, respectively).
  The present fit results suggest that the mean $\beta$'s for galaxies
  at very low luminosities, i.e., $M_{UV,AB}>-19$, show a weaker
  dependence on $UV$ luminosity than at higher luminosities.  This
  result is strikingly similar to the $z\sim4$ results of Oesch et al.
  (2013a) who also found that $\beta$ showed a steep dependence on the
  rest-frame optical luminosity for the most luminous galaxies and
  then a dramatic flattening to this relationship at lower
  luminosities.  Comparing the best-fit results at $z\sim4$
  (\textit{dotted black line in right panel}) with those at $z\sim6$,
  we see clear evidence for evolution in the mean $\beta$'s for faint
  ($M_{UV,AB}>-18.8$) galaxies (see also
  Figure~\ref{fig:faintevol}).\label{fig:brokenline}}
\end{figure*}

We bin galaxies as a function of their rest-frame $UV$ luminosity, as
we did for our $z\sim7$ samples.  We take the $UV$ luminosity to be
equal to the geometric mean of the magnitude measurements used to
derive $\beta$ to minimize the effect of noise in introducing an
artificial correlation between $\beta$ and $UV$ luminosity, as we did
in Bouwens et al.\ (2009) and Bouwens et al.\ (2012).  To ensure that
our prescription for deriving the $UV$ luminosity did not
substantially bias our $\beta$ results, we also examined the $\beta$
vs. $UV$ luminosity results defining the $UV$ luminosity in terms of
the flux in bands used to select the sources, i.e., the $Y_{105}$-band
flux for $z\sim4$ sources and the $J_{125}$ or $J_{125}+JH_{140}$ flux
for sources in our $z\sim5$-6 samples.  The results are briefly
presented in Appendix E; we find no significant differences relative
to our primary results.

In mapping out the mean $\beta$ versus $M_{UV}$ relationship, it is
important to have a sufficient number of sources in each luminosity
bin to determine the mean given the considerable scatter in the
intrinsic $\beta$ distribution (Bouwens et al.\ 2009, 2012; Castellano
et al.\ 2012; Rogers et al.\ 2014).  We therefore rely on our
CANDELS+ERS samples brightward of $-19$ mag and XDF/HUDF09-Ps samples
faintward of these limits.  Similar to the exercise shown in Figure 25
of Bouwens et al.\ (2012), we carefully compared the $\beta$'s derived
from the deeper data sets with the shallower data sets to ensure that
no large systematics are present between our data sets.

\subsection{$\beta$ Results for $z\sim4$-6 Samples}

Our biweight mean $\beta$ results for our $z\sim4$, $z\sim5$, and
$z\sim6$ samples are presented in Figure~\ref{fig:colmag456} and
Table~\ref{tab:medianbeta}.  Very small ($\Delta\beta\sim0.01$,
$\Delta\beta\sim0.02$, and $\Delta\beta\sim0.01$) redward corrections
were made to our biweight mean, median, and inverse-variance-weighted
mean $\beta$ results at $z\sim4$, $z\sim5$, and $z\sim6$,
respectively, to correct for the fact that our color criteria are less
efficient at selecting sources with intrinsically red $\beta$'s (see
Appendix F).  We also made a small ($\Delta\beta\sim0.02$) blueward
correction to $z\sim5$ $\beta$ results from our wide-area ERS+CANDELS
samples to correct for a slight coupling bias between our ERS+CANDELS
$z\sim5$ selections and the $\beta$ measurements we make in these data
sets (see Appendix B.1.2 of Bouwens et al.\ 2012 for a description of
the relevant simulations).

Current constraints on the mean $\beta$'s of galaxies extend to an
unprecedented $-15.5$ mag at $z\sim4$, $-16.5$ mag at $z\sim5$, and
$-17.0$ mag at $z\sim6$ thanks to the superb depth of the
optical+near-IR observations over our XDF data set.  Median and
inverse-variance-weighted mean $\beta$ are also provided for our
$z\sim4$-6 samples in Table~\ref{tab:medianbeta}, as a function of
$UV$ luminosity.  For context, the recent results of Bouwens et
al.\ (2012), Wilkins et al.\ (2011), Dunlop et al.\ (2012), and
Finkelstein et al.\ (2012) are also included in
Figure~\ref{fig:colmag456}.

Based on our $\beta$ results for our galaxy samples at $z\sim4$,
$z\sim5$, and $z\sim6$, we present the best-fit slope and intercept to
the $\beta$ vs. $M_{UV}$ relationship in Figure~\ref{fig:slope} and
Table~\ref{tab:slopes}.  In Figure~\ref{fig:slope}, we compare the
observed trends with our new $\beta$ results at $z\sim7$ (\S4), the
$\beta$ results at $z\sim1$-6 from the literature (Labb{\'e} et
al.\ 2007; Bouwens et al.\ 2009; Bouwens et al.\ 2012), and also with
the hydrodynamical simulation results from Finlator et al.\ (2011).
Encouragingly enough, almost identical trends are found between
$\beta$ and $UV$ luminosity in large cosmological hydrodynamical
simulations (e.g., Finlator et al.\ 2011).

The observed correlation of $\beta$ with luminosity is found to be
even stronger when viewed at rest-frame optical wavelengths with
Spitzer/IRAC (Oesch et al.\ 2013a).  This is likely a manifestation of
the well-known mass-metallicity relation (e.g., Tremonti et al.\ 2004;
Erb et al.\ 2006a; Maiolino et al.\ 2008) in galaxies at $z\gtrsim4$,
with $UV$ luminosity roughly tracing with mass and $\beta$ tracing
dust extinction and metallicity (Meurer et al.\ 1999; Bouwens et
al.\ 2009; Bouwens et al.\ 2012).

\begin{deluxetable}{ccccc}
\tablewidth{0cm}
\tabletypesize{\footnotesize}
\tablecaption{The Best-Fit Slope and Intercept of $\beta$ - $M_{UV}$ color-luminosity relationship (see also Figure~\ref{fig:slope}).\label{tab:slopes}}
\tablehead{\colhead{Sample} & \colhead{$<z>$} & \colhead{$\beta_{M_{UV}}=-19.5$\tablenotemark{a}} & \colhead{$d\beta/dM_{UV}$} &  \colhead{Ref\tablenotemark{b}}}
\startdata
$U_{300}$ & 2.5 & $-1.70\pm0.07\pm0.15$ & $-0.20\pm0.04$ & [1]\\
$B_{435}$ & 3.8 & $-1.85\pm0.01\pm0.06$ & $-0.11\pm0.01$ & [2]\\
$V_{606}$ & 5.0 & $-1.91\pm0.02\pm0.06$ & $-0.14\pm0.02$ & [2]\\
$i_{775}$ & 5.9 & $-2.00\pm0.05\pm0.08$ & $-0.20\pm0.04$ & [2]\\
$z_{850}$ & 7.0 & $-2.05\pm0.09\pm0.13$ & $-0.20\pm0.07$ & [2]\\
$Y_{105}$ & 8.0 & $-2.13\pm0.44\pm0.27$ & $-0.15$ (fixed) & [2]
\enddata
\tablenotetext{a}{Both random and systematic errors are quoted
  (presented first and second, respectively).}
\tablenotetext{b}{References: [1] Bouwens et al. 2009, [2] This Work}
\end{deluxetable}

\begin{deluxetable}{cccc}
\tablewidth{0cm}
\tabletypesize{\footnotesize}
\tablecaption{The Best-Fit Slope and Intercept to $\beta$ - $M_{UV}$ color-luminosity relationship derived here using our piecewise-linear fits (see also Figure~\ref{fig:brokenline} and \S3.4).\label{tab:bslopes}}
\tablehead{\colhead{Sample} & \colhead{$<z>$} & \colhead{$\beta_{M_{UV}}=-18.8$\tablenotemark{a}} & \colhead{$d\beta/dM_{UV}$\tablenotemark{b}}}
\startdata
$B_{435}$ & 3.8 & $-1.95\pm0.03\pm0.06$ & $-0.13\pm0.02$\\
$V_{606}$ & 5.0 & $-2.05\pm0.05\pm0.06$ & $-0.17\pm0.04$\\
$i_{775}$ & 5.9 & $-2.22\pm0.12\pm0.08$ & $-0.24\pm0.07$
\enddata
\tablenotetext{a}{Both random and systematic errors are quoted
  (presented first and second, respectively).}
\tablenotetext{b}{The $d\beta/dM_{UV}$ parameter provided here gives
  the dependence of $\beta$ on the $UV$ luminosity brightward of the
  break luminosity $-18.8$ mag.  Faintward of this, the dependence of
  $\beta$ on the $UV$ luminosity is taken to be equal to
  $-0.08$ which is the average best-fit value found for this
  dependence for our $z\sim4$, $z\sim5$, and $z\sim6$ samples.}
\end{deluxetable}

We also include our $z\sim8$ results from \S6, though we note that
they are quite uncertain and do not provide a useful constraint.
Consistent with our previous findings (Bouwens et al.\ 2009, 2012), we
observe a similar slope to the $\beta$-$M_{UV}$ relationship at all
redshifts under examination, with the most luminous galaxies being the
reddest while the lowest luminosity galaxies are generally the bluest
(as shown directly in Figure~\ref{fig:colmag456}).

\subsection{Comparison with Previous Results}

As can be seen in Figure~\ref{fig:colmag456}, the new $\beta$
determinations and those in the literature are broadly consistent
(Wilkins et al.\ 2011; Bouwens et al.\ 2012; Finkelstein et al.\ 2012;
Dunlop et al.\ 2012).  The Dunlop et al.\ (2012) $z\sim5$ $\beta$
determinations are clearly quite a bit bluer for the highest
luminosity galaxies, but Dunlop et al.\ (2012) do not likely probe
sufficient area with their study to include a statistically
representative number of the reddest, high luminosity galaxies at
$z\sim5$.

The strong correlation we observe between $\beta$ and $UV$ luminosity
is in excellent agreement with what was previously found in Bouwens et
al.\ (2012: see also Bouwens et al.\ 2009), Wilkins et al.\ (2011) and
Finkelstein et al.\ (2012: using their HUDF measurements to define the
$\beta$ dependence to lower luminosities).

\begin{figure}
\epsscale{1.15}
\plotone{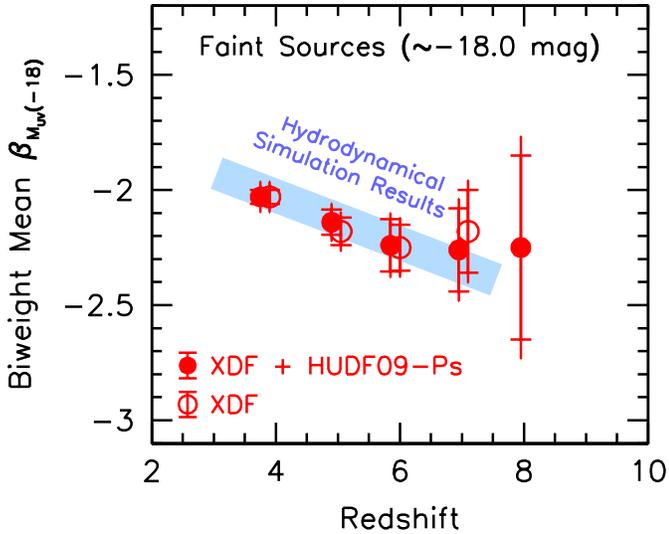}
\caption{The biweight mean $UV$-continuum slope $\beta$ observed for
  faint galaxies in our XDF + HUDF09-Ps $z\sim4$, $z\sim5$, $z\sim6$,
  $z\sim7$, and $z\sim8$ samples (red solid circles).  Only sources in
  the luminosity range $-19.0<M_{UV,AB}<-17.0$ are included in
  computing the means for the $z\sim4$, $z\sim5$, and $z\sim6$
  samples.  For our $z\sim7$ selection, sources in the somewhat more
  extended luminosity range $-19.3<M_{UV,AB}<-16.7$ are considered (to
  increase the signal-to-noise on our measurement somewhat).
  $1\sigma$ uncertainties on each of the determinations are also
  shown, for the statistical uncertainties alone (\textit{including
    hashes at the ends}) and including the systematic uncertainties
  (\textit{not including hashes}).  The $z\sim8$ results that are
  presented are based on an extrapolation of the $\beta$-$M_{UV}$
  relationship that we determine (see Table~\ref{tab:slopes}).  The
  blue open circles show our biweight mean $\beta$ determinations from
  the XDF data set alone.  The expectations from hydrodynamical
  simulations (Finlator et al.\ 2011) are shown for comparison with
  the thick light-blue line.  $\gtrsim$60\% of the evolution in
  $\beta$ in the Finlator et al.\ (2011) simulations occurs due to
  changes in the dust extinction (Bouwens et al.\ 2012; Finkelstein et
  al.\ 2012).\label{fig:faintevol}}
\end{figure}

Nevertheless, we do note a clear offset in the $\beta$ vs. $M_{UV}$
relationship relative to what we reported earlier in Bouwens et
al.\ (2012).  At fixed luminosity and redshift, the $\beta$'s we find
are systematically redder by $\Delta\beta\sim0.13$-0.19.  While our
new results are consistent with our older results given the large
systematic uncertainties we quoted on the derive $\beta$'s (Bouwens et
al.\ 2012), the present results do represent a modest shift in our
best-fit $\beta$ determinations.  In general, this shift brings our
derived $\beta$ measurements into better agreement with other
determinations in the literature (Figure~\ref{fig:colmag456}).

The offset in the $\beta$ vs. $M_{UV}$ relationship arises from the
systematically redder $\beta$'s ($\Delta\beta\sim0.10$-0.15) we
measure at all redshifts.  As we explain in Appendix B.3, this
occurred due to the empirical PSFs from Bouwens et al.\ (2012) not
containing sufficient light on the wings.  As a result, during the
PSF-matching process, light in the bluer bands was not sufficiently
smoothed to match the $H_{160}$-band PSF.  This resulted in a
systematic bias towards bluer $\beta$'s at all redshifts.  Other
potentially contributing factors are (1) the fact that the mean
$\beta$'s we measure for bright galaxies in the CANDELS-North field
are slightly redder ($\Delta\beta\sim0.05$) in the mean than what we
measure over the CANDELS-South and (2) the corrections that Bouwens et
al.\ (2012) performed to remove the photometric error coupling bias
appear to have been too large (although these corrections had a size
of just $\Delta\beta\sim0.025$ to $\Delta\beta\sim0.05$).  The latter
issue is no longer a concern for the present analysis, since we have
constructed our faint samples such that the photometric error coupling
bias is identically zero (see \S3.1 and \S4.2).

\subsection{Evidence for a Much Weaker Dependence of $\beta$ on $UV$ Luminosity for Faint $z\sim4$-6 Galaxies?}

The superb depth of the optical+near-IR observations and improved
techniques allow us to study the $UV$ slopes $\beta$ of faint $z\geq4$
galaxies at even lower luminosities than in previous work, reaching to
$-15.5$ mag ($0.006L_{z=3}^{*}$) at $z\sim4$, $-16.5$ mag
($0.014L_{z=3}^{*}$) at $z\sim5$, and $-17.0$ mag ($0.025L_{z=3}^{*}$)
at $z\sim7$.  Key questions for these faint galaxy samples include
what is the mean $\beta$ and how does $\beta$ depend on $UV$
luminosity (and therefore likely mass).

In our previous work and in \S3.2, we modeled the dependence of
$\beta$ on $UV$ luminosity using a simple two-parameter linear
relationship.  Such a model was required to capture the clear trend in
the mean $\beta$'s from very red values for the highest luminosity
systems to very blue values at lower luminosities.  As
Figure~\ref{fig:colmag456} illustrates, a simple linear relation
largely captures the observed trends in $\beta$ with $UV$ luminosity.

However, looking at the relationship between the observed $\beta$'s
and $M_{UV}$ more closely, we can see clear evidence for deviations
from such a simple linear relationship.  In particular, brightward of
$-19$ mag, the mean $\beta$'s show a strong dependence on the $UV$
luminosity, while faintward of $\sim-19$ mag, these $\beta$'s show
almost no dependence on the luminosity.  This is illustrated in
Figure~\ref{fig:brokenline}.  Oesch et al.\ (2013a) have already
observed a similar behavior in $\beta$ as a function of luminosity,
but at rest-frame optical wavelengths.

Based on simple physical considerations, we might expect such a
dependence of $\beta$ on $UV$ luminosity if the dust extinction in
galaxies shows a significant correlation with the mass or the
luminosity of galaxies, and indeed such has been shown (e.g., Figure
18 from Reddy et al. 2010 and Figure 5 from Pannella et al. 2009).
For galaxies where the luminosities or dust extinction is large,
modest changes in luminosity (or dust extinction) would have a large
impact on $\beta$.  For galaxies where the $UV$ luminosity or dust
extinction is smaller, the impact would be much less.  The theoretical
predictions of Dayal et al.\ (2013) also suggest only a mild
dependence of $\beta$ on luminosity in this regime, though at
sufficiently low luminosities Dayal et al.\ (2013) predict somewhat
bluer $\beta$'s (e.g., $\beta\sim-2.5$).

One way of attempting to model the observed $\beta$'s given this
situation is to use a piecewise-linear model, where we allow for a
different dependence on the $UV$ luminosity brightward of some break
than faintward of this break.  Utilizing such a four-parameter model
(adding some break luminosity and slope faintward of the break to
standard two-parameters linear fits), we recover $-19.0\pm0.4$ mag,
$-18.8\pm0.2$ mag, and $-18.8\pm0.8$ mag for the break luminosity at
$z\sim4$, $z\sim5$ and $z\sim6$, respectively, and find a best-fit
$d\beta/dM_{UV}$ dependence of $-$0.08$\pm$0.03, $-$0.08$\pm$0.07,
$-$0.02$\pm$0.17, respectively, faintward of this break luminosity.
Almost identical break luminosities (i.e., $-19.3$ to $-18.8$) are
found fitting only to our binned $\beta$ results from the XDF and
HUDF09-Ps data sets, so the position of this break is not an artifact
of some offset between the wide and deep field $\beta$'s and our only
making use of wide-area samples brightward of $-19$ mag.  It is
remarkable how similar our results are for break luminosities and
faint-end slopes at all redshifts, suggesting that the break
luminosity we uncovered has a fundamental physical origin.

It is interesting to compare the reduced $\chi^2$'s we derive from
these piecewise-linear fits with those obtained using simple fits to a
line.  For the piecewise-linear fits, we fix the break luminosity to
$-18.8$ (the average break luminosity we find at $z\sim4$, $z\sim5$,
and $z\sim6$) and the $d\beta/dM_{UV}$ slope faintward of this
luminosity to $-0.08$ (the average slope faintward of the break for
our $z\sim4$, $z\sim5$, and $z\sim6$ samples).  This reduces the
dimensionality from four ($\beta$ intercept, break luminosity, slope
brightward of the luminosity break, slope faintward of the break) to
two ($\beta$ intercept, slope brightward of the luminosity break).
For this two-parameter piecewise linear model, we find best-fit
$\chi^2$'s that are $\Delta\chi^2$ $=\,$3.6, 1.3, and 2.1 lower than
the equivalent linear fits in representing our $z\sim4$, $z\sim5$, and
$z\sim6$ $\beta$ vs. $M_{UV}$ determinations, respectively.  These two
models for representing the observed $\beta$ vs. $M_{UV}$ relationship
are shown in Figure~\ref{fig:brokenline}.  The best-fit parameters we
find for our piecewise-linear model (with fixed faint-end slope and
break luminosity) are provided in Table~\ref{tab:bslopes}.  Based on
the differences in $\chi^2$ values for these two-parameter models, it
is clear that our piecewise-linear model (with only a weak dependence
on $UV$ luminosity faintward of $-19$ mag) provides a noticeably
superior representation for the observed $\beta$ vs. $M_{UV}$
relationship at $z\sim4$, $z\sim5$ and $z\sim6$ (96\%, 75\%, and 85\%
confidence, respectively).\footnote{While Rogers et al.\ (2014) find
  no evidence in their $z\sim5$ sample for a clear change in the slope
  of the $\beta$ vs. $M_{UV,AB}$ relationship at $M_{UV,AB}\sim-19$,
  our samples of faint $z\sim4$-6 galaxies provide us with much better
  statistics to test for such a change in slope.  Not only do we
  quantify the $UV$ slopes for faint galaxies in three deep fields at
  $z\sim5$ (i.e., XDF, HUDF09-1, and HUDF09-2) as opposed to the one
  that Rogers et al.\ (2014) consider, but we also have similar
  samples of galaxies at $z\sim4$ and $z\sim6$ with which to look for
  such a change in slope.}

\subsection{Extrapolating $\beta$ Results from Faint Galaxies at $z\sim4$-6 
to Higher Redshifts}

As Figure~\ref{fig:colmag456} illustrates, the mean $UV$-continuum
slopes $\beta$ for $z\sim4$, $z\sim5$, and $z\sim6$ galaxies are now
quite well established on the basis of current observations and
exhibit a similar dependence on $UV$ luminosity independent of
redshift (see also Figure~\ref{fig:slope}).  This dependence on $UV$
luminosity is likely weaker faintward of $-19$ mag for galaxies at all
redshifts under consideration here (Figure~\ref{fig:brokenline}).  The
quality of the $z\sim4$-6 constraints directly follows from the much
larger sample sizes available at these redshifts, the substantial
wavelength leverage available to constrain $\beta$ for individual
sources, and the superb depth of data sets like the XDF.

The $z\sim4$-6 $\beta$ results provide us with a solid baseline on
which to establish expectations for the $\beta$ distribution at even
higher redshifts.  An important question from the previous sections
regards the mean value of $\beta$ for lower-luminosity galaxies in the
$z\sim7$-8 universe.  For our $z\sim4$, $z\sim5$, and $z\sim6$
samples, we can set strong constraints on the mean $\beta$'s using
existing observations.  In Figure~\ref{fig:faintevol}, we present the
biweight mean $\beta$'s we derive as a function of redshift from
$z\sim4$, $z\sim5$, and $z\sim6$ XDF+HUDF09-Ps samples.

Only galaxies in the luminosity range $-19<M_{UV,AB}<-17$ are
considered in these comparisons for our $z\sim4$-6 samples.  Also
shown on Figure~\ref{fig:faintevol} are the expectations from the
hydrodynamical simulations of Finlator et al.\ (2011).  The best-fit
trend we find at $z\gtrsim4$ for the biweight mean $\beta$ for faint
($-19<M_{UV,AB}<-17$) galaxies is $-2.14\pm0.06 -
(0.10\pm0.06)(z-4.9)$.  Here we include in our error budget both the
statistical errors and our conservative systematic error estimates.
Extrapolating the results from our $z\sim4$-6 samples to $z>6$, we
predict the mean $\beta$ for lower-luminosity galaxies at $z\sim7$ and
$z\sim8$ to be $-2.35\pm0.16$ and $-2.45\pm0.23$, respectively
(conservatively accounting for possible systematic errors on our
$\beta$ determinations).  The mean $\beta$'s we derive for faint
$z\sim7$-8 galaxies in \S4 and \S6 are in excellent agreement with
these extrapolations.

\begin{deluxetable*}{ccccccc}
\tablewidth{0cm}
\tabletypesize{\scriptsize}
\tablecaption{Filters used to measure $\beta$ for $z\sim7$  galaxies (approximate orbit totals in parentheses).\label{tab:strategy}}
\tablehead{
\colhead{} & \multicolumn{3}{c}{Source Detection and Selection\tablenotemark{a}} & \multicolumn{3}{c}{} \\
\colhead{} & \multicolumn{2}{c}{Color Redward of Break} & \colhead{Contributes to} & \multicolumn{3}{c}{$\beta$ Measurement}\\
\colhead{Field} & \colhead{Blue Anchor} & \colhead{Red Anchor\tablenotemark{b}} & \colhead{Detection Image\tablenotemark{c}} & \colhead{Blue Anchor\tablenotemark{b}} & \colhead{Central Anchor} & \colhead{Red Anchor}}
\startdata
XDF & $Y_{105}$ (100) & $J_{125,1}$ (20) & $H_{160,1}$ (43) & $J_{125,2}$ (20) & $JH_{140}$ (30) & $H_{160,2}$ (42) \\
    & $Y_{105}$ (100) & $J_{125,2}$ (20) & $H_{160,2}$ (42) & $J_{125,1}$ (20) & $JH_{140}$ (30) & $H_{160,1}$ (43) \\
HUDF09-1 & $Y_{105}$ (8) & $J_{125,1}$ (6) & $H_{160,1}$ (7) & $J_{125,2}$ (6) & --- & $H_{160,2}$ (6) \\
         & $Y_{105}$ (8) & $J_{125,2}$ (6) & $H_{160,2}$ (7) & $J_{125,1}$ (6) & --- & $H_{160,1}$ (7) \\
HUDF09-2 & $Y_{105}$ (11) & $J_{125,1}$ (9) & $H_{160,1}$ (10) & $J_{125,2}$ (9) & --- & $H_{160,2}$ (9)\\
         & $Y_{105}$ (11) & $J_{125,2}$ (9) & $H_{160,2}$ (9) & $J_{125,1}$ (9) & --- & $H_{160,1}$ (10)\\
CANDELS & $Y_{105}$ (3) & $J_{125,1}$ (2) & $H_{160,1}$ (2) & $J_{125,2}$ (2) & --- & $H_{160,2}$ (2)\\
        & $Y_{105}$ (3) & $J_{125,2}$ (2) & $H_{160,2}$ (2) & $J_{125,1}$ (2) & --- & $H_{160,1}$ (2)\\
ERS & $Y_{098}$ (2) & $J_{125,1}$ (1) & $H_{160,1}$ (1) & $J_{125,2}$ (1) & --- & $H_{160,2}$ (1)\\
    & $Y_{098}$ (2) & $J_{125,2}$ (1) & $H_{160,2}$ (1) & $J_{125,1}$ (1) & --- & $H_{160,1}$ (1)
\enddata
\tablenotetext{a}{Our $z\sim7$ selections also
  depend on the flux measurements in the five optical/ACS bands
  $B_{435}V_{606}i_{775}I_{814}z_{850}$ to identify the presence of a
  strong Lyman break in the spectrum with no flux blueward of the break
  (\S4.3).  However, flux at these wavelengths are never used for
  estimates of $\beta$ and so have no impact on the noise-driven
  systematic biases we discuss here (\S4.1: see also Appendix B.1.2 of
  Bouwens et al.\ 2012).}
\tablenotetext{b}{$J_{125,1}$ and $J_{125,2}$ indicate the first and
  second half of the F125W observations for fields considered in this
  study.  Since noise in $J_{125,1}$ and $J_{125,2}$ are
  independent (being composed of disjoint exposures of the same
  field), one does not need to be concerned that the measurement of
  $\beta$ will be coupled to source selection (see \S4.2).}
\tablenotetext{c}{$H_{160,1}$ and $H_{160,2}$ indicate the first and
  second half of the F160W observations for fields considered in this
  study.}
\end{deluxetable*}

\begin{figure}
\epsscale{1.15}
\plotone{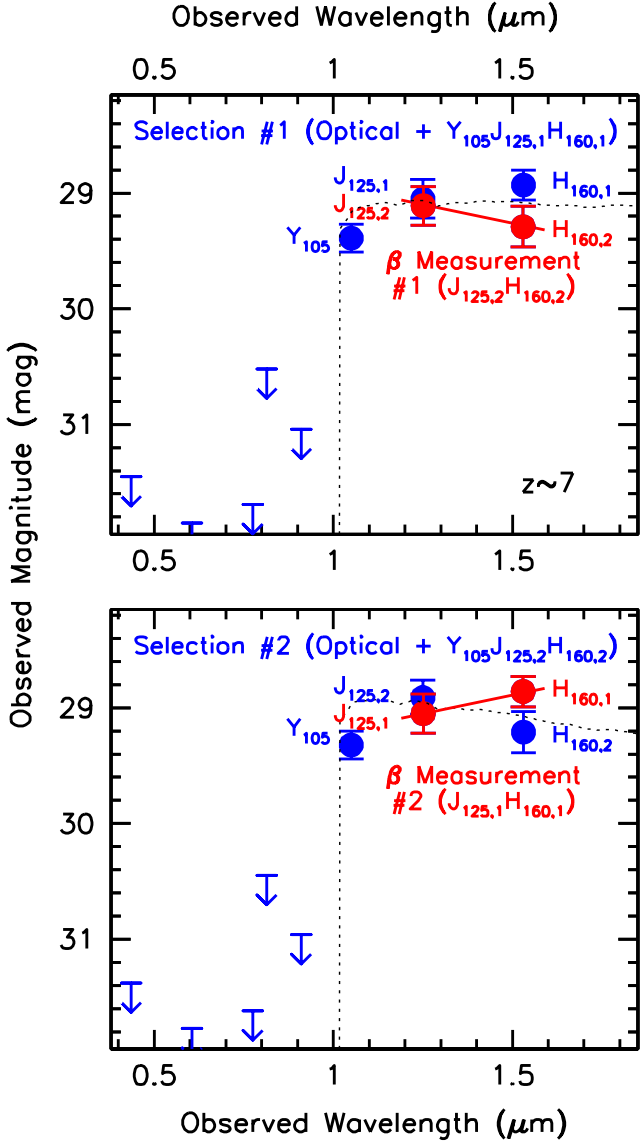}
\caption{ (\textit{upper}) An illustration of the information we use
  for selecting $z\sim7$ sources from our fields and measuring
  $\beta$.  The fluxes and $1\sigma$ uncertainties we derive for a
  $z\sim7$ galaxy in the XDF.  Two separate flux measurements are
  derived for this source at both $1.25\mu$m and $1.6\mu$m, from the
  two $\sim$50\% splits of the $J_{125}$ and $H_{160}$-band exposures
  for this field (i.e., $J_{125,1}$, $J_{125,2}$, $H_{160,1}$, and
  $H_{160,2}$).  Of importance for the selection (blue circles and
  $1\sigma$ upper limits) are the optical observations, $Y_{105}$, and
  half of the $J_{125}$ and $H_{160}$ band observations -- which allow
  us to demonstrate that the selected sources are real, show a sharp
  Lyman break in the spectrum across the $z_{850}$ and $Y_{105}$
  bands, show absolutely no flux blueward of the break (in the
  $B_{435}V_{606}i_{775}$ data), and exhibit a blue color redward of
  the break (using the flux in the $Y_{105}$ band and the first half
  of the $J_{125}$-band observations).  Of importance for the
  measurement of $\beta$ are the measured fluxes (red circles) in the
  other half of the $J_{125}$ and $H_{160}$-band observations.  By
  clearly separating source selection from the measurement of $\beta$,
  we can be sure that our $\beta$ measurements remain unbiased to
  faint magnitudes.  (\textit{lower}) To make full use of the
  available $J_{125}$ and $H_{160}$-band observations to maximize the
  accuracy of our $\beta$ measurements, we also consider a second
  selection of $z\sim7$ galaxies over each of our fields, but
  utilizing the second half of the $J_{125}$ and $H_{160}$-band
  observations for the selection and the first half of the $J_{125}$
  and $H_{160}$-band observations for the measurement of $\beta$.  The
  photometry we obtain for sources is consistent with that from our
  first selection (upper panel), but is nonetheless slightly different
  (due to the impact of noise on the derived
  apertures).\label{fig:examp}}
\end{figure}

\section{$\beta$ Results for $z\sim7$ Samples}

A key part of the discussion regarding $\beta$ measurements has
concerned the extent to which $\beta$ measurements may or may not be
subject to biases.  While this question is not new (e.g., see Meurer
et al.\ 1999; Figure 4 from Bouwens et al.\ 2009), essentially all
studies of $\beta$ at high redshift have fallen short in some regard
in their handling of systematic errors.

We begin this section with a brief motivation and summary of our
procedure for obtaining a measurement of the $\beta$ distribution for
$z\sim7$ galaxies, over a wide range in $UV$ luminosity while
remaining free of systematic errors (we use the word ``bias'' here for
simplicity).  We use the phrase ``noise-driven biases'' to describe
systematic biases that result from the impact of noise on two or more
coupled quantities (such as when the same noise fluctuations can
affect both source selection and the measurement of $\beta$).  Then,
we describe our selection and measurement procedures for $z\sim7$
galaxies and give our results.

\subsection{Why Noise-Driven Systematic Biases are A Particular Concern 
for $\beta$ Measurements at $z\sim7$}

While there are many potentially important biases for determinations
of the mean $\beta$, one of the most prominent sources of bias that
has received much discussion recently regards the interplay between
redshift selection and $\beta$ measurement (see Appendix B.1.2 of
Bouwens et al.\ 2012 or Dunlop et al.\ 2012).  Sizeable biases in the
measured $\beta$'s can occur, if the same information is used to
select galaxies as is used to measure their $UV$-continuum slopes
$\beta$.  Since sources with bluer observed colors (redward of an
apparent Lyman break) are more readily identified as $z\sim7$ galaxies
than sources with redder observed colors, one would be biased towards
selecting sources which also have the bluest-apparent $\beta$, if
the high-redshift selection is performed in the conventional way.

To overcome this bias, we must ensure that the information we have on
sources relevant to their selection as $z\sim7$ galaxies (detection
significance, color redward of the break) is completely independent of
that required to measure the $UV$-continuum slope $\beta$.
Accomplishing this, however, requires that we be able to define two
colors redward of the break that are entirely independent of each
other.  This appears to be challenging for $z\sim7$ sources, due to
the availability of only three different near-IR passbands with deep
observations in the typical legacy field.  For example, only three
near-IR bands are available with deep WFC3/IR coverage in public
surveys like CANDELS (Grogin et al.\ 2011; Koekemoer et al.\ 2011),
the Early Release Science (ERS) program (Windhorst et al.\ 2011), or
the HUDF09 program (Bouwens et al.\ 2011).  This challenge has
prompted researchers to correct for this bias based on simulations
(Bouwens et al.\ 2012; Finkelstein et al.\ 2012) or to argue that
observations in a fourth WFC3/IR band (e.g., F140W) may be required to
obtain unbiased measurements of $\beta$ at $z\sim7$ (Rogers et
al.\ 2013; Dunlop et al.\ 2012).  While this latter approach, with
more bands has merit, very few data sets have deep observations in so
many filters, and restricting $z\sim7$ $\beta$ studies to those that
do would restrict our ability to build up maximally-sized samples over
a wide range of luminosity.

\subsection{Bias-Free Procedure for Deriving $\beta$ at $z\sim7$} 

Fortunately, we have a way forward that allows for essentially
bias-free measurements of $\beta$ for $z\sim7$ samples using just the
available three-band WFC3/IR imaging in the many public HST fields.
Observations in a fourth WFC3/IR band are not required to obtain
bias-free results for $\beta$.  The approach is simply to construct
two independent, equal-depth reductions of both the $J_{125}$ and
$H_{160}$-band data by splitting the full dataset on each field in
half.  Each dataset consists of dithered images that are reduced using
our standard WFC3/IR pipeline (e.g., as in Bouwens et al.\ 2011).  The
result is two independent $J_{125}$ and $H_{160}$-band images of
comparable depth covering the same field, but with independent noise
properties.

We use the reduced image from the first half of the
$J_{125}$/$H_{160}$-band data set to do the $z\sim7$ selection, while
the reduced image of the second half of the $J_{125}$/$H_{160}$-band
data set is used for the $\beta$ measurements.  We then reverse the
roles of the $\sim$50\% splits of the $J_{125}$/$H_{160}$-band data
set, so that the first half is used for the $\beta$ measurements and
the second half is used for the $z\sim7$ selection.
Figure~\ref{fig:examp} illustrates the algorithm.  Putting together
the $\beta$ measurements from the two different selections on each
field, we can derive the mean $\beta$ with comparable S/N to what we
would achieve, had we made use of the full (unsplit) $J_{125}$ and
$H_{160}$ data sets to derive $\beta$.  Previously, we introduced a
less sophisticated version of this approach in \S4.8 of Bouwens et
al.\ (2012), but had only used this as a simple cross check on the
Bouwens et al.\ (2012) $z\sim7$ results for faint galaxies.  Here we
have systematically applied this procedure across all data sets under
study (HUDF/XDF, HUDF09-1, HUDF09-2, CANDELS-North, CANDELS-South,
ERS).

This more sophisticated procedure explicitly ensures (by construction)
that our derived $\beta$'s are completely robust against the
photometric error coupling bias described by Dunlop et al. (2012) and
also by Bouwens et al.\ (2012: Appendix B.1.2).  The photometric error
coupling bias is the term Bouwens et al.\ (2012) used to describe the
noise-driven systematic bias in $\beta$ that occurs when the same
noise fluctuation can have an effect on both the selectability of a
source and its measured $\beta$.  More details regarding this
procedure are given in \S4.5.  

\begin{figure*}
\epsscale{1.15}
\plotone{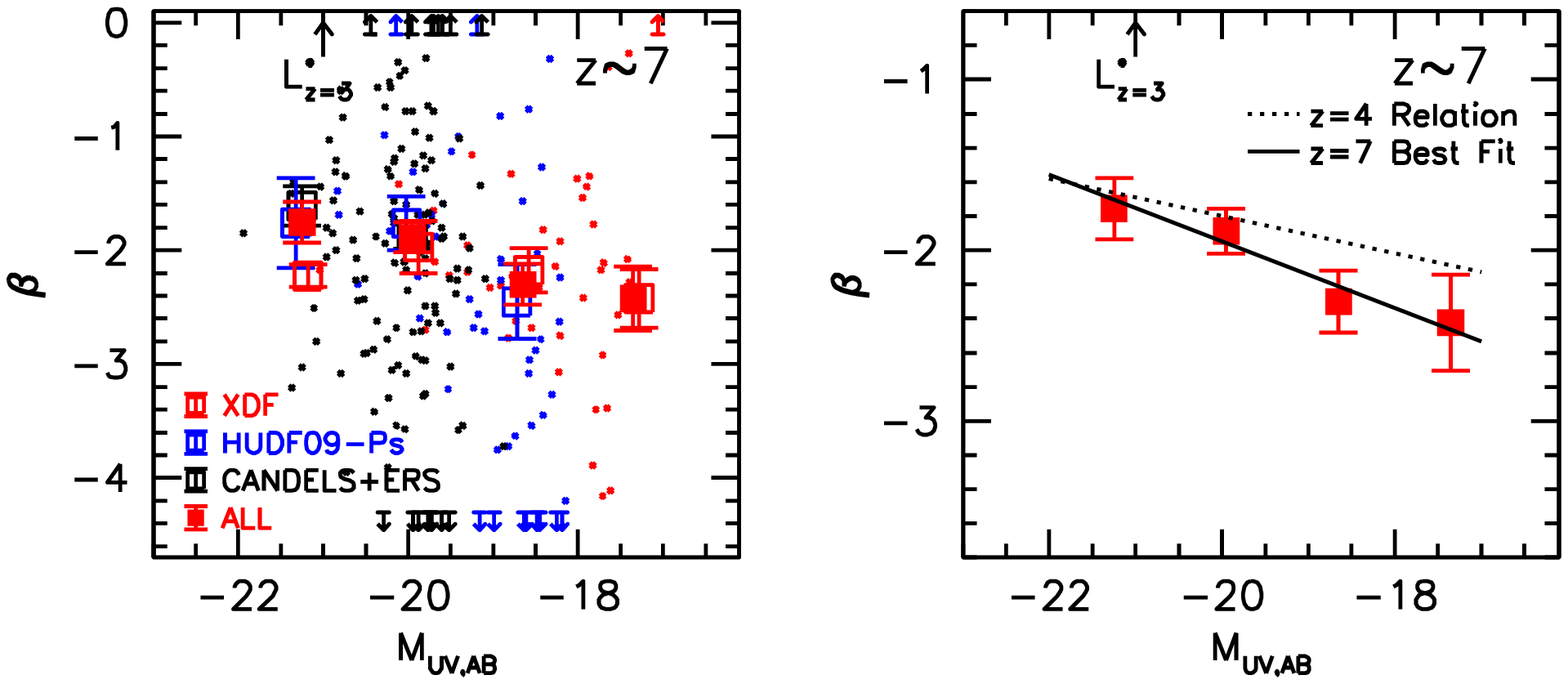}
\caption{(\textit{left}) Measured $\beta$ of $z\sim7$ galaxies in our
  XDF samples (\textit{red points}), HUDF09-Ps samples (\textit{blue
    points}), and ERS+CANDELS-S+CANDELS-North samples (\textit{black
    points}).  The points show sources at the average of the two
  $\beta$'s measured for an object if the same source is selected
  twice and at a source's only $\beta$ determination if it is selected
  once.  The large squares with $1\sigma$ error bars indicate the
  biweight mean $\beta$'s we measure in various $UV$ luminosity
  ($M_{UV}$) intervals from the XDF (\textit{red open squares}), from
  the HUDF09-Ps (\textit{blue open squares}), from the
  ERS+CANDELS-S+CANDELS-N (\textit{black open squares}), and from our
  entire field set (\textit{red solid squares}).  The different
  determinations are slightly offset horizontally to improve the
  clarity of this figure.  Error bars are derived from bootstrap
  resampling the $\beta$ results in each $UV$ luminosity interval.
  The biweight mean $\beta$'s that we derive from different data sets
  are generally in agreement to within their $1\sigma$ uncertainties,
  even at the detection limits of the shallowest ones.  This suggests
  that our results are not substantially impacted by noise-driven
  systematic biases.  These mean $\beta$'s are also presented in
  Table~\ref{tab:medianbeta}.  (\textit{right}) The mean $\beta$ we
  find in various $UV$ luminosity ($M_{UV}$) intervals (\textit{solid
    red squares}).  The solid black line indicates the best-fit linear
  relationship between $\beta$ and $UV$ luminosity ($M_{UV}$) for
  $z\sim7$ galaxies, while the dotted line shows the relationship
  Bouwens et al.\ (2012) found for $z\sim4$ star-forming galaxies.
  The relationship between $\beta$ and $UV$ luminosity exhibits a
  similar correlation at $z\sim7$ as it exhibits at lower redshifts
  (see Figure~\ref{fig:slope}).\label{fig:colmag7B}}
\end{figure*}

\subsection{Source Selection}

Many different datasets are used for our selection of $z\sim7$
galaxies to use in establishing $\beta$.  As indicated in the previous
section, we select $z\sim7$ candidates from each of our data sets
twice (see Table~\ref{tab:strategy}).

We base each of these selections on an independent photometric catalog
we construct from each field.  Object detection is performed on the
basis of the square root of $\chi^2$ image generated from the
$Y_{105}$, $J_{125}$, and $H_{160}$-band observations.  For each
catalog, we construct the $\chi^2$ image from the $\sim$50\% $J_{125}$
and $H_{160}$-band stacks not utilized for the $\beta$ measurements.
Photometry on the two $\sim$50\% $J_{125}$-band stacks for a given
field is carried out separately and kept completely separate
throughout the selection and measurement process, as if the two stacks
were flux measurements in different bands.  An identical procedure is
used for photometry on the two $\sim$50\% stacks of the $H_{160}$-band
observations.  The reductions we use for our $z\sim7$ $\beta$
determinations were done on a $0.06''$-pixel scale.

Redshift $z\sim7$ galaxies are then selected from these catalogs using
the same Lyman-break criteria we previously employed in Bouwens et
al. (2012).  For the XDF, HUDF09-Ps, and CANDELS fields (where we have
$Y_{105}$-band data) the criteria we use are
\begin{eqnarray*}
(z_{850} - Y_{105} > 0.7)\wedge(Y_{105}-J_{125} < 0.8) \wedge \\
(z_{850}-Y_{105}>1.4(Y_{105}-J_{125})+0.42),
\end{eqnarray*}
while for the ERS field the availability of only $Y_{098}$-band data
lead us to use the criteria:
\begin{eqnarray*}
(z_{850}-J_{125}>0.9)\wedge~~~~~~~~~~~~ \\
(z_{850}-J_{125}>0.4+1.1(Y_{098}-J_{125})).
\end{eqnarray*}
When applying the above criteria, we set the fluxes of sources that
are undetected to their $1\sigma$ upper limits.  The $J_{125}$-band
fluxes and images we utilize for these criteria are based on the
$\sim$50\% splits not utilized for the $\beta$ measurements (see
Table~\ref{tab:strategy}).  To ensure source reality, we require
sources required to be detected at $5\sigma$, adding in quadrature the
$Y_{105}$-band image with the 50\% split of the $J_{125}$ and
$H_{160}$-band exposures not used for the $\beta$ measurements.  We
also require sources are detected at $>4.3\sigma$ in the 50\% splits of
the $J_{125}$ and $H_{160}$-band exposures not used for the $\beta$
measurements.  This is to ensure that the sources we use in deriving
the mean $\beta$ have sufficient S/N that their $\beta$'s are
well-defined (i.e., so the S/N of the F125W and F160W-band fluxes used
to measure $\beta$ is not typically less than 1).

To minimize potential contamination from lower redshift interlopers,
we require that sources in our selection show no detection
($<$2$\sigma$) in the $B_{435}$, $V_{606}$, or $i_{775}$ bands.  To
take advantage of the very deep $I_{814}$-band observations over the
CANDELS fields (see e.g. Oesch et al.\ 2012 for a discussion), we
require that sources be either undetected in the $I_{814}$-band in a
given field ($<1\sigma$) or have a $I_{814}-Y_{105}$ band color $>$2
mag (or $I_{814}-Y_{098}>2$ mag for the ERS field).

Finally, we require the $\chi_{opt}^2$ statistic we construct for each
source to be less than 3 for sources in the XDF, HUDF09-1, HUDF09-2,
CANDELS, and ERS fields.  The upper limits we set are similar but
nonetheless slightly stronger in general than those adopted in Bouwens
et al.\ (2011) and Bouwens et al.\ (2012).  These limits allow us to
select $z\gtrsim7$ galaxies with minimal contamination
($\lesssim$10\%).

As in Bouwens et al.\ (2011), we take the $\chi_{opt} ^2$ statistic to
be equal to $\Sigma_i SGN(f_i) (f_i/\sigma_i)^2$ where $f_{i}$ is the
flux in band $i$, $\sigma_i$ is the flux error in band $i$,
SGN($f_{i}$) is equal to 1 if $f_{i}>0$ and $-1$ if $f_{i}<0$, and
where we consider the $B_{435}$, $V_{606}$, and $i_{775}$ bands in the
sum.  We apply the above $\chi^2$ criterion in three different
apertures (0.18$''$-diameter apertures, small scalable apertures, and
$0.35''$-diameter apertures) to ensure candidate $z\sim7$ sources in
our selection show no evidence for positive flux blueward of the break
irrespective of morphology.  We estimate that our contamination rates,
while dependent on the field, are typically in the range $\sim$2\% to
10\% (see discussion in Bouwens et al.\ 2011 and Bouwens et
al.\ 2014).

\subsection{Methodology for Measuring $\beta$}

In deriving $\beta$'s for our candidate sources, we use the same
approach as in Bouwens et al.\ (2012: see also Castellano et
al.\ 2012).  Specifically, we fit the observed $J_{125}$-band and
$H_{160}$-band fluxes to a power law $f_{\lambda} \propto
\lambda^{\beta}$ to determine $\beta$.  The effective wavelengths we
use in performing the fit assume a flat $\beta\sim-2$ spectrum, and
are 1243 nm, 1383 nm, 1532 nm for the $J_{125}$, $JH_{140}$,
$H_{160}$.  Given that the typical galaxy in our programs have
$\beta$'s in the range $\beta\sim-1.5$ to $\beta\sim-2.5$, this is a
somewhat better approximation than using the pivot wavelength
$\lambda_p$ (Tokuanga \& Vacca 2005) for this purpose.\footnote{The
  pivot wavelength $\lambda_p$ is a measure of the effective
  wavelength of a filter and is defined to equal $((\int_{\lambda}
  S(\lambda) \lambda d\lambda) / (\int_{\lambda} S(\lambda)
  d\lambda/\lambda))^{1/2}$ where $S(\lambda)$ is the integrated system
  throughput for a filter.}  With the exception of the $z\sim7$
sources over the HUDF -- where fluxes are also available at
$\sim$1.4$\mu$m from deep $JH_{140}$ data -- only two flux
measurements are used in this fit (i.e., $J_{125}$ and $H_{160}$).
This approach therefore becomes equivalent to using the fitting
formula
\begin{equation}
\beta = -2.0+4.39(J_{125}-H_{160}).
\end{equation}
The 4.39 factor is 2\% higher than the factor used in Bouwens et
al.\ (2010) and 1\% lower than advocated by Dunlop et al.\ (2012).
The $J_{125}$-band fluxes that we use in these fits come from the
$\sim$50\% of the $J_{125}$-band observations not used in the
selection of $z\sim7$ candidates for a field (see \S4.2 and
Table~\ref{tab:strategy}).

In deriving $\beta$, we only make use of flux measurements which are
clearly not affected by the IGM absorption or Ly$\alpha$
emission. Some studies (e.g., Finkelstein et al. 2012) have attempted
to exploit the additional wavelength leverage provided by the
$z_{850}$ and $Y_{105}$-band flux measurements to further improve
their estimates of $\beta$. The difficulty with such approaches is
their sensitivity to a number of potentially large (and unknown)
systematics.  Differences between the SED shapes or Ly$\alpha$
prevalence assumed versus that present in the real observations can
have a significant impact on the results that cannot be readily
quantified or appropriately corrected. This point is illustrated in
Figure 12 of Rogers et al. (2013).  Given this concern we do not make
use of the $Y_{105}$-band flux for $z\sim 7$ candidates.

\begin{figure*}
\epsscale{1.15}
\plotone{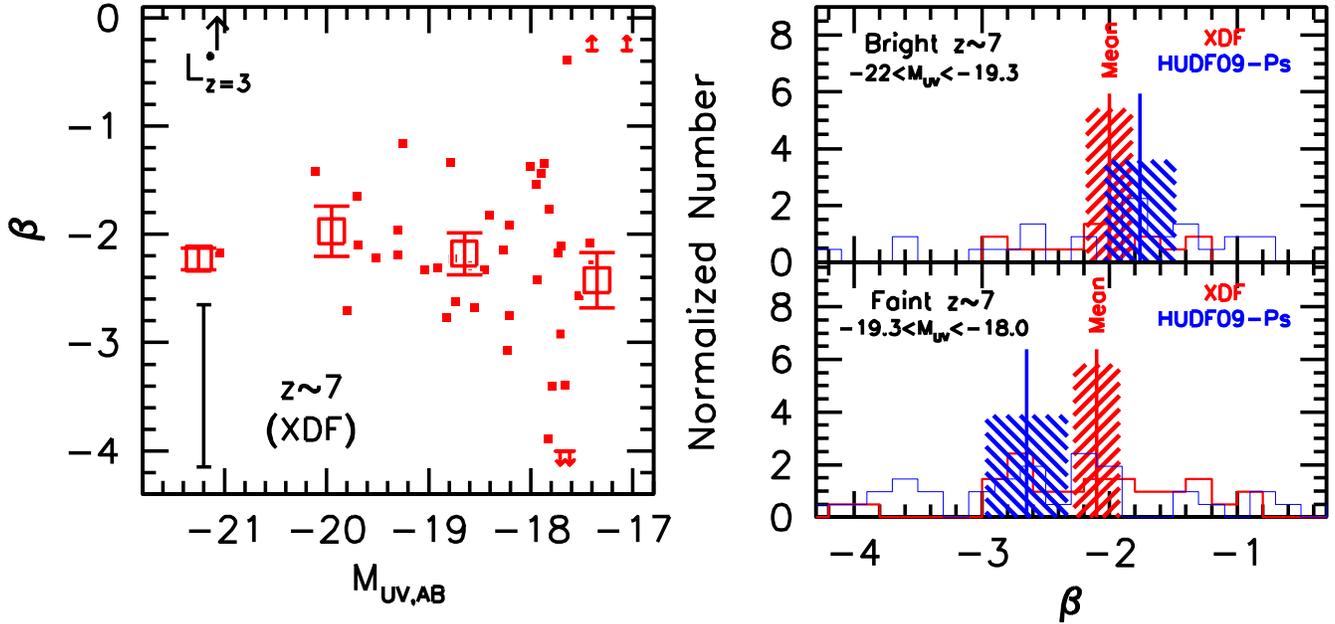}
\caption{(\textit{left panel}) Measured $\beta$'s for $z\sim7$
  star-forming galaxy candidates in our XDF samples (\textit{small red
    squares}).  $1\sigma$ upper and lower limits are plotted for
  sources whose nominal $\beta$ measurements are redder than $-0.3$ or
  bluer than $-4.0$ (as one would expect for the lowest S/N sources in
  each data set).  The error bar in the lower left-hand corner
  indicates the typical $1\sigma$ uncertainty on an individual $\beta$
  determination in the luminosity range $-19.3<M_{UV,AB}<-18.0$.  The
  large open red squares with $1\sigma$ error bars indicate the
  biweight mean $\beta$'s we measure in various $UV$ luminosity
  ($M_{UV}$) intervals from the XDF.  As in previous studies
  (Labb{\'e} et al.\ 2007; Bouwens et al.\ 2009; Bouwens et al.\ 2010;
  Bouwens et al.\ 2012), our measurement of $\beta$ for galaxies in
  the XDF provide evidence that lower luminosity galaxies have bluer
  $\beta$'s.  (\textit{upper right}) The distribution of measured
  $UV$-continuum slopes $\beta$ for bright ($-22<M_{UV,AB}<-19.3$)
  $z\sim7$ galaxies in the XDF (\textit{red histogram}) and HUDF09-Ps
  (\textit{blue histogram}) selections.  While individual sources may
  appear as often as two times in each panel (due to our repeating our
  selection twice on each field), the number of sources in the two
  selections has been divided by two to give a realistic indication of
  the numbers of sources included in our $\beta$ estimates.  The red
  vertical bar indicates the biweight mean $\beta$ for each luminosity
  subsample and the hashed region indicates the $1\sigma$ uncertainty.
  (\textit{lower right}) Similar to the upper right panel, but for
  fainter ($-19.3<M_{UV,AB}<-18.0$) $z\sim7$ sources over our fields.
  Clearly the fainter $z\sim7$ sources have bluer $\beta$'s than the
  brighter sources.\label{fig:colmag7A}}
\end{figure*}

\subsection{Deriving a Bias-free Sample}

As outlined in the introduction and at the beginning of this section,
our procedure is to select sources from our fields twice,
alternatively using the first and second half of the available
$J_{125}$/$H_{160}$-band data in each field.  Measurements of $\beta$
are made using the other half of the $J_{125}$/$H_{160}$-band
observations not included in the $z\sim7$ selection.

We systematically applied this procedure to all of the data sets and
search fields considered in this study.  The $\beta$'s we measure for
candidate $z\sim7$ galaxies in our fields are shown in
Figure~\ref{fig:colmag7B} as the blue, red, and black points.  

\begin{figure}
\epsscale{1.15}
\plotone{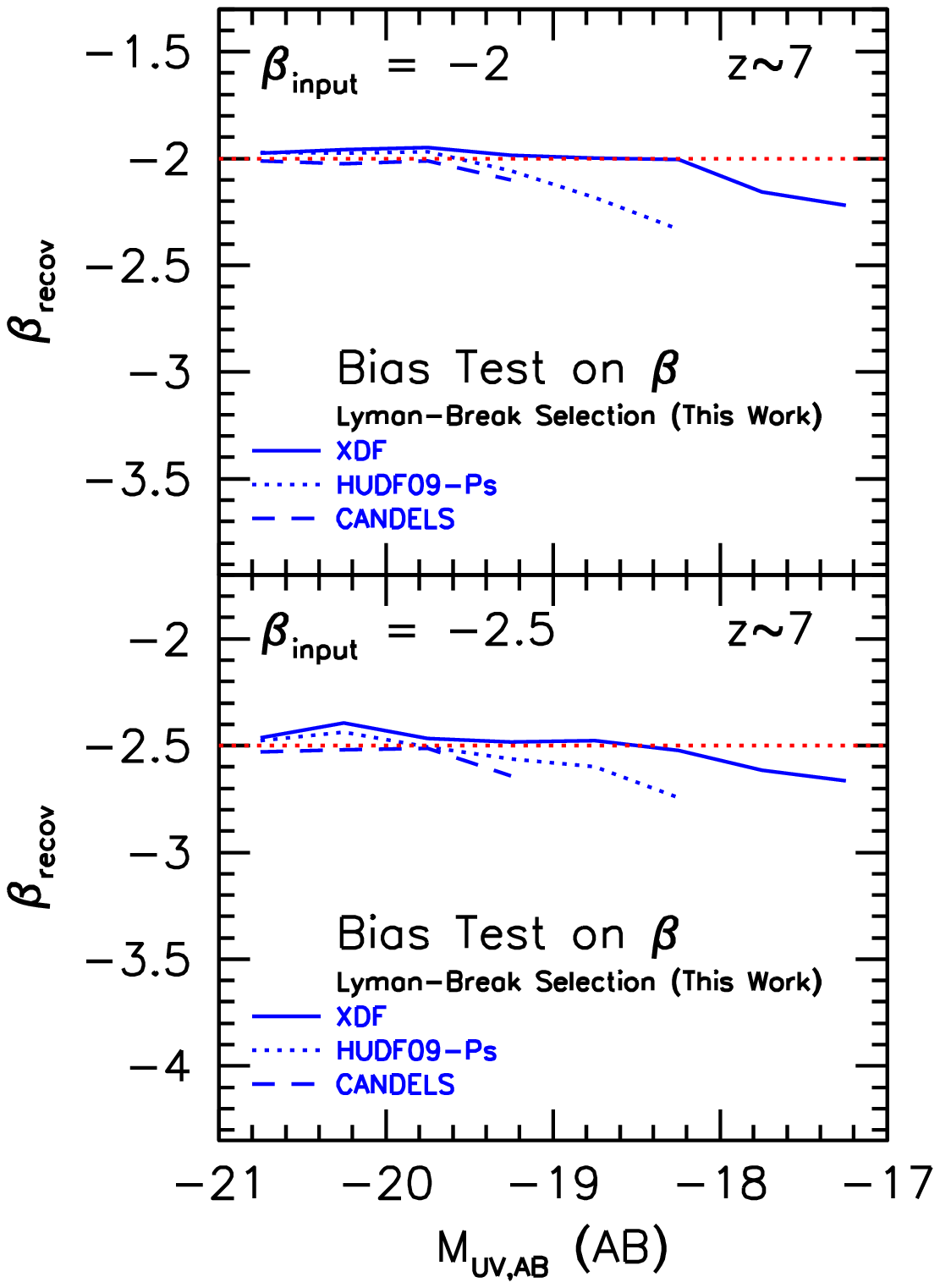}
\caption{The mean $\beta$'s we recover for a $z\sim7$ galaxy
  population vs. the near-IR magnitude assuming an input $\beta$ of
  $-2$ (\textit{top}) and $-2.5$ (\textit{bottom}).  Shown are the
  results for our $z\sim7$ selections from the XDF (\textit{solid
    lines}), the HUDF09-Ps fields (\textit{dotted lines}), and the
  CANDELS-South/Deep region (\textit{dashed lines}).  The only source
  of bias in our recovered $\beta$'s is due to the fact that galaxies
  with bluer $\beta$'s are slightly easier to select than galaxies
  with redder $\beta$'s.  This is the selection volume bias discussed
  in Appendix B.1.1 of Bouwens et al.\ (2012), and it affects both
  Lyman-Break (Bouwens et al.\ 2012; Wilkins et al.\ 2011) and
  photometric redshift selections (Dunlop et al.\ 2012; Finkelstein et
  al.\ 2012).  Fortunately, the calculated selection volumes for our
  $z\sim7$ samples from the XDF only exhibit a weak dependence on the
  input $\beta$'s, and therefore the expected bias is very small,
  i.e., $\Delta \beta \lesssim 0.03$ for galaxies with $UV$
  luminosities brighter than $-18$ mag.  Faintward of $-18$ mag, we
  expect a small bias in $\beta$ ($\sim$0.1-0.15).  The other
  significant bias in $\beta$ is the photometric error coupling bias
  (described by Dunlop et al.\ (2012) and Bouwens et al.\ [2012:
    Appendix B.1.2]), but this bias is identically zero here, due to
  our using completely different information for selecting sources
  than is used for our $\beta$ measurements.  While small, corrections
  for all biases shown here have been performed in the results we
  report.\label{fig:bias}}
\end{figure}

In deriving the mean $\beta$ for $z\sim7$ galaxies as a function of
luminosity, we incorporate the $\beta$ measurements made for sources
in each of our two $z\sim7$ selections over each field.  We can
include up to two measurements of $\beta$ for the same source in our
calculated mean, if this source makes it into both of our selections.
Sources that are only selected as part of one of the two samples are
counted once (and always such that $\beta$ used for the mean is
derived based on a different $\sim$50\% split of the
$J_{125}$/$H_{160}$-band data set than was used in the selection of
the source).  Typically, we observe only a modest variation in the
measured value of $\beta$ for a source between the two selections, as
we illustrate in Appendix C for faint $z\sim7$ sources from the XDF.
As we show in Appendix D, deriving the mean $\beta$'s with this
weighting naturally removes any significant systematic biases
resulting from noise.

Sources are then binned as a function of their $M_{UV,AB}$ magnitude,
and then the biweight mean (Beers et al.\ 1990) $\beta$ is calculated.
In computing the biweight mean, we use the same procedure as in
Bouwens et al.\ (2012), except that we also weight each data point
according to the inverse variance.  The inverse variance for each
source accounts for the intrinsic $1\sigma$ scatter in $\beta$
(assumed to be 0.35, similar to what is observed for bright $z\sim4$
and $z\sim5$ galaxies: Bouwens et al.\ 2009, 2012; Castellano et
al.\ 2012; Rogers et al.\ 2014) and the observational error in
deriving $\beta$.  For the purposes of computing the inverse variance
weighting to apply to individual sources, the maximum error we allow
on the flux of individual sources is 20\% of the flux measurement.
This is to minimize the impact that noise can have on the biweight
mean results through the weighting scheme.\footnote{To ensure that
  this weighting scheme had no significant impact on our results, we
  also computed the biweight mean $\beta$ for faint sources in each of
  our $z\sim4$-7 samples without applying this inverse-variance
  weighting and observed no significant change in the results.}  For
each luminosity bin, we also compute the median $\beta$ and
inverse-variance-weighted mean $\beta$ to illustrate how the central
value for $\beta$ can depend on the statistic used to quantify it.

Care is required regarding the $UV$ luminosity $M_{UV}$ we assign to
individual sources in our samples.  We derive the $UV$ luminosity
$M_{UV}$ for individual sources from the geometric mean of the
$J_{125}$ and $H_{160}$-band flux measurements made from the 50\%
stacks used in their selection.  Since these flux measurements are
completely independent of $J_{125}$ and $H_{160}$-band flux
measurements made from the 50\% stacks used in the measurement of
$\beta$, we ensure that our luminosity determinations are completely
independent of our $\beta$ measurements.  While the current procedure
differs from the procedure we had earlier used in Bouwens et
al.\ (2012: there we derive $M_{UV}$ from the same $J_{125}$ and
$H_{160}$ fluxes as we used to derive $\beta$), we have done so only
because the current procedure is cleaner and robust against the small
biases that arise in $\beta$ (i.e., $\Delta\beta \sim 0.1$-0.2) when
the depth of the $J_{125}$ and $H_{160}$ band observations
differ.\footnote{In any case, our simulations (Appendix D) suggest
  that the Bouwens et al.\ (2012) methodology for deriving the $UV$
  luminosity should have resulted in no large biases in the mean
  $\beta$ ($\Delta\beta \lesssim 0.1$).}  Galaxies in our $z\sim7$
selection were binned in 1.3-mag magnitude intervals to minimize the
importance of the considerable scatter in the $\beta$ measurements in
a given magnitude interval (see Figure~\ref{fig:colmag7A}).

Uncertainties on the median $\beta$'s are computed based on the
observed dispersion in $\beta$ using a bootstrap resampling procedure.
When combining the mean $\beta$ derived from data sets of
significantly different depths, the results are weighted according to
inverse square error -- where this error includes both the intrinsic
dispersion in $\beta$ (taken to be 0.35 similar to that found at lower
redshifts) and the typical measurement error in $\beta$ added in
quadrature, divided by the square root of the number of sources.

A small correction to the derived $\beta$'s is made to account for the
effect of the intrinsic $\beta$ on the selectability of sources.  The
required correction is shown in Figure~\ref{fig:bias} and is typically
just $\Delta \beta \lesssim$0.1 in size.  The correction is derived
from simulations we ran where we added sources to our search fields
and then attempted to reselect the sources (see Appendix F here or
Appendix B.1.1 of Bouwens et al.\ 2012 for a description of this small
bias).  Such a bias is unavoidable and must be corrected for in both
Lyman-Break redshift selections (Bouwens et al.\ 2012; Wilkins et
al.\ 2011) and photometric redshift selections (e.g., Dunlop et
al.\ 2013; Finkelstein et al.\ 2012).  While this bias has been
frequently quantified in the context of Lyman-break selections (e.g.,
Bouwens et al.\ 2009; Wilkins et al.\ 2011), a similar quantification
and illustration of this bias in the context of photometric redshift
selections would be useful to see.\footnote{We note, however, that the
  simulations in \S4 of Dunlop et al.\ (2013), which include a range
  of intrinsic $\beta$'s, should succeed in implicitly correcting for
  this bias.  From the results shown in Figure 5 from Rogers et
  al.\ (2014), it would appear that this bias is likely small.}

\subsection{Estimated Systematic Uncertainties}

There are a large number of small systematic uncertainties that can
contribute to the overall error on the derived $\beta$'s for sources.
These include uncertainties in the effective PSFs on the HST
observations, errors in accurately registering the observations with
each other, errors in deriving the PSF kernel to match the
observations across multiple bands, uncertainties in the HST
zeropoints, light from neighboring sources, and possible systematics
in the subtration of the background.  We estimate that the typical
systematic errors in the measured colors for individual sources from
each of these issues are not large (and not greater than $2\%$), and
are likely to be around $1\%$, as we show for the PSF kernel matching
in Appendix A.  We therefore allow for total 3\% systematic
uncertainties on our measured colors.

A 3\% systematic uncertainty in the measured colors translates into
systematic uncertainties in our derived $\beta$'s of
$\Delta\beta\sim0.13$ at $z\sim7$.  For our samples at other
redshifts, i.e., $z\sim4$-5, $z\sim5$, and $z\sim8$ (see \S3.2, \S6.2,
and \S6.3), the equivalent systematic uncertainties are
$\Delta\beta\sim0.06$, $\Delta\beta\sim0.08$, and
$\Delta\beta\sim0.27$, respectively.

\subsection{Results}

The results of our searches for $z\sim7$ galaxies and measurements of
the biweight mean $\beta$ across our many deep, wide-area fields are
shown in Figure~\ref{fig:colmag7B} for galaxies from the
ERS+CANDELS-S+CANDELS-N fields alone (\textit{black open squares}),
from the HUDF09-Ps alone (\textit{red open squares}), from the XDF
alone (\textit{red open squares}), and from the entire data set
(\textit{solid red squares}).  Our biweight mean $\beta$ results are
also presented in Table~\ref{tab:medianbeta}, along with the median
and inverse-variance-weighted mean $\beta$'s we derive in the same
magnitude intervals.  Overall the results from the different data sets
are in excellent agreement within the errors, even for our lowest
luminosity $z\sim7$ samples, where there has been much debate.

In particular, the availability of mean $\beta$ measurements over the
HUDF09-Ps are quite important, since they provide us with valuable
cross checks on results from the XDF data set.  This emphasizes again
the value of our development of an approach that eliminates systematic
error when only three near-IR bands are available.  Such cross checks
would not be possible without this approach which is robust against
noise-driven systematic biases.  This capacity to cross check our
results with faint galaxies from the HUDF09-Ps fields distinguishes
the current analysis from that of Dunlop et al.\ (2013) who only
examine the $\beta$ properties of faint galaxies in fields with deep
F140W observations, i.e., the XDF.

The best-fit linear relationship between the mean $\beta$ and $UV$
luminosity is
\begin{eqnarray*}
\beta = (-2.05\pm0.09\pm0.13) + ~~~~~~~~~~\\
~~~~~~~~~~ (-0.20\pm0.07)(M_{UV}+19.5)
\end{eqnarray*}
and is indicated by the black line in Figure~\ref{fig:colmag7B}.  The
second set of errors here are systematic and assume a $\sim0.03$ mag
systematic uncertainty in the $J_{125}-H_{160}$ colors to be
conservative.  The biweight mean $\beta$ is clearly bluer for lower
luminosity $z\sim7$ galaxies than it is for more luminous galaxies,
both for the XDF and the HUDF09-Ps data sets
(Figure~\ref{fig:colmag7A}).  These trends in $\beta$ as a function of
$UV$ luminosity are essentially identical to those shown by Bouwens et
al.\ (2012) for their $z\sim4$-7 selections.

Finally, it is useful to look at what our new $z\sim7$ results imply
for the evolution of $\beta$ with redshift or cosmic time.  As in
\S3.5, we will focus on the evolution of lower luminosity galaxies
($M_{UV,AB}>-19$) as seen in our deepest data sets, the XDF+HUDF09-Ps
fields.  However, instead of considering galaxies over the luminosity
range $-19<M_{UV,AB}<-17$, we consider galaxies over a somewhat larger
luminosity range $-19.3<M_{UV,AB}<-16.7$ to increase the
signal-to-noise on this measurement somewhat.

The biweight $\beta$ for this luminosity interval is presented in
Figure~\ref{fig:faintevol} and shown in relation to similar
lower-redshift measurements.  Is there evidence for an evolution in
$\beta$ with redshift?  If we combine our new $\beta$ results at
$z\sim7$ with our lower-redshift $\beta$ results at $z\sim4$-6, we
find a best-fit relation of $-2.14\pm0.05 - (0.10\pm0.05)(z-4.9)$
(again adopting conservative systematic error estimates on our $\beta$
determinations).  This argues for a slow but moderately significant
reddening of $\beta$ with cosmic time.  Evidence for such an evolution
was previously presented by Stanway et al.\ (2005), Bouwens et
al.\ (2006), Bouwens et al.\ (2009), Bouwens et al.\ (2012), and
Finkelstein et al.\ (2012).

\subsection{Cross-Checking Our $\beta$ Results Using Fixed-Aperture Color
Measurements}

To ensure that the results we obtained earlier in this section are as
robust as possible, we repeated the above analysis, but using
photometry on the $z\sim7$ candidates with fixed $0.32''$-diameter
apertures (after PSF-matching the data).  One advantage of
fixed-aperture photometry for the faintest sources is that it allows
for a very consistent measurement of the colors in these sources
(where there is no concern that the aperture may not be optimally
defined due to the low S/N of the sources under
examination).\footnote{Despite this possible disadvantage to using
  Kron-style photometry on the faintest sources, the apertures chosen
  for most sources are nonetheless reasonably optimal.  In addition,
  with Kron-style photometry, one can naturally cope with different
  source sizes, allowing for more optimal photometry for sources over
  a wide range of magnitudes.}  Using fixed-aperture photometry to
define the colors (while retaining Kron-style photometry to estimate
the total magnitudes for sources), we find a mean $\beta$ of
$-$2.20$\pm$0.20 and $-$2.31$\pm$0.24 for galaxies in our faintest two
magnitude bins $-19.3<M_{UV,AB}<-18.0$ and $-18.0<M_{UV,AB}<-16.7$,
respectively.  These results are quite consistent within the $1\sigma$
errors (the mean offset $\Delta\beta\sim0.2$) to the mean $\beta$'s
obtained using just the $JH_{125}$ and $H_{160}$ fluxes.

\subsection{Cross-Checking Our Results Using a $JH_{140}$-band Flux Selection}

For the sake of completeness, we also utilize a similar strategy for
selecting $z\sim7$ sources as that of Dunlop et al.\ (2013), taking
special advantage of the $JH_{140}$ observations.  Similar to our use
of the 50\% splits of the $J_{125}$/$H_{160}$-band observations, the
deep $JH_{140}$-band data allow us to select $z\sim7$ star-forming
galaxies based on different information than is used to derive the
$UV$-continuum slopes $\beta$.  Such a procedure should ensure that
the results will be robust against noise-driven systematic biases.
The selection criteria we use for this $z\sim7$ sample are analogous
to the criteria for our primary selection, but use a
$Y_{105}-JH_{140}<0.8$ color criterion (instead of a
$Y_{105}-J_{125}<0.8$ color criterion).  We also require sources to be
detected at $5\sigma$ in the $JH_{140}$ band (small scalable
apertures) to be included in our sample.  $\beta$ is estimated using
the measured $J_{125}$-band and $H_{160}$-band flux from the full XDF
observations.  There is no need to repeat the $z\sim7$ selection twice
(as for our primary selection), since the deep $JH_{140}$ observations
provide us with a constraint on the color of the source redward of the
break that is not used in the measurement of $\beta$.  Based on the 26
sources that make it into this $z\sim7$ selection, we find a biweight
mean $\beta$ of $-2.61\pm0.21$ and $-2.03\pm0.49$ in the luminosity
intervals $-19.3<M_{UV,AB}<-18.0$ and $-18.0<M_{UV,AB}<16.7$,
respectively.  These results are consistent with the results from our
primary selection (having a mean offset $\Delta\beta\sim0.2$).

\begin{deluxetable}{ccc}
\tabletypesize{\footnotesize}
\tablecaption{Mean or Median $\beta$ determinations for Faint $z\sim7$ Galaxies in the
Approximate Luminosity Range $-19.3\lesssim M_{UV,AB}\lesssim -18$ from the Literature\label{tab:disput7}}
\tablehead{\colhead{} & \multicolumn{2}{c}{Mean or Median $\beta$}\\
 \colhead{Reference} & \colhead{Uncorrected} & \colhead{Corrected/Final}}
\startdata
This Work\tablenotemark{a} & & $-2.30\pm0.18\pm0.13$ \\
Dunlop et al.\ (2013)\tablenotemark{b} & $-2.23\pm0.15$ & $-2.36\pm0.15$\tablenotemark{d}\\
Finkelstein et al.\ (2012)\tablenotemark{c} & $-2.68_{-0.24}^{+0.39}$ & $-2.45_{-0.24}^{+0.39}$\tablenotemark{e} \\
Wilkins et al.\ (2011)\tablenotemark{b} & & $-2.3\pm0.2$ \\
Bouwens et al.\ (2012)\tablenotemark{a} & $-2.68\pm0.19$ & $-2.46\pm0.19\pm0.13$\tablenotemark{f} \\
 & $\pm0.27$ \\
Bouwens et al.\ (2010)\tablenotemark{b} & $-3.0\pm0.2$ & $-2.68\pm0.24$\tablenotemark{g} \\
Simulations\tablenotemark{h} & & $\sim-2.3$ \\
Lensed $z\sim6.2$ Galaxy\tablenotemark{i} & & $-2.5\pm0.06$
\enddata
\tablenotetext{a}{A biweight mean is used to characterize the center of the $\beta$ distribution.}
\tablenotetext{b}{A mean is used to characterize the center of the $\beta$ distribution.}
\tablenotetext{c}{A median is used to characterize the center of the $\beta$ distribution}
\tablenotetext{d}{The uncorrected measurement for $\beta$
  $-2.23\pm0.15$ appears to be biased to redder values due to Dunlop
  et al.\ (2013)'s assuming that $z\sim7$ galaxies are point sources.
  Using the measured half-light radii for faint $z\sim7$ galaxies in
  the luminosity range $-19<M_{UV,AB}<-18$ from the XDF
  ($\sim$0.074$\pm$0.013$''$), we estimate that one would derive
  $J_{125}-H_{160}$ colors that are $\sim$0.03 mag too red
  (equivalent to a $\Delta\beta \sim 0.13$ systematic error), if one treats faint
  $z\sim7$ galaxies as point sources as Dunlop et al.\ (2013) do with
  their photometric procedure (see \S5.4).}
\tablenotetext{e}{Finkelstein et al.\ (2012) estimate that due to the
  effect of noise on both their selection of sources and their $\beta$
  measurements, their $\beta$ measurements are biased blueward by
  $\Delta\beta$ $\sim0.23$, resulting in a corrected $\beta$ of
  $-2.45_{-0.24}^{+0.39}$.  It also seems likely that the results of
  Finkelstein et al.\ (2012) may be slightly biased due to utilizing
  flux constraints from passbands contaminated by Ly$\alpha$.  Scaling
  the simulation results of Rogers et al.\ (2013) to the observed
  prevalence of Ly$\alpha$ emission in $z\sim7$ galaxies (e.g.,
  Schenker et al.\ 2012), we estimate that Finkelstein et al.\ (2012)
  results could be biased blueward by $\Delta\beta\gtrsim0.1$,
  suggesting a mean $\beta$ measurement closer to $\sim-2.35$.}
\tablenotetext{f}{The $J_{125}-H_{160}$ colors of Bouwens et
  al.\ (2012) appear to have been $\sim$0.05 mag too blue as a result
  of small systematics in the empirical PSFs extracted by Bouwens et
  al.\ (2012) from the HUDF (utilized for PSF-matching the $J_{125}$
  and $H_{160}$-band observations: see \S5.3).  Correcting for this
  effect would make the Bouwens et al.\ (2012) $\beta$ determinations
  $\Delta\beta$ $\sim0.22$ redder.}
\tablenotetext{g}{Similar to Bouwens et al.\ (2012), the mean $\beta$
  reported by Bouwens et al.\ (2010) for faint $z\sim7$ galaxies is
  also likely too blue by $\Delta\beta\sim0.2$ due to a 0.05 mag bias
  in the measured $J_{125}-H_{160}$ colors (see Appendix B.2).  The
  original measurement was also subject to a slight noise-driven
  selection/measurement bias (Bouwens et al.\ 2012; Dunlop et
  al.\ 2012).  This bias appears to be significantly smaller in size
  than estimated by Rogers et al.\ (2013).  We were able to estimate
  the size of the noise-driven systematic bias by imposing the same
  5.5$\sigma$ S/N limit on the $H_{160}$-band flux of $z\sim7$
  galaxies in the Bouwens et al.\ (2010) selection as had been imposed
  on the $J_{125}$-band flux (Appendix B.2).}
\tablenotetext{h}{The cosmological hydrodynamical
  simulations of Finlator et al.\ (2011) yield a median $\beta$ of
  $-2.3$.  The Finlator et al.\ (2011) simulations have proven to be
  quite successful in forecasting a wide range of different
  observables for galaxies in the $z\gtrsim4$ universe, such as the
  evolution of the UV LF (see e.g. Bouwens et al.\ 2008) or the
  evolution of the $UV$-continuum slopes $\beta$ with cosmic time
  (e.g. Bouwens et al. 2012; Finkelstein et al. 2012).}

\tablenotetext{i}{The CLASH program reveals one highly-magnified
  $z\sim6.2$ source (Zitrin et al.\ 2012) which has a redshift and
  $UV$ luminosity very close to the $z\sim7$ selections considered in
  this table.  The existence of this source demonstrates that some
  lower luminosity, $z\sim6.2$ galaxies do have $\beta$'s as blue as
  $\sim-2.5$.  See also the quadruply-lensed $z=6.107$ source behind
  RXJ2248 which has a reported $\beta$ of $-2.89\pm0.38$ (Monna et
  al.\ 2014) and the doubly-lensed $z=6.4$ source behind MACS0717 with
  a reported $\beta$ of $-3.0\pm0.5$ (Vanzella et al.\ 2014).}
\end{deluxetable}

\begin{figure}
\epsscale{1.15} \plotone{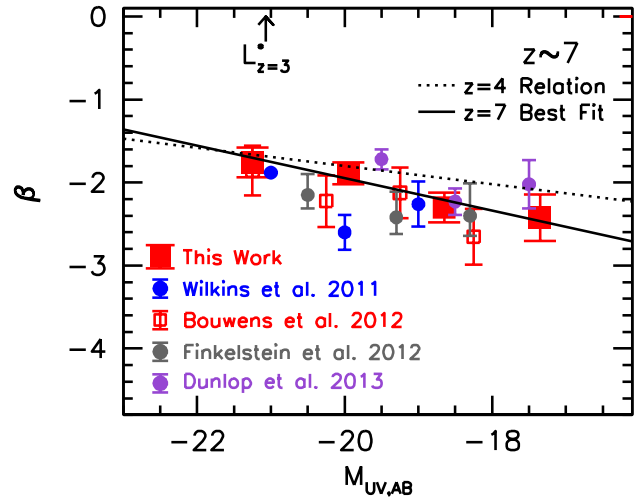}
\caption{Comparison of the mean $\beta$'s we find in various $UV$
  luminosity ($M_{UV}$) intervals (\textit{solid red squares}) with
  those from Wilkins et al.\ (2011: \textit{green circles}), Bouwens
  et al.\ (2012: \textit{open red squares}), and Finkelstein et
  al.\ (2012: \textit{gray circles}).  The solid black line indicates
  the best-fit linear relationship between $\beta$ and $UV$ luminosity
  ($M_{UV}$) for $z\sim7$ galaxies, while the dotted line shows the
  relationship we derived for $z\sim4$ star-forming galaxies (\S3.2).
  Even with the systematic biases now known to exist in previous work
  (\S5 and Appendix B), broad agreement is observed between the
  present $\beta$ determinations and earlier determinations in the
  literature.\label{fig:colmag7l}}
\end{figure}

\section{Comparison with Previous Results}

\subsection{$z\sim7$ Results: Basic Comparisons}

In \S4, we have made use of the ultra-deep XDF, HUDF09-Ps, ERS,
CANDELS-North, and CANDELS-South observations to obtain the best
available constraints on the value of the $UV$-continuum slope $\beta$
for galaxies at $z\sim7$.  The biweight mean $\beta$ we find for
sources in our faintest $z\sim7$ luminosity subsamples is
$-2.30\pm0.18\pm0.13$.  For our brightest $z\sim7$ subsamples, we find
a mean $\beta$ of $-1.75\pm0.18\pm0.13$.

Given the comprehensive nature of our analysis, large sample sizes,
and essentially bias-free methodology, we would expect our results to
be an excellent baseline for evaluating the many different
determinations from the literature (Bouwens et al.\ 2012; Wilkins et
al.\ 2011; Finkelstein et al.\ 2012; Dunlop et al.\ 2013).  A
comparison of the present results with other results in the literature
is provided in Figure~\ref{fig:colmag7l}.

\subsection{$z\sim7$ Results: Ascertaining the Nature of the Tension between 
$\beta$ Results in the Literature}

Small, but rather clear differences have existed between the $\beta$
results in the literature, particularly for the lowest luminosity
galaxies at $z\sim7$.  This has made for quite a colorful debate, with
various groups arguing strongly that the results of other groups may
be subject to one or more biases (e.g., Dunlop et al.\ 2012; Bouwens
et al.\ 2012; Finkelstein et al.\ 2012; Rogers et al.\ 2013).

In order to understand the nature of the differences between the many
$\beta$ determinations, we have conducted a comprehensive set of
comparisons with the large number of $\beta$ measurements already
presented in the literature for specific $z\sim7$ sources (Bouwens et
al.\ 2010, 2012; Finkelstein et al.\ 2012; Dunlop et al.\ 2013).
These comparisons are presented in great detail in Appendix B and in
Figures~\ref{fig:compfink}-\ref{fig:compd12} and are performed on an
object-by-object basis.  While the focus of these comparisons has been
on the $\beta$ measurements for individual $z\sim7$ galaxies, detailed
comparisons between the $\beta$ measurements for $z\sim4$, $z\sim5$,
and $z\sim6$ galaxies have also been performed (based on the results
in \S3.2) to obtain the best possible perspective on the types of
differences and systematic errors that can occur.

It is not particularly surprising given the debate in the literature
on $z\sim7$ $\beta$ results to observe modest differences in the
measured $\beta$'s values for individual $z\sim7$ sources.  In
general, the Dunlop et al.\ (2013) $\beta$ measurements are redder
than those found in Finkelstein et al.\ (2012) which are redder in
general than those found in Bouwens et al.\ (2012).  Some differences
between the measured $\beta$'s are also evident in the $z\sim4$-6
results, but in general the differences are smaller.

What is striking in the comparisons
(Figures~\ref{fig:compfink}-\ref{fig:compd12}) is that the observed
offsets between the derived $\beta$'s are more dependent on the
\textit{study} making the measurement than on the \textit{luminosity}
or the \textit{S/N} of the sources where the measurements are made.
What this suggests is that the primary explanation for the tension
that has existed between $\beta$ results in the literature are
systematics in the measurements of the $J_{125}-H_{160}$ colors for
individual sources.  Given the relatively short lever arm in
wavelength one has to establish $\beta$ from the $J_{125}$ and
$H_{160}$ photometry, even $\sim$0.04 mag systematics in the measured
$J_{125}-H_{160}$ colors are sufficient to explain the discrepancies
between the different results in the literature, since such a
systematic bias would translate into changes of $\sim0.18$ in $\beta$.

\begin{figure}
\plotone{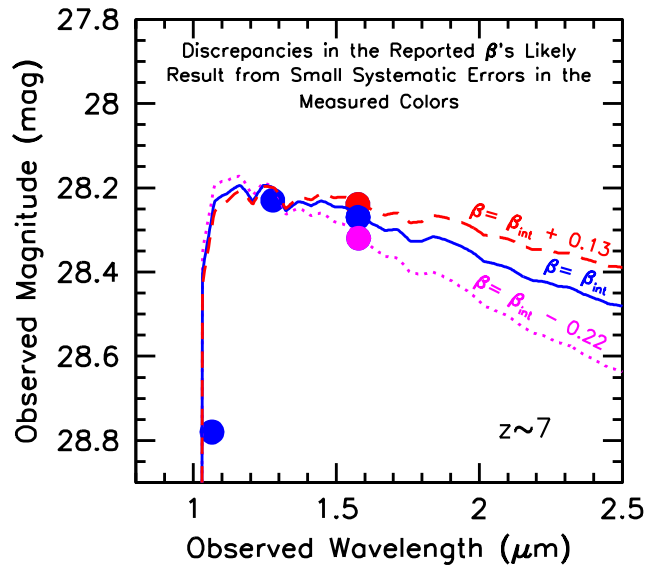}
\epsscale{1.15}
\caption{Illustration of how small $\sim$0.03-0.05-mag systematic
  errors in the measured $J_{125}-H_{160}$ colors for $z\sim7$
  galaxies can result in large differences in the measured $\beta$'s
  (\S5.2-5.4).  Each of the measured $\beta$’s shown in this figure
  relies on a similar $J_{125}$-band flux measurement, but with slight
  differences in the $H_{160}$-band flux measurements in three
  ``studies.''  Compared to the first study, the second study measures
  a $H_{160}$-band flux 0.03 mag fainter than the first study, while
  the third study measures a $H_{160}$-band flux 0.08 mag fainter than
  the first.  Both the first and third set of flux measurements only
  show very minor differences relative to the second set of flux
  measurements, yet the measured $\beta$’s for the first and third
  studies differ by $\Delta\beta\sim 0.35$, due to small ($\sim$3-5\%)
  systematics in the measured colors. The change in $\beta$ from very
  small differences in colors is dramatic.  Changes this large,
  especially if the systematic errors in the measured colors from
  competing studies are in different directions, can actually
  reconcile all previous $\beta$ measurements for faint galaxies at
  $z\sim7$ (see the text in \S5.2).\label{fig:illust}}
\end{figure}

This is illustrated in Figure~\ref{fig:illust} for three different
measurements of $\beta$ at $z\sim7$.  In this example, while each of
these measured $\beta$'s relies on a similar $J_{125}$-band flux
measurement, slight differences in the $H_{160}$-band flux
measurements are observed.  For the sake of illustration, the
$H_{160}$-band flux is increased and decreased by small amounts and an
assessment of the impact on $\beta$ is made.  While the first and
third set of flux measurements only show very minor differences
relative to the second set of flux measurements, the measured
$\beta$'s for the first and third studies differ by $\Delta\beta\sim
0.35$, due to small ($\sim$3-5\%) systematics in the measured colors.

\begin{figure}
\plotone{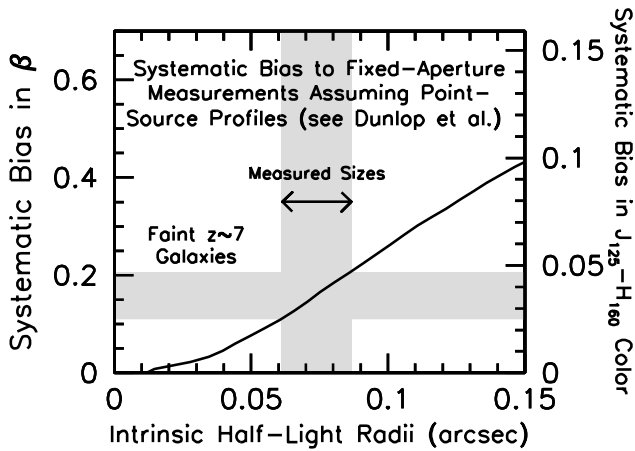}
\epsscale{1.15}
\caption{Systematic bias in the estimated $UV$-continuum slopes
  $\beta$ (and $J_{125}-H_{160}$ color) of $z\sim7$ galaxies one would
  expect as a function of the intrinsic half-light radii of sources,
  if one treats all $z\sim7$ candidates as point sources and performs
  photometry in fixed filter-dependent apertures enclosing 70\% of the
  light for a point source (e.g., Dunlop et al.\ 2013).  For sources
  with $UV$ luminosities in the $-19.3\lesssim M_{UV,AB}\lesssim -18$
  range, half-light radii estimates have ranged from 0.06$''$ to
  0.09$''$ (Oesch et al.\ 2010; Ono et al.\ 2013).  We find a
  half-light radius of 0.074$\pm$0.013$''$, by using \texttt{galfit}
  to fit a stack of the sources in this magnitude range.  Based on
  these size estimates, we find that the above photometric procedure
  would result in a systematic bias of $\sim$0.03 to $\sim$0.04 mag to
  the red in the $J_{125}-H_{160}$ color and a bias of $\sim$0.11-0.18
  in $\beta$ (see also Figure~\ref{fig:sims2}).  Correcting the Dunlop
  et al.\ (2013) measurements for this bias, we can fully reconcile
  discrepancies between their measurements and our own (see \S5.4,
  Appendix B.4.4, and
  Figure~\ref{fig:reconcile}).\label{fig:sizebias}}
\end{figure}

If the systematic errors in the measured colors from competing studies
are in different directions, we can reconcile all previous $\beta$
measurements for faint galaxies at $z\sim7$.  In particular and as we
will show in \S5.3-\S5.4, we find that the $\beta$ measurements from
Bouwens et al.\ (2012), and likely those from Bouwens et al.\ (2010),
were too blue by $\Delta\beta\sim0.22$ (see Appendix B.2-B.3) while
those from Dunlop et al.\ (2013) are too red by
$\Delta\beta\sim0.11$-0.18.  The Dunlop et al.\ (2013) bias depends on
the size or luminosity of the source (see Figure~\ref{fig:sizebias},
Appendix B.4, and Figure~\ref{fig:sims2} from Appendix E).  See also
discussion in Table~\ref{tab:disput7}.

\subsection{Systematic Biases due to Errors in the PSF}

The extensive testing we describe in Appendix B provides strong
evidence that systematics in the measured colors can successfully
explain the moderately discrepant $\beta$ results obtained in the
literature for faint ($-19\lesssim M_{UV,AB}\lesssim -18$) $z\sim7$
candidates.  Of particular relevance for the discussion are the bluer
measurements of $\beta$ provided by Bouwens et al.\ (2012), i.e.,
$\beta=-2.68\pm0.19\pm0.28$ for lower-luminosity
($-18.75<M_{UV,AB}<-17.75$) $z\sim7$ galaxies.

The tests in Appendix B.3 clearly indicate a $\sim0.05$ mag bias in
the measured $J_{125}-H_{160}$ colors from Bouwens et al.\ (2012).
The source of this systematic error appears to occur as a result of
the PSF-matching that Bouwens et al.\ (2012) perform.  To investigate
this issue in detail, we conducted detailed comparisons between the
encircled energy distributions implied by the Bouwens et al.\ (2012)
empirically-derived PSFs and standard determinations of these
encircled-energy distributions (Dressel et al.\ 2012).  We found that
the $J_{125}$-band PSF from Bouwens et al.\ (2012) showed slightly
more energy ($\sim$3\%) at intermediate radii, i.e.,
$\sim0.15$-$0.20$$''$, relative to the Dressel et al.\ (2012)
encircled-energy distributions in the $J_{125}$ band.  The wings of
the empirically-derived $J_{125}$-band PSF also contained 2\% less
flux than in the wings of the $H_{160}$-band PSF (relative to what
should have been the case given the Dressel et al.\ 2012
encircled-energy distributions).  Together these effects would cause
the measured $J_{125}$-band fluxes from Bouwens et al.\ (2012) to be
systematically bright by $\sim$0.05 mag.  This would result in $\beta$
determinations from Bouwens et al.\ (2012) that are too blue by
$\Delta\beta\sim$0.22.

\subsection{Biases Caused by Applying Point-Source Aperture Photometry to Sources with Non-Zero Size}

The tests that we describe in Appendix B.4 also revealed small
systematic biases in the photometry of Dunlop et al.\ (2013),
particularly for the most luminous sources.  We found that the
magnitude of the bias was a function of source size.  Source size does
not have an impact on the accuracy of color measurements when
performing photometry in identical apertures, after PSF-matching the
observations, as was done by Bouwens et al.\ (2012) and Finkelstein et
al.\ (2012).  However, Dunlop et al.\ (2013) utilize a different
approach for performing photometry, measuring the flux in fixed
filter-dependent circular apertures that enclose 70\% of the light for
a point source.  Dunlop et al.\ (2013) find that the 70\% encloded
flux occurred in 0.44$''$-diameter, 0.47$''$-diameter, and
0.50$''$-diameter circular apertures for the $J_{125}$, $JH_{140}$,
and $H_{160}$ band observations, respectively.\footnote{We find that
  these apertures enclose 70$\pm$1\% of the flux for the PSFs we
  derive for these bands, consistent with Dunlop et al.\ (2013).}
While this approach would provide accurate colors \textit{if} faint
$z\sim7$ galaxies were point sources, small systematics will be
present in the $J_{125}-H_{160}$ color measurements since $z\sim7$
galaxies are spatially extended.  The effect of assuming point sources
instead of using the actual profiles leads to a surprisingly
significant bias.

This bias occurs because non-zero sizes of sources result in a larger
percentage of the light being pushed to larger radii, relative to the
expectations for a point source.  We can estimate the size of this
bias by stacking galaxies from our own $z\sim7$ samples and then
deriving the encircled energy distribution.  For stacks of the
galaxies from different luminosity subsamples in our XDF data set, we
found 11-14\% more flux in 0.50$''$-diameter apertures than in
0.44$''$-diameter apertures, with the largest excesses being present
for the most extended sources.  This compares with just $\sim$8\% more
light in the larger aperture for stars.  We found essentially
identical results in all WFC3/IR bands, indicating a likely 3-6\%
systematic error in the $J_{125}-H_{160}$ colors performing the color
measurement in this way.  While small, this is sufficient to have a
noticeable effect on $\beta$.  We repeated these tests on the
Koekemoer et al.\ (2013) HUDF12 reductions and found essentially
identical results.

An alternative approach to deriving the color bias is by fitting the
$z\sim7$ stack to a model profile, measuring the half-light radii, and
then computing the bias one would expect from that half-light radius.
Let us focus on sources with UV luminosities in the $-19.3\lesssim
M_{UV,AB} \lesssim -18$ range.  Using \texttt{galfit} (Peng et
al.\ 2002) to fit the stacked profile (assuming an exponential
profile), we find a half-light radius of $0.074\pm0.013''$.  Both the
mean half-light radius and error on this measurement were made by
repeating this measurement in all four WFC3/IR bands and computing the
mean and variance (the smallest size measurement was in the $Y_{105}$
band).  Oesch et al.\ (2010) and Ono et al.\ (2013) find similar
half-light radii, i.e., 0.06$''$ to 0.09$''$.

We computed the implied encircled energy distribution for a galaxy
with this scale length by first creating a model profile with the same
scale length, convolving it with the $J_{125}$ and $H_{160}$ band
PSFs, and then computing the fraction of the total flux inside a
$0.44''$-diameter aperture in the $J_{125}$ band and $0.50''$-diameter
aperture in the $H_{160}$ band.  Then, by comparing these fractions to
the expected fraction (i.e., 70\%) for a point source, we estimated
the likely bias in the Dunlop et al.\ (2013) $J_{125}-H_{160}$ color
measurements.  Using this procedure, we concluded that there would be
a 0.03 mag redward bias in the $J_{125}-H_{160}$ colors for
$M_{UV,AB}\sim-18.6$ galaxies.  This would be equivalent to a
$\Delta\beta$ $\sim$0.13 bias in $\beta$.  The expected biases for
$z\sim7$ galaxies with other half-light radii are presented in
Figure~\ref{fig:sizebias}.  We would expect the most substantial
biases to occur for the brightest (and typically largest) sources.
The comparisons presented in Appendix B.4 and Figure~\ref{fig:compd12}
also show the trends expected based on these tests (and similarly for
the simulations presented in Figure~\ref{fig:sims2} from Appendix E).

\subsection{Reconciling $\beta$ Results for Faint Sources at $z\sim7$}

In \S5.3-5.4, we identified small systematic biases in the measured
$\beta$'s derived by Bouwens et al.\ (2012) and Dunlop et al.\ (2013).
These biases would cause the measured $\beta$'s from Bouwens et
al.\ (2012) to be $\Delta\beta\sim0.22$ too blue and the measured
$\beta$'s of Dunlop et al.\ (2013) to be $\Delta\beta\sim0.13$ too
red.  The comparisons we performed in Appendix B.1 show no
statistically-significant offset between our current results and those
of Finkelstein et al.\ (2012), though there are concerns that the
Finkelstein et al.\ (2012) $\beta$ determinations could be affected by
the use of broadband fluxes affected by Ly$\alpha$ emission (e.g., see
Figure 12 from Rogers et al.\ 2013).\footnote{While we find no
  statistically significant biases in the $z\sim7$ $\beta$
  measurements from Finkelstein et al.\ (2012), the $\beta$ results
  that Finkelstein et al.\ (2012) derive for their $z\sim6$ selections
  are $\Delta\beta\sim0.2$ bluer than what we derive (e.g., see
  Figure~\ref{fig:compfink}).  The faintest $z\sim4$ and $z\sim5$
  CANDELS $\beta$ measurements from Finkelstein et al.\ (2012) are
  redder (typically by $\Delta\beta\sim0.2$) than our own $\beta$
  measurements in this regime and similarly relative to their own
  $\beta$ results from the HUDF.  See Appendix B.1 and \S4.7 of
  Bouwens et al.\ (2012) for a discussion of this issue.}

Correcting the $\beta$ results of the Bouwens et al.\ (2012) and
Dunlop et al.\ (2013) for the biases identified above, we can
approximately reconcile all the results in the literature on $\beta$
for faint $z\sim7$ galaxies.  Our suggested revisions to published
measurements are presented in Table~\ref{tab:disput7} for galaxies in
the luminosity range $-19\lesssim M_{UV,AB} \lesssim-18$ which has
been the focus of recent debate.  The corrected results extend over
the range $\beta\sim-2.5$ to $\beta\sim-2.3$.  One can see that the
results are all consistent with each other within the $1\sigma$
statistical errors of each study.

It is worthwhile noting that this range in $\beta$, i.e., $-2.3$ to
$-2.5$, also includes the $\beta=-2.5\pm0.06$ measurement recently
obtained for one highly-magnified, lower-luminosity $z\sim6.2$ galaxy
found in the CLASH program.  The existence of this source clearly
demonstrates that some lower-luminosity galaxies in the redshift range
$z\sim6$-8 do have $\beta$'s as blue as $-2.5$.  Also found within
this range is the median $\beta=-2.3$ measurement from Jiang et
al.\ (2013) for their sample of $z\sim6.6$ LAEs (although the
intrinsic luminosities of the Jiang et al.\ 2013 LAE sample are
clearly larger, one would expect a Ly$\alpha$-emitting population to
be bluer on average than the typical luminous galaxy: Stark et
al.\ 2010).

Based on these comparisons, we can also understand why the mean
$\beta$ for faint sources measured in Bouwens et al.\ (2010) differ
from those given here or from others in the literature.  The shift
from a $\beta$ of $\sim-3.0$ to the $\sim-2.4$ (the mean $\beta$ we
measure for $z\sim7$ galaxies from the HUDF using the same binning
scheme) resulted from three effects: (1) a $\sim$0.05 mag systematic
error in the measured $J_{125}-H_{160}$ color to the blue (resulting
in a $\Delta\beta \sim 0.2$ systematic bias), (2) a
$\Delta\beta\sim0.1$ noise-driven systematic bias (see Appendix B.2),
and (3) statistical noise in the measurements due to the limited depth
of the early WFC3/IR observational data.

\subsection{Other Considerations In Comparing $\beta$ Results}

In addition to the detailed issues just discussed, other effects can
be important in comparing the central values for $\beta$.  For
example, whether one uses a mean or median to define the center of the
$\beta$ distribution needs to be considered, given that the mean
values for $\beta$ are typically $\Delta\beta \sim 0.1$ redder than
the median values, due to the tail in the $\beta$ distribution to
redder values (e.g., compare the biweight means, medians, and
inverse-variance-weighted means reported in
Table~\ref{tab:medianbeta}).  The biweight mean is our preferred
measure.

Finally, the value of $\beta$ one derives can also show some
dependence ($\Delta\beta \sim 0.2$) on the precise filter set or
wavelength range one uses to estimate $\beta$.  This is due to the
fact that the $UV$ continuum in star-forming galaxies is almost
certainly not a precise power law.  The inset to
Figure~\ref{fig:beta8} provides a direct illustration on how
significant such effects can be (and how these effects even depend on
the precise star formation history one assumes for individual
galaxies).  See also Figure 2 from Rogers et al.\ (2013) who noted
that $\Delta\beta$ differences as large as 0.2 in the measured values
for $\beta$ depending on how one made the measurement.

\textit{The point here is that $\beta$ results, particularly at the
  highest redshifts where the available leverage in wavelength is
  small, can be extraordinarily sensitive to small systematics in the
  photometry and model dependencies in one's analysis technique.}

\subsection{Summary}

In summary, by conducting a comprehensive set of comparisons with the
source-by-source $\beta$ measurements from previous studies, we have
succeeded in identifying a number of small but important systematic
errors in previous measurements of $\beta$ at $z\sim7$.  We find that
the $\beta$ measurements from Bouwens et al.\ (2012) were too blue by
$\Delta\beta\sim0.22$.  The $\beta$ measurements of Dunlop et
al.\ (2013), by contrast, are too red by $\Delta\beta\sim0.11$-0.18 in
the luminosity interval $-19<M_{UV,AB}<-18$.  The $\beta$ results of
Dunlop et al.\ (2013) likely show larger biases brightward of $-19$
and smaller biases faintward of $-18$.  While no statistically
significant systematic errors were identified in the $z\sim7$ $\beta$
measurements from Finkelstein et al.\ (2012), their results are
nonetheless susceptible to small measurement biases due to their using
flux information contaminated by Ly$\alpha$ emission (e.g., see Rogers
et al.\ 2013).  After accounting for these systematic errors, we are
able to successfully reconcile all previous results for $\beta$ at
$z\sim7$.

\section{$\beta$ Results for $z\sim8$-8.5 Samples}

The availability of very deep $JH_{140}$ observations over the HUDF
allows for the possibility of establishing the mean value of the
$UV$-continuum slope $\beta$ to $z\sim8$ and even $z\sim8.5$.  These
new $\beta$ determinations should provide us with additional leverage
for constraining the evolution of the $UV$-continuum slope $\beta$ to
early times.

\subsection{Methodology for $z\sim8$ Sample}

\begin{figure}
\epsscale{1.15}
\plotone{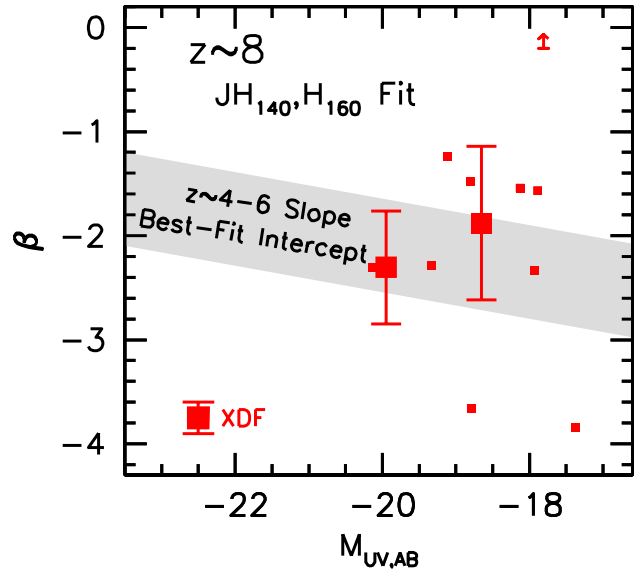}
\caption{(\textit{left}) Measured $\beta$'s versus the rest-frame $UV$
  luminosity $M_{UV}$ for galaxies in our $z\sim8$ sample from the
  XDF.  The $\beta$'s we present here are based on the
  $JH_{140}-H_{160}$ colors of $z\sim8$ sources in our samples.
  Sources which yield a $\beta$ in the range of this figure are shown
  as the small red squares, while the one source with derived
  $\beta$'s beyond the vertical bounds of this plot are shown as a
  lower limit.  The large red squares give the weighted-mean
  $\beta$'s observed in various bins of $UV$ luminosity.  Despite the
  small amount of wavelength leverage available using $JH_{140}$ and
  $H_{160}$-band photometry, the $JH_{140}-H_{160}$ color allows for a
  clean measurement of $\beta$ in $z\sim8$ sources.  It is not
  possible to obtain an entirely clean measurement of $\beta$ for
  $z>8$ galaxies, if one includes the $J_{125}$-band flux in
  estimating $\beta$ (see Figure~\ref{fig:beta8}).  The gray-shaded
  band shows the $1\sigma$ range allowed for a fit to the
  $\beta$-$M_{UV}$ relation fixing the slope of this relation to the
  average value ($-0.13$) found by Bouwens et al.\ (2012) for
  $z\sim4$-6 galaxies.\label{fig:colmag8a}}
\end{figure}

Our $z\sim8$ selection criteria are as follows:
\begin{displaymath}
(Y_{105}-J_{125}>0.8)\wedge(J_{125}-H_{160}<0.5)
\end{displaymath}
We constructed these criteria to be complementary to those used to
select sources in our $z\sim7$ sample, so these criteria are useful in
selecting galaxies with redshifts just higher than those selected by
our $z\sim7$ criteria.  Sources must also by necessity be undetected
at $<2\sigma$ in each optical/ACS band.  Finally, we require the
computed $\chi^2$ statistic for the sources (computed from their
$B_{435}V_{606}i_{775}I_{814}z_{850}$ fluxes) not be greater than 3 in
any of the three apertures we use ($0.18''$-diameter aperture, scaled
aperture, $0.35''$-diameter aperture).\footnote{Since our selection
  criteria at $z\sim8$ include the $H_{160}$-band flux (also used to
  derive $\beta$), we would in principle expect a small bias in our
  $z\sim8$ $\beta$ measurements.  However, since our
  $J_{125}-H_{160}<0.5$ limit would only be relevant for an
  exceedingly small fraction of the sources to the limit of our
  $z\sim8$ probe (i.e., $\sim$$-$18 mag), the bias would be very small
  ($\Delta\beta\lesssim0.1$).}

Our primary method for deriving $UV$-continuum slopes for $z\sim8$
galaxies is based on a power-law fit to the $JH_{140}$ and $H_{160}$
band fluxes.  Given that we have only two constraints on the power law
slope using fluxes in these bands, this approach is functionally
equivalent to using a fitting formula
\begin{equation}
\beta = -2.0+8.98(JH_{140}-H_{160}).
\end{equation}
The coefficients in this fitting formula are derived assuming galaxies
have a $\beta=-2$ spectrum.\footnote{This is a slightly more accurate
  prescription than the formula given in Dunlop et al.\ (2013) based
  on the pivot wavelengths.}  Given that Ly$\alpha$ emission or the
Lyman break only begin to enter the $JH_{140}$ band at $z\gtrsim9$,
the $JH_{140}-H_{160}$ color should allow for a totally clean
measurement of the spectral slope $\beta$ in $z\sim8$ sources.

The challenge with this approach is the very small effective
wavelength difference between the $JH_{140}$ and $H_{160}$ filters.
This makes the derived $\beta$'s extremely sensitive to any
uncertainties or systematics in the measured $JH_{140}-H_{160}$
colors.  Minimizing the systematics is therefore essential.  For this
reason, care was taken not only in PSF-matching the observations
(Appendix A), but in testing the accuracy of the $JH_{140}-H_{160}$
color measurements.  The most relevant test here was comparing the
$UV$-continuum slopes $\beta$ we derived based on the
$JH_{140}-H_{160}$ colors for $z\sim7$ galaxies with $UV$-continuum
slopes $\beta$ derived based on the $J_{125}$, $JH_{140}$, and
$H_{160}$ photometry.  A median difference of $\Delta\beta\sim 0.05$
was found between the two sets of $\beta$'s, with sources with the
$JH_{140}$+$H_{160}$-derived $\beta$'s being slightly redder overall.
Since such differences are well within the expected statistical
uncertainties, this test strongly suggested that our
$JH_{140}-H_{160}$ colors did not suffer from large systematic biases.

One alternate approach for deriving the $UV$-continuum slopes $\beta$
at $z\sim8$ involves leveraging the available $J_{125}$, $H_{160}$,
and $JH_{140}$ (where available) photometry for sources.  Such an
approach has previously been utilized by Dunlop et al.\ (2013).  The
primary advantage of using such an approach is the much greater
leverage in wavelength one has taking advantage of the $J_{125}$-band
photometry.  $J_{125}$-band observations are also more widely
available over legacy fields (e.g., CANDELS and HUDF09-2) than are
$JH_{140}$-band observations.  This approach, unfortunately, suffers
from one major systematic drawback.  The measured $\beta$'s can be
subject to large redward biases, e.g., $\Delta\beta\sim1$, if the
galaxies one is examining have redshifts in excess of $z\gtrsim8$.
The large biases result from the effect of the IGM on the
$J_{125}$-band flux.  Figure~\ref{fig:beta8} from Appendix G
illustrates how large such biases can be.  While these large biases
may seem straightforward to avoid, uncertainties in the photometric
redshifts of individual sources make it challenging to determine the
precise correction that should be applied to the $J_{125}$-band fluxes
of individual sources.

Because of these biases and the significant assumptions required to
correct for them, no use of the $J_{125}$-band fluxes is made in
quantifying $\beta$ for our baseline $z\sim8$ study.  For
completeness, however, we consider such an approach in Appendix G.

\subsection{$\beta$ Results for $z\sim8$ Sample}

The $\beta$ measurements we obtain for individual galaxies in our
$z\sim8$ samples are presented in Figure~\ref{fig:colmag8a}.  Also
shown are the biweight mean $\beta$'s for the $z\sim8$ galaxies in
specific bins in $UV$ luminosity.  The results for the median
$\beta$'s are similar, albeit slightly redder in the $M_{UV}=-18.5$
bin.  We determine the linear relationship which best fits the mean
$\beta$'s we find at a given $M_{UV}$ luminosity.  Due to the large
uncertainties on the binned $\beta$ determinations, no attempt is made
to determine the slope of the $\beta$ vs. $M_{UV}$ relationship at
$z\sim8$.  We simply assume that the slope of this relationship at
$z\sim8$ is the same as the average slope we find for this
relationship at $z\sim4$-6, i.e., $-0.15$ (e.g.,
Table~\ref{tab:slopes}).  The binned $\beta$'s and coefficients to the
best-fit $\beta$ vs. $M_{UV}$ relationship are presented in
Table~\ref{tab:medianbeta} and \ref{tab:slopes}.

The inverse-variance-weighted mean $\beta$ we measure for
lower-luminosity ($\sim-18$ mag) $z\sim8$ galaxies is
$-$1.88$\pm$0.74$\pm$0.27 using only the $JH_{140}$+$H_{160}$-band
flux information.  In Appendix G, using the flux information in the
$J_{125}$+$JH_{140}$+$H_{160}$ bands (after correcting for the impact
of the IGM), we find $-2.39\pm0.35\pm0.13$ for faint $z\sim8$ galaxies
in this same magnitude interval.  The current measurements are
consistent with the $\beta=-2.03_{-0.38}^{+0.46}$ and
$\beta=-1.88_{-0.56}^{+1.03}$ measurements obtained by Finkelstein et
al.\ (2012) for $z\sim8$ galaxies found over a similar range of
luminosities, though the errors are so large in both data sets that
consistency does not indicate robust agreement.  The
$\beta=-1.9\pm0.3$ measurements from Dunlop et al.\ (2013) are also
consistent with the $\beta$'s we derived here.  However, for the
Dunlop et al.\ (2013) study, a slight revision to the $z\sim8$ $\beta$
results is required to correct for the slight bias in $\beta$ due to
non-zero sizes of $z\sim8$ sources (e.g., see
Figure~\ref{fig:sizebias}, \S5.4, and Appendix B.4) and to correct for
the effect of the IGM on the $J_{125}$-band fluxes of $z\sim8$
galaxies in their samples (see Figure~\ref{fig:beta8} and Appendix G).
Correcting for both effects results in a revision of their mean
$\beta$ measurement at $z\sim8$ to $\beta\sim-2.1\pm0.3$.

Regardless of the approach and dataset the current $z\sim8$
derivations of $\beta$ are all highly uncertain and do not add
particularly strong constraints because of the small sample sizes,
sensitivity to photometric uncertainties because of the limited filter
separation, and systematic biases.  The most useful constraints in the
reionization epoch are still to be found at $z\sim7$.

\subsection{$\beta$ Results for a Small $z\sim8.5$ Sample}

Finally, it is possible to try to take advantage of the small number
of $z\sim8.5$-9 candidates that have recently been identified in the
literature (Zheng et al.\ 2012; Bouwens et al.\ 2013; Ellis et
al.\ 2013; Oesch et al.\ 2013b) to see what constraints can be set on
$\beta$ for galaxies in this redshift range.

Measuring $\beta$ for galaxies at $z\sim8.5$ and higher is extremely
challenging at present.  There are a number of reasons for this: (1)
the small number and extreme faintness of most $z\sim9$ candidates,
(2) the short lever arm in wavelength to constrain $\beta$, and (3)
the rather significant effect the IGM would have on the observed
fluxes in the $JH_{140}$-band, if any of the candidates had redshifts
in excess of $z\sim9$.  This final effect is particularly important,
since even a 10\% lower flux in the $JH_{140}$-band due to absorption
from the IGM (expected for sources at $z\sim9.3$) would bias the
measured $\beta$'s by the large factor $\Delta\beta\sim0.9$.  To
successfully correct for this bias would require accurate constraints
on the redshifts for individual sources.  Unfortunately, almost every
$z\sim9$ candidate known has redshift uncertainties of $\sigma(z)
\gtrsim 0.7$, so even corrections for the average source would not be
particularly effective at eliminating this bias.

Due to the very large random and potential systematic uncertainties in
estimates of $\beta$ at $z\sim9$, it is perhaps best if we only
consider those $z\sim8.5$ candidates which are the brightest and
photometrically well-constrained.  Given that the majority of the
$z\sim9$ candidates in the Oesch et al.\ (2013b) and Ellis et
al.\ (2013) samples have a S/N of $\sim$3-4 in individual bands, this
leaves us with just two $z\sim9$ galaxy candidates that we can examine
where our $\beta$ measurements would be likely clean, i.e.,
XDFyj-38135540 and XDFyj-39478076 from Oesch et al.\ (2013b).
Unfortunately, all three lensed candidate $z\sim9$ galaxies from the
CLASH program (Postman et al.\ 2012), i.e., MACS1149-JD ($z\sim9.7$),
MACSJ1115-JD1 ($z\sim9.2$), and MACSJ1720-JD1 ($z\sim9.0$: Zheng et
al.\ 2012; Bouwens et al.\ 2013) do not qualify since their estimated
redshifts are quite likely in excess of $z\sim9$ (particularly
MACS1149-JD), and therefore even their $JH_{140}$-band fluxes are
substantially affected by IGM absorption.

For these two sources, we estimate $\beta$ using the
$JH_{140}-H_{160}$ color using the same formula
$\beta=8.98(JH_{140}-H_{160})-2$ we presented in \S6.2 for $z\sim8$
sources in our samples.  Based on these two $z\sim8.5$ sources, we
derive an inverse-variance-weighted mean $\beta$ of $\sim-2.1\pm0.9$.
The mean $\beta$ is $-1.8$.  Given the large uncertainties, this
determination is consistent with our $z\sim7$ and $z\sim8$
determinations.

Previously, Dunlop et al.\ (2013) have presented an approximate
measurement of the mean $\beta$ for galaxies at $z\sim9$ based on a
small number of sources from the XDF.  While the present measurement
is, of course, consistent with the Dunlop et al.\ (2013) estimate, the
present estimate of the mean $\beta$ at $z\sim8.5$ is more reliable,
though it is not particularly meaningful given the uncertainties.
First of all, Dunlop et al.\ (2013) have included a large number of
galaxies with photometric redshifts consistent with being at $z\sim9$.
Including these sources is dangerous, since as we emphasized earlier
any candidate with a redshift in excess of $z\sim9$ would be strongly
biased (e.g., $\Delta\beta\gtrsim1$) to redder $\beta$ values (see
e.g. Figure~\ref{fig:beta8}).  It is therefore concerning that the
$z\sim9$ sample of Dunlop et al.\ (2013) includes one and possibly two
$z\sim9$ candidates generally agreed to be at $z>9$ (Ellis et
al.\ 2013; Oesch et al.\ 2013b).  This will result in a clear redward
bias.

\begin{figure}
\epsscale{1.15}
\plotone{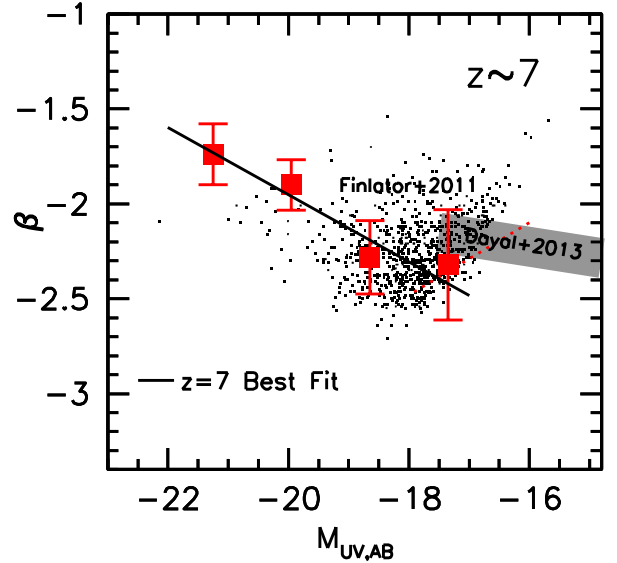}
\caption{A comparison of the mean $UV$-continuum slopes $\beta$
  vs. $UV$ luminosity relationship found here for our $z\sim7$ sample
  (\textit{red solid squares}) with that predicted in the Finlator et
  al.\ (2011) hydrodynamical simulations (\textit{black points}).  The
  dotted red line shows where incompleteness becomes important for the
  results from the simulations.  The mean $\beta$ expected in the $UV$
  luminosity range $-19<M_{UV,AB}<-18$ by the simulations is
  $\sim-2.3$.  Also shown on this plot are the predicted trends at
  $M_{UV}>-18$ from the simulations of Dayal et al.\ (2013:
  \textit{gray shaded region}).\label{fig:finlator}}
\end{figure}

Second, as Oesch et al.\ (2013b) have demonstrated, the $z\sim9$
sample used by Dunlop et al.\ (2013: see also Ellis et al.\ 2013)
includes a source HUDF12-4106-7304 whose flux is substantially boosted
(i.e., factor of $\sim$2) by a diffraction spike from a bright nearby
galaxy (Figure 9 from Oesch et al.\ 2013b).  Not only does this cast
considerable doubt on the reality of this source (now only $2.8\sigma$
significance), but it would bias its photometry as well.

For the above reasons, we consider the current measurement of the mean
$\beta$ at $z\sim8.5$ to be the highest redshift, reliable estimate of
$\beta$ at present (although given the large uncertainties on this
measurement it is not really meaningful at present).

\section{Discussion and Physical Implications}

In this present analysis, we have made use of the extremely deep
WFC3/IR observations over the XDF, HUDF09-Ps, CANDELS-North,
CANDELS-South, and ERS fields to establish the distribution of
$UV$-continuum slopes $\beta$ for $z\sim4$-8 galaxies while ensuring
that the systematic errors are minimized.

As in our previous study, we found strong evidence for a correlation
between $\beta$ and $UV$ luminosity for star-forming galaxies at
$z\sim4$, $z\sim5$, $z\sim6$, and $z\sim7$, with higher luminosity
galaxies being redder and lower luminosity galaxies being bluer at all
redshifts.  The existence of such a relationship is likely largely
driven by changes in the mean metallicity or dust extinction of
galaxies as galaxies grow in stellar mass and $UV$ luminosity (Bouwens
et al.\ 2009; Bouwens et al.\ 2012; Finkelstein et al.\ 2012).  The
slope of the $\beta$ vs. $UV$ luminosity relationship appears to be
roughly constant as a function of redshift or cosmic time (e.g., see
Figure~\ref{fig:slope}).

\begin{figure}
\epsscale{1.15}
\plotone{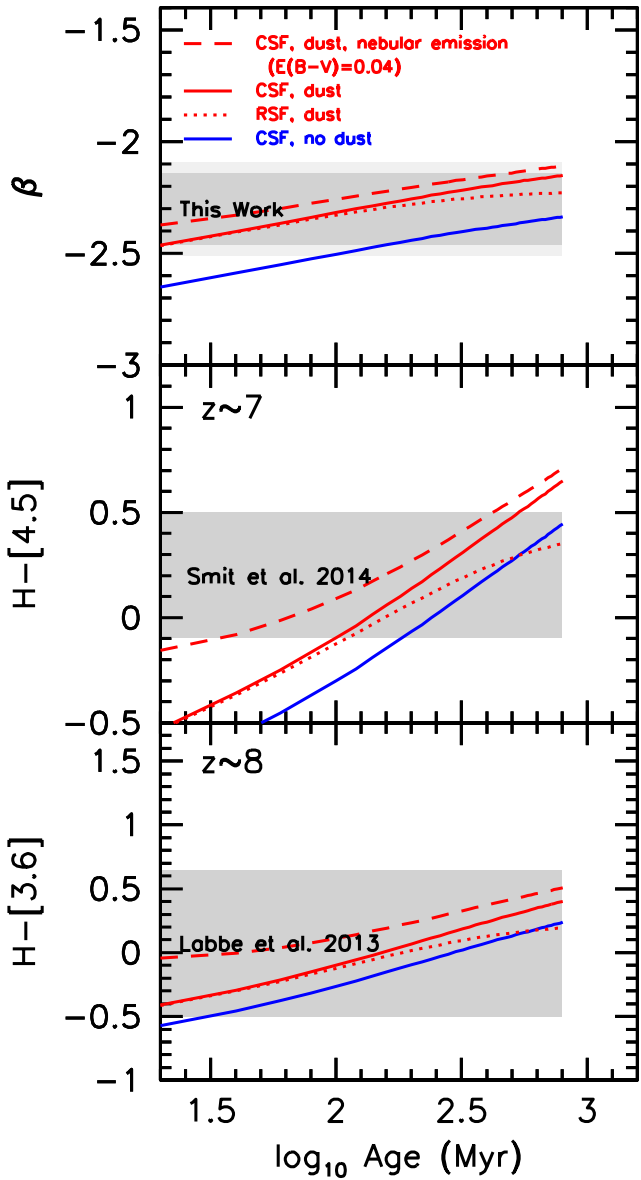}
\caption{(\textit{upper}) Predicted $UV$-continuum slope $\beta$ for a
  constant star formation model versus age.  Shown are the predictions
  assuming no dust extinction (\textit{blue line}) and for a
  $E(B-V)=0.04$ model (\textit{red line}).  Model tracks are shown
  from zero age to the age of the universe at $z\sim7$.  0.5
  $Z_{\odot}$ metallicity is assumed, similar but slightly higher than
  the predictions for faint galaxies at high redshift (Finlator et
  al.\ 2011; Dayal \& Ferrara 2011; Wise et al.\ 2012).  The dark
  shaded gray region shows the 68\% confidence interval on the mean
  value for $\beta$ observed in faint ($-19.3\lesssim M_{UV,AB}
  \lesssim -18.0$) $z\sim7$ galaxies.  Also shown are the predictions
  for an exponentially-increasing star formation history (\textit{red
    dotted line}) with $\tau=5\times10^8$ years and including
  nebular-continuum emission (\textit{red dashed line}: see Oesch et
  al.\ 2013a).  (\textit{middle}) Predicted $H_{160}-[4.5]$ colors for
  the same constant star formation models as shown in the upper panel.
  The 68\% confidence interval on the average $H_{160}-[4.5]$ color of
  $z\sim7$, $M_{UV}\sim -20$ galaxies from the CLASH program (Smit et
  al.\ 2014) is indicated by the shaded region.  (\textit{lower})
  Predicted $H_{160}-[3.6]$ colors for the same constant star
  formation models as shown in the upper panel.  The 68\% confidence
  interval on the average $H_{160}-[3.6]$ color of $z\sim8$,
  $M_{UV}\sim -19$ galaxies from the HUDF09 fields (Labb{\'e} et
  al.\ 2013) is indicated by the shaded region.  The mean $\beta$'s
  and $H_{160}-IRAC$ colors we derive here for lower luminosity
  $z\sim7$ galaxies are broadly consistent with stellar population
  ages from 50 Myr to 600-800 Myr (age of the universe at $z=7$-8) and
  dust extinction from $E(B-V)=0.0$ to $E(B-V)\sim0.06$.
\label{fig:interp}}
\end{figure}

The primary focus of this study was to establish reliable and
bias-free measures of $\beta$ for $z\sim7$ galaxies at very low
luminosities.  To support this goal, we also leveraged the
deepest-ever optical+near-IR HST observations to establish the mean
$UV$ slope $\beta$ for faint galaxies in four other redshift intervals
$z\sim4$, $z\sim5$, $z\sim6$, and $z\sim8$ to determine the trend with
redshift.  Mean $\beta$'s of $-2.03\pm0.03\pm0.06$,
$-2.14\pm0.06\pm0.06$, $-2.24\pm0.11\pm0.08$, and
$-2.30\pm0.18\pm0.13$ are found for faint galaxy selections at
$z\sim4$, $z\sim5$, $z\sim6$, and $z\sim7$ galaxies, respectively.
The mean $UV$ slopes of the faint galaxies at $z\sim8$ and $z\sim8.5$
are consistent with the values found above, but are too uncertain to
play a role in any discussion of the trends.

Similar to the previous results presented in Bouwens et al.\ (2012),
these results are consistent with the mean $UV$-continuum slope
$\beta$ of galaxies only evolving gradually as a function of cosmic
time, if one examines galaxies at the same $UV$ luminosity or stellar
mass at all epochs.  Figure~\ref{fig:faintevol} provides a good
illustration of the observed trends.

The $UV$ slopes of the faint galaxies we find in $z\sim8$ and
$z\sim8.5$ selections are consistent with the blue values found above,
but are nonetheless quite uncertain.  Consistent with these trends,
the faintest galaxies at $z\sim7$ show evidence for exhibiting
somewhat bluer $\beta$'s, i.e., $\beta\sim-2.3$-2.4 at $z\sim7$-8.
This is evident for our results from the XDF and HUDF09-Ps fields and
for our $-19.3<M_{UV,AB}<-18.0$ and $-18.0<M_{UV,AB}<-16.7$ subsamples
from those fields.

At lower luminosities, our mean $\beta$ results at $z\sim7$ are in
excellent agreement with the results from cosmological hydrodynamical
simulations, such as those performed in Finlator et al.\ (2011), as
shown in Figure~\ref{fig:finlator}.  The agreement is less easy to
discern in the faintest luminosity bin, due to incompleteness in the
simulation results at low masses.  However, in the luminosity bin
$-19.3<M_{UV,AB}<-18.0$, the mean $\beta$ from the simulations is
$\sim-$2.33, consistent with the mean $\beta$ we report
$-2.30\pm0.18\pm0.13$.  Also shown on this diagram are the results
from Dayal et al.\ (2013) for very faint $M_{UV,AB}>-18$ galaxies at
$z\sim7$ and again the agreement is excellent.

What do the present results imply for the stellar populations of faint
$z\sim7$-8 galaxies?  In Figure~\ref{fig:interp}, we include a figure
showing the predicted $UV$-continuum slope $\beta$ for galaxies as a
function of the stellar population age assuming a constant rate of
star formation.  The results are shown for both the case with no dust
extinction ($E(B-V)=0.0$) and a small amount of dust extinction
($E(B-V)=0.04$).  Dust extinction is implemented using the Calzetti et
al.\ (2000) prescription.  The metallicity of galaxies is assumed to
be 0.5 $Z_{\odot}$, which is similar but slightly higher than the
expectations of simulations (Dav{\'e} et al.\ 2006; Finlator et
al.\ 2011).

Comparing the mean $\beta$'s we derive for the faintest galaxies at
$z\sim7$ with the model predictions, the observed $\beta$'s are
suggestive of a dust extinction that is not substantially higher than
$E(B-V)=0.06$.  No significant constraints can be set on the mean age
of the stellar population.  On the basis of a redder derived $\beta$,
Dunlop et al.\ (2013) favored a model with slightly higher dust
extinction than what we prefer here (equivalent to $E(B-V)\sim0.05$).

We can also potentially gain insight into the stellar populations of
$z\sim7$-8 galaxies by comparing against the observed rest-frame $UV$
to optical colors of $z\sim7$ and $z\sim8$ galaxies.  This requires we
make use of flux measurements with IRAC to constrain the rest-frame
optical fluxes.  This is somewhat challenging due to the strong
emission lines in these sources that contaminate the observed IRAC
fluxes (e.g., Schaerer \& de Barros 2009; de Barros \& Schaerer 2014;
Gonz{\'a}lez et al.\ 2012, 2014; Stark et al.\ 2013; Labb{\'e} et
al.\ 2013).  Fortunately, there are specific redshift windows where
one can obtain relatively clean measurements of the rest-frame optical
flux (e.g., Stark et al.\ 2013).  Smit et al.\ (2014) exploited one
such redshift window $z\sim6.6$-7.0 to obtain a constraint on the
$H_{160}-[4.5]$ color for moderately faint ($\sim-20.0$ mag) $z\sim7$
galaxies (albeit slightly brighter than we consider here).  Meanwhile,
Labb{\'e} et al.\ (2013) exploited another such redshift window
$z\gtrsim7.1$ to obtain a relatively clean constraint on the
$H_{160}-[3.6]$ color for faint $H_{160,AB}\sim28$ galaxies.

The observational constraints are shown in the middle and lowest
panels of Figure~\ref{fig:interp} and can be compared with the results
from stellar population models.  No strong constraints on the age of
the stellar population can be set.  Ages from $\sim$50 Myr to 800 Myr
(age of the universe) are formally allowed, though the real upper
limit is set by the onset of the earliest significant star formation
in the universe (which is currently thought to occur at ages around
150-200 Myr after the Big Bang).  For stellar population ages of
$\sim$200 Myr, the joint constraints on $\beta$ and $UV$-to-optical
colors suggest a non-zero dust extinction in these sources, with a
best-fit value of $E(B-V)\sim0.02$.  Unfortunately, the available IRAC
observations are simply not deep enough at present to set strong
constraints on UV-to-optical colors of $z\sim7$-8 galaxies and hence
their stellar population models.  Dunlop et al.\ (2013) also drew
somewhat similar conclusions to these, despite modest differences in
their preferred values for $\beta$ (see also the simulation results
from Wilkins et al.\ 2013).

As in our previous study (e.g., Bouwens et al.\ 2012), the present
results provide no significant evidence for exotic or unusual stellar
populations, consistent with the results from many recent studies
(Finkelstein et al.\ 2010, 2012; Wilkins et al.\ 2011; Dunlop et
al.\ 2012).

\begin{deluxetable}{ccc}
\tablewidth{0cm}
\tabletypesize{\footnotesize}
\tablecaption{Systematic Biases that have likely affected previous
  studies of $\beta$ at
  $z\sim7$.\tablenotemark{a}\label{tab:biassummary}}
\tablehead{\colhead{} & \colhead{Systematic} & \colhead{Magnitude
    of}\\
\colhead{} & \colhead{(potential \&} & \colhead{Bias}\\
\colhead{Paper} & \colhead{demonstrated)} & \colhead{($\Delta\beta$)}}
\startdata
Bouwens et al.\ 2010 & Photometric Error & $-$0.1\tablenotemark{b} \\
 & Coupling Bias\tablenotemark{c} & \\
 & PSF-Matching Bias\tablenotemark{d} & $-$0.22 \\
Finkelstein et al.\ 2010 & Photometric Error & $-$0.5(?)\tablenotemark{e} \\
                         & Coupling Bias\tablenotemark{c} & \\
   & Selection Volume Bias\tablenotemark{f} & Uncorrected\tablenotemark{g} \\ 
 Wilkins et al.\ 2011 & Photometric Error & Uncorrected\tablenotemark{g}\\
 & Coupling Bias\tablenotemark{c} & \\
Dunlop et al.\ 2012 & Photometric Error & Uncorrected\tablenotemark{g,h}\\
 & Coupling Bias\tablenotemark{c} & \\
 & Selection Volume Bias\tablenotemark{f} & Uncorrected\tablenotemark{g} \\ 
Bouwens et al.\ 2012 & PSF-Matching Bias\tablenotemark{d} & $-$0.22 \\
Finkelstein et al.\ 2012 & Ly$\alpha$ contamination & Uncorrected\tablenotemark{g,j} \\ 
& Bias\tablenotemark{i} \\
 & Selection Volume Bias\tablenotemark{f} & Uncorrected\tablenotemark{g} \\
Dunlop et al.\ 2013 & Non-Zero Size Bias\tablenotemark{k} & $+$0.13 \\
& Selection Volume Bias\tablenotemark{f} & Uncorrected\tablenotemark{g}
\enddata
\tablenotetext{a}{Another possible bias one could consider is a
  contamination bias (the bias one would expect from the small
  fraction of contaminants present at low levels in high-redshift
  selections).  Because corrections for this bias would be highly
  model dependent (though likely small), no study (including the
  present one) has corrected for it.}
\tablenotetext{b}{We estimate the size of this bias in Appendix B.2 by applying
a similar S/N cut in the $H_{160}$ band as was applied in the $J_{125}$ band to the original catalog from Bouwens et al.\ (2010).  See Appendix B.2}
\tablenotetext{c}{Biases will
  be present in the measurement of $\beta$ if noise in the measured
  flux of $z\sim7$ sources can affect both the selection of sources
  and the measurement of $\beta$.  For example, if the same noise
  fluctuation can cause a source to look both bluer (for the purposes
  of its being selected as a high-redshift candidate) and bluer (for
  the purposes of measuring its $\beta$), then the $\beta$ results
  will be biased.  See Appendix B.2.  See also Dunlop et al.\ (2012)
  and Appendix B.1.2 of Bouwens et al.\ (2012).}
\tablenotetext{d}{Small systematics in the PSFs Bouwens et al.\ (2010,
  2012) used to do the PSF matching resulted in $\beta$ measurements
  that were too blue.  See \S5.3, Appendix B.2, and Appendix B.3.}
\tablenotetext{e}{This is a rough estimate of the bias, drawing on simulations presented in Dunlop et al.\ (2012), Bouwens et al.\ (2012: Appendix D), and Rogers et al.\ (2013).}
\tablenotetext{f}{The selection
  volume bias (Appendix B.1.1 of Bouwens et al.\ 2012) occurs because sources with certain
  intrinsic colors are less likely to be selected as high-redshift
  candidates and therefore the mean $\beta$ from a high-redshift
  selection is biased.  Corrections for this bias are very uncertain
  at present and highly model dependent since they require an accurate
  knowledge of the volume density of high-redshift galaxies with
  intrinsically red $\beta$'s.  The selection volume bias is likely
  small (particularly at lower luminosities) since there is little
  evidence that a modest fraction of faint $z\sim4$-7 galaxies are
  especially red.  See also Appendix F.}
\tablenotetext{g}{The
  indicated results are subject to this bias, but we do not estimate
  the size of the bias here.}
\tablenotetext{h}{This bias should not
  be especially large, due to Dunlop et al.\ (2012) restricting their
  analysis to the highest-significance $z\sim7$ candidates.}
\tablenotetext{i}{Contribution of Ly$\alpha$ emission to the
  $Y_{105}$-band flux of $z\sim7$ sources could bias $\beta$
  measurement procedures that make use of the $Y_{105}$-band flux. See
  Appendix B.1 and Figure 12 from Rogers et al.\ (2013).}
\tablenotetext{j}{The $\beta$ measurements for $z\sim6$ galaxies by
  Finkelstein et al.\ (2012) show a consistent $\Delta\beta\sim0.2$
  blueward offset relative to our own measurements comparing sources
  both in the HUDF/XDF and in the CANDELS-South field (see
  Figure~\ref{fig:compfink} and \ref{fig:compfinkg}).  This bias
  appears to be significant and could result from Ly$\alpha$ emission
  contaminating the broadband fluxes of $z\sim6$ galaxies, though it
  can also arise from other issues.}
\tablenotetext{k}{The Dunlop et al.\ (2013) $\beta$'s are measured to 
be too red, since faint $z\sim7$ sources have non-zero sizes and the 
Dunlop et al.\ (2013) photometric procedure assumes point sources.  
See \S5.4, Figure~\ref{fig:sizebias}, and Appendix B.4.}
\end{deluxetable}

\begin{figure}
\epsscale{1.15}
\plotone{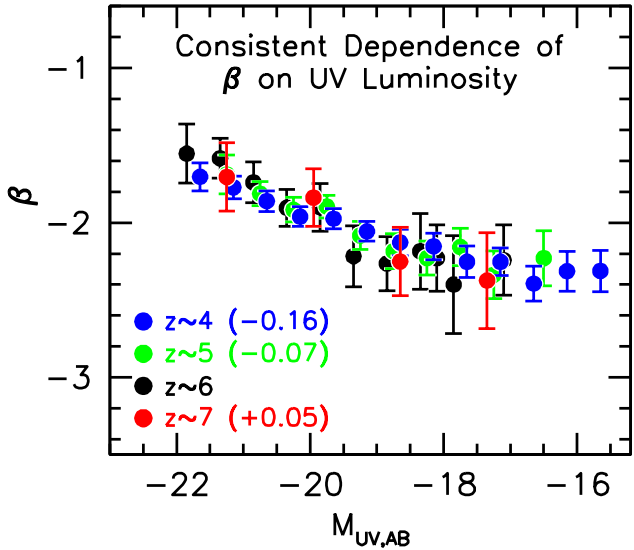}
\caption{Biweight mean $UV$ continuum slope $\beta$ derived versus
  $UV$ luminosity for our $z\sim4$, $z\sim5$, $z\sim6$, and $z\sim7$
  galaxy samples (see \S3.2 and \S4.7).  Slight offsets to the $\beta$
  results from our $z\sim4$, $z\sim5$, and $z\sim7$ samples have been
  applied, i.e., $\Delta\beta\sim-0.16$, $\Delta\beta\sim-0.07$,
  $\Delta\beta\sim0.05$, to better illustrate the consistent
  dependence on luminosity.  Only a small amount of evolution in the
  $\beta$ vs. $M_{UV}$ relationship is observed as a function of
  redshift.  It is clear that $\beta$ shows a consistent dependence on
  $UV$ luminosity at all redshifts, with a clear trend to bluer colors
  at lower luminosities.  The dependence of $\beta$ on luminosity
  becomes weaker at the lowest luminosities (see
  Figure~\ref{fig:brokenline} and \S3.4).\label{fig:summary4567}}
\end{figure}

\begin{figure*}
\epsscale{1.15}
\plotone{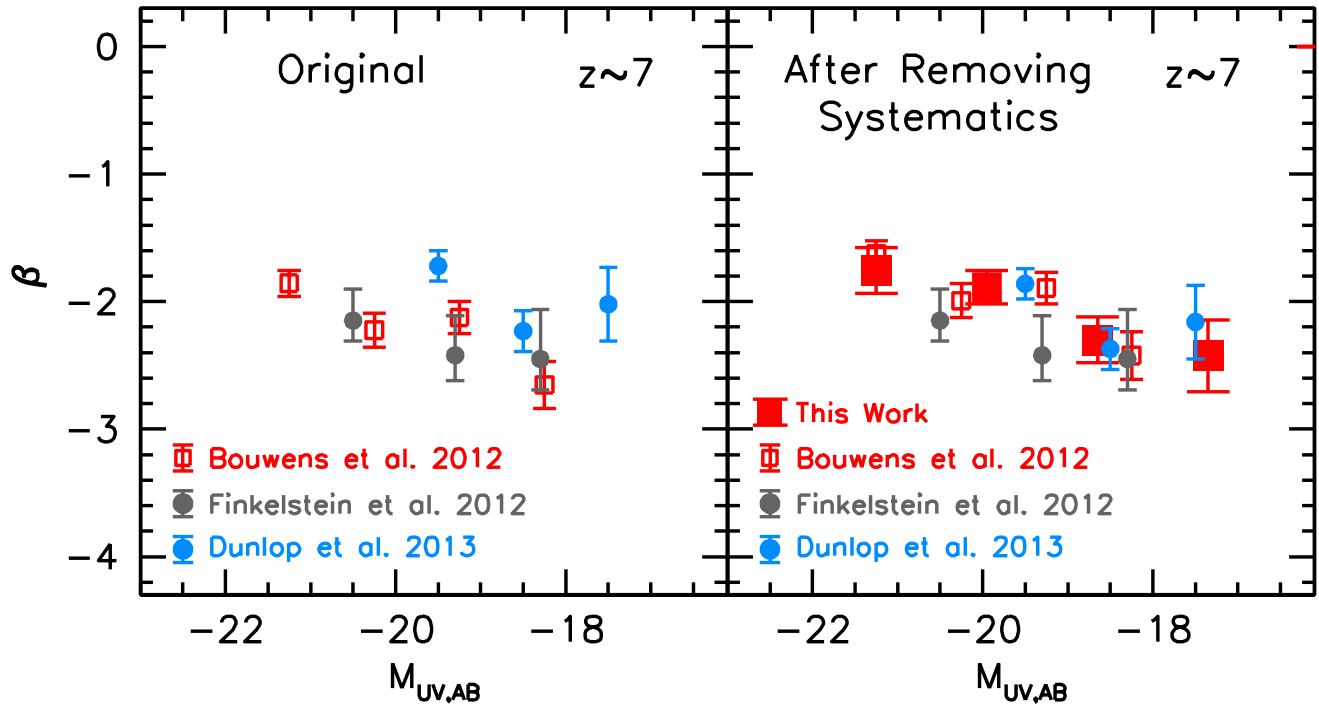}
\caption{Mean $\beta$ as a function of $UV$ luminosity, before
  correction for the systematic biases we identified in this study
  (\textit{left}) and after correction (\textit{right}).  The two
  prominent biases that we have identified in the $z\sim7$ $\beta$
  measurements are a $\Delta\beta\sim0.22$ blueward bias in the
  Bouwens et al.\ (2012) measurements (see Appendix B.3) and
  $\Delta\beta\sim0.13$ redward bias in the Dunlop et al.\ (2013)
  measurements (\S5.4 and Appendix B.4).  That Dunlop et al.\ (2013)
  suffer from a $\Delta\beta\gtrsim0.13$ bias in their $\beta$
  measurements can be demonstrated using the measured sizes for faint
  $z\sim7$ galaxies (see Figure~\ref{fig:sizebias}) and from detailed
  object-by-object comparisons with the $\beta$ results from
  Finkelstein et al.\ (2012) and the present study (see
  Figure~\ref{fig:compd12}).  After correcting for these systematic
  biases, we are able to reconcile all $z\sim7$ $\beta$ results in the
  literature (see also \S5.3-5.5 and Tables~\ref{tab:disput7} and
  \ref{tab:biassummary}).\label{fig:reconcile}}
\end{figure*}

\section{Summary}

In this paper, we have utilized the deepest-ever set of ACS and
WFC3/IR observations over the HUDF/XDF as well as the very deep ACS
and WFC3/IR observations over the HUDF09-Ps to establish the mean
$UV$-continuum slope $\beta$'s for galaxies at $z\sim7$, giving
particular attention to the issue of systematic errors that have
plagued $\beta$ measurements for the last few years.  We have also
made use of the wide-area WFC3/IR observations over the CANDELS-South
and CANDELS-North fields to better quantify the $\beta$ distribution
over a wide range of luminosity.  We expanded our comprehensive
$z\sim7$ study by using the same data sets to extend $\beta$
measurements at $z\sim4$, $z\sim5$, and $z\sim6$ to fainter limits and
larger samples, and also by carrying out a careful analysis at
$z\sim8$ and $z\sim8.5$.

The present $z\sim7$ analysis is an improvement on the recent analysis
of Dunlop et al.\ (2013) in that we are able to make full use of the
faintest sources in the HUDF09-Ps fields to map out the $\beta$
distribution at $z\sim7$ to very low luminosities.  We also make full
use of the CANDELS observations to obtain the best available
constraints on the mean $\beta$ for $z\sim7$ galaxies at bright $UV$
luminosities.

We are able to provide for a more comprehensive analysis of the
available observations to establish the mean $\beta$'s for
lower-luminosity $z\sim7$ galaxies by taking advantage of a new
technique we have developed in this paper (\S4.2: Appendix D).  This
technique is useful in that it allows us to establish the
$UV$-continuum slope $\beta$ distribution for $z\sim7$ galaxies in a
way that is completely robust against systematic biases using the
existing WFC3/IR observations over legacy fields.  Such biases can
arise due to a coupling between the noise affecting the measurement of
$\beta$ and that affecting source selection (Dunlop et al.\ 2012;
Bouwens et al.\ 2012).  This is particularly an issue for $z\sim7$
samples with near-IR coverage in only three bands (typically
$Y_{105}J_{125}H_{160}$), given the difficulty in forming two
independent colors from three bands, where one color is required for
our LBG selections and a separate color is required for the
measurement of $\beta$.

Fortunately, this issue can be solved, by splitting observations in
the reddest two filters (here the $J_{125}$ and $H_{160}$ bands) into
two subsets and using one half of the observations for source
selection and the other half of the observations for the measurement
of $\beta$.  By repeating the source selection and $\beta$ measurement
a second time swapping the splits, one can take advantage of the total
information content in the $J_{125}$ and $H_{160}$-band imaging data
to measure $\beta$.  The primary value of this approach is that we do
not require observations in a fourth WFC3/IR filter (i.e., F140W) to
obtain accurate or robust measurements of $\beta$ to lower
luminosities, although it is obviously valuable to have observations
in a fourth WFC3/IR band where possible.  Only a small fraction of HST
legacy fields, i.e., the HUDF and CLASH, have deep data in four
near-IR bands.

We also have taken advantage of the very deep WFC3/IR observations
over the XDF to refine our measurements of the mean $\beta$'s for the
lowest luminosity $z\sim4$-6 galaxies.  Our new measurements provide
us with our best constraints on the mean $UV$-continuum $\beta$ slopes
for faint galaxies at high redshift and how these $\beta$'s change as
a function of cosmic time.  These measurements represent an
improvement on those given in Bouwens et al.\ (2012) and are provided
in Table~\ref{tab:medianbeta} and \ref{tab:slopes}.  The additional
bright sources over CANDELS-North allow us to further refine our
determination of $\beta$ for the rarer, brighter sources.  A complete
set of the $z\sim4$-8 sources and $\beta$ measurements used in this
study is provided in Tables~\ref{tab:mrtable} and \ref{tab:phottable},
along with the photometry used to derive $\beta$.

As a cross check on our results and to obtain the best possible
perspective from which to minimize systematic errors, we have compared
the present $\beta$ measurements against a comprehensive set of
previous $\beta$ determinations from the literature (Bouwens et
al.\ 2010; Bouwens et al.\ 2012; Finkelstein et al.\ 2012; Dunlop et
al.\ 2013).  This has provided us with an unparalleled view on how
systematic errors have affected previous $\beta$ determinations in the
literature.  Appendix B discusses these comparisons in great detail
(see also Figures~\ref{fig:compfink}-\ref{fig:compd12} and
\S5.3-\S5.5).

Here are our findings:
\begin{itemize}
\item{\textit{The XDF and HUDF09-Ps datasets allow us to significantly
    improve the $\beta$ measurements for faint galaxies at $z\sim4$,
    $z\sim5$, and $z\sim6$.}  The deeper WFC3/IR observations also
  allow for modest improvements in our measurements of the
  $UV$-continuum slopes $\beta$ for a faint subsample
  ($-19<M_{UV,AB}<-17$) of galaxies at $z\sim4$, $z\sim5$, and
  $z\sim6$ (\S3.5).  The mean $\beta$ we measure for faint galaxies at
  $z\sim4$, $z\sim5$, and $z\sim6$ is $-2.03\pm0.03\pm0.06$ (random
  and systematic errors), $-2.14\pm0.06\pm0.06$, and
  $-2.24\pm0.11\pm0.08$, respectively.}
\item{\textit{The dependence of $\beta$ on $UV$ luminosity becomes
    flatter faintward of $M_{UV,AB}\sim-19$.}  The deeper WFC3/IR and
  ACS observations over the XDF also allow us to probe the dependence
  of the mean $\beta$ on $UV$ luminosity at very low luminosities
  (\S3.4).  Our study takes advantage of the faintest-ever probe of
  the $\beta$ at $z\sim4$ ($-15.5$ mag: $\sim$0.006 $L_{z=3}^{*}$),
  $z\sim5$ ($-16.5$ mag: $\sim$0.014 $L_{z=3}^{*}$), and $z\sim6$
  ($-17.0$ mag: $\sim$0.025 $L_{z=3}^{*}$).  The mean $\beta$ of
  galaxies shows a substantially weaker dependence on $UV$ luminosity
  faintward of $\sim-19$ mag than it does brightward of this
  luminosity.  A similar change in the dependence of $\beta$ on
  luminosity was previously found by Oesch et al.\ (2013a) in terms of
  the rest-frame optical luminosities of $z\sim4$ galaxies.  The large
  changes in $\beta$ at high luminosities is likely due to the changes
  in the dust content and the dust-mass correlation (Reddy et
  al.\ 2010; Pannella et al.\ 2009).  Making use of a new
  two-parameter fitting formula, where $\beta$ shows a steep linear
  dependence on $M_{UV}$ brightward of $-19$ mag and a fixed shallow
  dependence on $M_{UV}$ ($d\beta/dM_{UV}=-0.08$) faintward of $-19$
  mag, we obtain significantly improved fits to the $\beta$
  vs. $M_{UV}$ results (96\%, 75\%, and 85\% confidence for the
  $z\sim4$, $z\sim5$, $z\sim6$ results, respectively).}
\item{\textit{We have derived robust estimates of $\beta$ at $z\sim7$
    to low luminosities.}  The mean $\beta$ we derive for galaxies at
  $z\sim7$ from the XDF is $-2.30\pm0.18\pm0.13$ and
  $-$2.42$\pm$0.28$\pm$0.13 over the luminosity range
  $-19.3<M_{UV,AB}<-18.0$ and $-18.0<M_{UV,AB}<-16.7$, respectively
  (\S4.7).  The first set of uncertainties are random and second set
  are systematic.  We tabulate the results using three different
  measures, biweight means, median, and inverse-variance-weighted
  means (see Table~\ref{tab:medianbeta}), but use the biweight mean
  because of the stability of the results.  We also obtain similar
  results using simple aperture photometry (0.32$''$-diameter
  apertures) to derive the $J_{125}$, $JH_{140}$, $H_{160}$ fluxes
  (\S4.8) and binning our $z\sim7$ samples as a function of the
  $JH_{140}$-band magnitude (\S4.9).}
\item{\textit{We have reconciled the $\beta$ measurements in different
    studies by determining their systematic biases.}  Through detailed
  object-by-object comparisons between the measured $\beta$'s from
  many different studies (Bouwens et al.\ 2010; Bouwens et al.\ 2012;
  Finkelstein et al.\ 2012; Dunlop et al.\ 2013), we have succeeded in
  reconciling the many different $\beta$ determinations in the
  literature for faint galaxies at $z\sim7$ (\S5;
  Figure~\ref{fig:reconcile}).  See also \S5.2-\S5.5 and
  Table~\ref{tab:disput7}.  A large part of the differences can be
  explained as a result of systematic errors in the measurement of the
  $J_{125}-H_{160}$ colors (see Table~\ref{tab:biassummary}).  The
  primary explanation is due to a blueward bias ($\sim$0.05 mag) in
  the measured $J_{125}-H_{160}$ colors of Bouwens et al.\ (2012).
  This bias resulted from small systematics in the empirical $J_{125}$
  and $H_{160}$-band PSFs used to PSF-match the observations (\S5.3).
  Another part of the explanation also appears to be due to the
  $J_{125}-H_{160}$ colors of Dunlop et al.\ (2013) being too red (by
  $\sim$0.03-0.04 mag), due to their treating $z\sim7$ galaxies as
  point sources when faint $z\sim7$ galaxies actually have small, but
  non-zero half-light radii (see Figure~\ref{fig:sizebias}, \S5.4,
  Appendix B.4, and Figure~\ref{fig:sims2} from Appendix E: Oesch et
  al.\ 2010; Ono et al.\ 2013).  While we find no statistically
  significant difference between the Finkelstein et al.\ (2012)
  $z\sim7$ $\beta$ results and the present results
  (Figure~\ref{fig:compfink}), the use of flux measurements affected
  by Ly$\alpha$ emission could bias their derived $\beta$
  determinations to bluer values (Rogers et al.\ 2013).  $\beta$
  results at $z\sim7$ are extraordinarily sensitive to small
  systematics in the photometry due to the limited leverage in
  wavelength to constrain $\beta$ (Figure~\ref{fig:illust}).  The
  biases that have affected or can affect prior papers are identified
  in Table~\ref{tab:biassummary}.}
\item{\textit{There is a clear correlation between $\beta$ and $UV$
    luminosity at $z\sim7$.}  Similar to previous work (Bouwens et
  al.\ 2009, 2010, 2012; Wilkins et al.\ 2011; Labb{\'e} et al.\ 2007;
  Finkelstein et al.\ 2012), we find strong evidence for a correlation
  between $\beta$ and $UV$ luminosity for $z\sim7$ galaxies (99.7\%
  confidence), in the sense that brighter galaxies are redder and
  fainter galaxies are bluer.  The apparent slope to this relation we
  determine at $z\sim7$ is $-0.20\pm0.07$ and is similar to what we
  find at lower redshifts (see Figure~\ref{fig:slope} and
  Table~\ref{tab:slopes}).  This relationship is remarkably similar
  for all four of our lowest redshift samples $z\sim4$, $z\sim5$,
  $z\sim6$, and $z\sim7$ (Figure~\ref{fig:summary4567}).  }
\item{\textit{For galaxies of a given luminosity, $\beta$ becomes
    bluer, as the redshift increases.}  Comparing the mean $\beta$'s
  measured at $z\sim4$ through $z\sim8$, we can assess the evolution
  in the $\beta$ vs. $M_{UV}$ relationship as a function of cosmic
  time.  We find evidence for a small but clear evolution in $\beta$
  as a function of redshift at fixed $UV$ luminosity.  The change in
  $\beta$ per unit redshift we find is $-0.10\pm0.05$ (conservatively
  accounting for possible systematic errors).  This is similar to that
  found in the Bouwens et al.\ (2012) and Finkelstein et al.\ (2012)
  studies and also to what is predicted in the hydrodynamical
  simulations of Finlator et al.\ (2011).  Extrapolating the trend in
  the mean $\beta$ seen from $z\sim4$ to $z\sim6$ to $z\sim7$ and
  $z\sim8$ suggests mean $\beta$'s of $-2.35\pm0.16$ and
  $-2.45\pm0.23$, respectively.  These expectations are consistent
  with the mean $\beta$'s we derive for faint galaxies at these
  redshifts.}
\item{\textit{We measure the mean $\beta$ at $z\sim8$, but the
    uncertainty is large.}  The inverse-variance-weighted mean $\beta$
  we find at $z\sim8$ using just the $JH_{140}$ and $H_{160}$-band
  photometry is $-$1.88$\pm$0.74$\pm$0.27 for galaxies in the
  luminosity range $-19.3<M_{UV,AB}<-18.0$ (\S6.2).  Alternatively, we
  can attempt to derive $\beta$ for $z\sim8$ galaxies using their
  observed $J_{125}$, $JH_{140}$, and $H_{160}$-band fluxes, but we
  emphasize that in this case a correction to the $J_{125}$-band
  fluxes must be performed to account for absorption by both the IGM
  and the Ly$\alpha$ damping wing (see Figure~\ref{fig:beta8} and
  Appendix G).  Using this alternate approach, we derive an
  inverse-variance-weighted $\beta$ of $-2.39\pm0.36\pm0.13$ over the
  same luminosity interval (Appendix G.2).}
\item{\textit{We examine the mean $\beta$ at $z\sim8.5$, but find that
    current data sets are inadequate for useful measurements.}  Taking advantage of the availability of the new
  ultra-deep $JH_{140}$ and $H_{160}$ observations over the XDF, we
  explore the value of the $UV$-continuum slope $\beta$ at $z>8$
  (\S6.3).  Because of the sensitivity of $\beta$ measurements at
  $z\sim9$ to any attenuation of the $JH_{140}$-band flux by the IGM
  (which would occur if the redshift of any source was greater than
  9), we restrict our sample to the two brightest $z\sim8.5$ galaxies
  from Oesch et al.\ (2013b).  The inverse-variance-weighted mean and
  simple mean $\beta$ we find for our $z\sim8.5$ sample is
  $-2.1\pm0.9$ and $\sim-1.8$, respectively.  Given our focus on the
  highest signal-to-noise $z\sim8.5$ sources, this represents the most
  reliable measurement of the mean $\beta$ for $z>8$ galaxies to date.
  Nevertheless, the current uncertainties are too large for this
  measurement to be especially useful.}
\item{\textit{The observed $\beta$'s at $z\sim7$ are in excellent
    agreement with cosmological hydrodynamical simulations by Finlator
    et al.\ (2011).}  The mean $\beta$'s we derive here for lower
  luminosity $z\sim7$-8 galaxies suggest low (but non-zero)
  $E(B-V)\sim 0.02$-0.03 dust extinction assuming stellar population
  ages of $\sim$200 Myr and 0.5 $Z_{\odot}$ metallicity (see \S7).
  See also Dunlop et al.\ 2013 who favor non-zero dust extinction
  (albeit slightly higher than we prefer here).  For stellar
  population ages $\sim$200 Myr and this dust content, the implied
  $H_{160}-[4.5]$ and $H_{160}-[3.6]$ colors for $z\sim7$ galaxies and
  $z\sim8$ galaxies, respectively, are in good agreement with current
  constraints on the $UV$-optical colors (Smit et al.\ 2014; Labb{\'e}
  et al.\ 2013: see \S7 and Figure~\ref{fig:interp}).  The measured
  $\beta$'s are also in excellent agreement with the expectations from
  the hydrodynamical simulations by Finlator et al.\ (2011: see
  Figure~\ref{fig:finlator}).}
\end{itemize}
The availability of even deeper observations over the HUDF/XDF and an
improved methodology to maximally leverage deep HST observations from
other fields like the two HUDF09-Ps fields, have allowed us to obtain
the best available constraints on the $UV$-continuum slopes $\beta$ of
faint galaxies at $z\sim7$-8 and also at $z\sim4$-6.  Together with
improved constraints on the rest-frame $UV$-to-optical colors of
$z\sim7$-8 galaxies (Labb{\'e} et al.\ 2013; Smit et al.\ 2014), the
deeper HST observations and improved analysis techniques (focusing on
minimizing systematic errors) allow us to place valuable constraints
on the stellar populations of faint galaxies in the early universe.

In the future, we can expect improvements in our determinations of the
$UV$-continuum slope $\beta$ distribution and mean $\beta$'s from the
HST Frontier Fields Initiative (Lotz et
al.\ 2014).\footnote{http://www.stsci.edu/hst/campaigns/frontier-fields/}$^{,}$\footnote{http://www.stsci.edu/hst/campaigns/frontier-fields/HDFI\_SWGReport2012.pdf}
The Frontier Fields initiative will provide very deep ($\sim$29 AB
mag) ACS+WFC3/IR observations over both strong lensing clusters and
blank fields.  This program will be very useful for increasing the
overall numbers for the $UV$-continuum slope $\beta$ distribution at
$z\sim7$-8.  Particularly important will be the small number of highly
magnified, faint $z\sim7$-9 galaxies that will be imaged to great
depths in these fields.  These sources will provide the best available
constraints on the $\beta$'s of faint galaxies during the reionization
epoch.

\acknowledgements

We acknowledge the support of NASA grant NAG5-7697, NASA grant
HST-GO-11563, ERC grant HIGHZ \#227749, and a NWO vrij competitie
grant.  PO acknowledges support from NASA through a Hubble Fellowship
grant \#51278.01 awarded by the Space Telescope Science Institute.
The authors would like to thank Steve Finkelstein and James Dunlop for
publishing their detailed source-by-source measurements of $\beta$
which greatly helped in evaluating the likely systematic errors that
has affected various $\beta$ measurements.  These source-by-source
measurements were key to resolving the discrepancies between the
various works.

\appendix

\begin{figure}
\epsscale{0.5}
\plotone{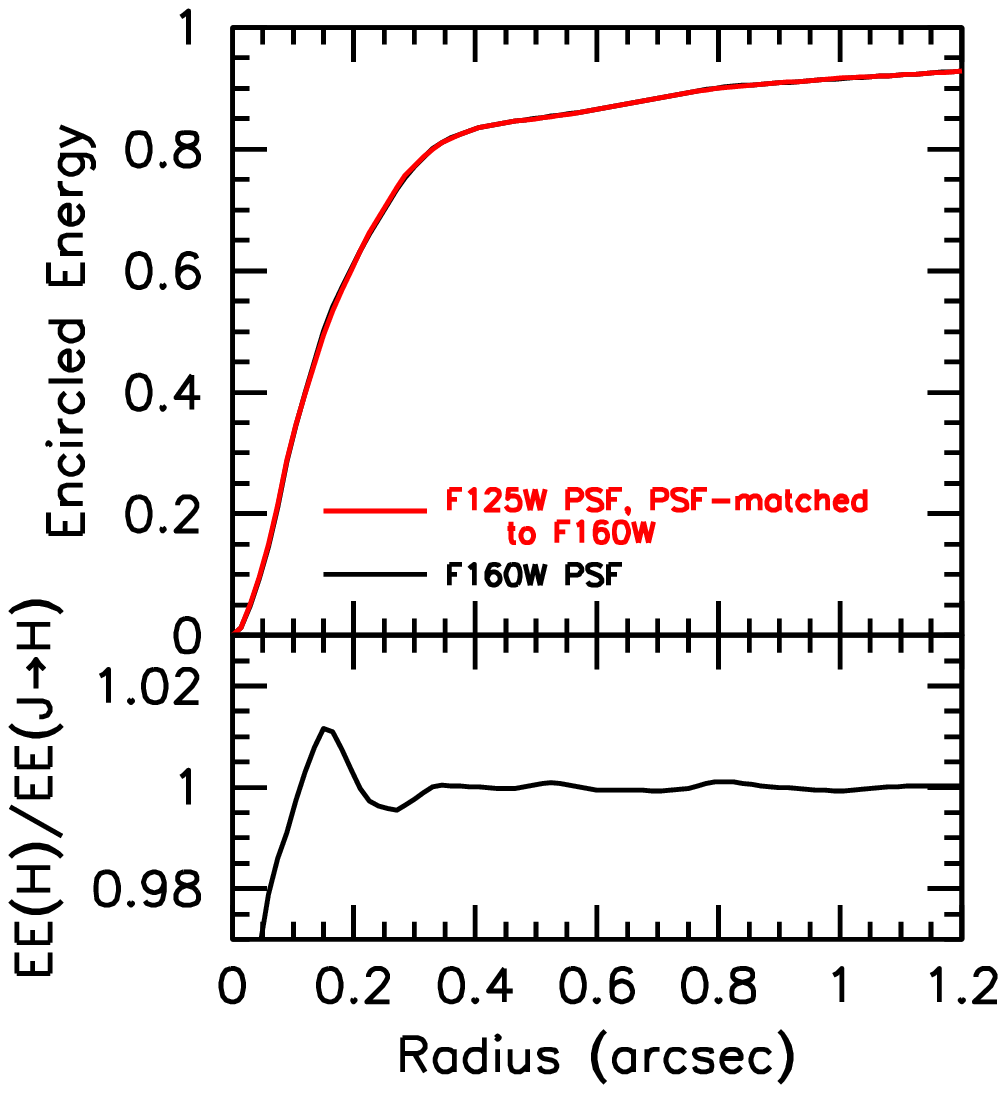}
\caption{(\textit{upper}) Encircled-energy distribution for our
  $H_{160}$-band PSF (\textit{red line}) and the $J_{125}$-band PSF
  (after PSF-matching to our $H_{160}$-band data: \textit{black
    line}).  (\textit{lower}) Ratio between the encircled-energy
  distributions for the $H_{160}$-band PSF and our $J_{125}$-band PSF,
  after PSF-matching to the $H_{160}$-band data.  The PSFs were
  matched so that the effective encircled-energy distributions would
  be identical in the two bands to $\lesssim1.2$\%.  As a result, we
  would expect any systematic errors in our $\beta$ measurements to be
  smaller than $\Delta\beta \sim 0.05$ though we quote systematic
  errors of $\Delta\beta\sim \pm0.13$ (3\% accuracy on the colors) to
  be conservative.\label{fig:ee}}
\end{figure}

\begin{figure}
\epsscale{1.10}
\plotone{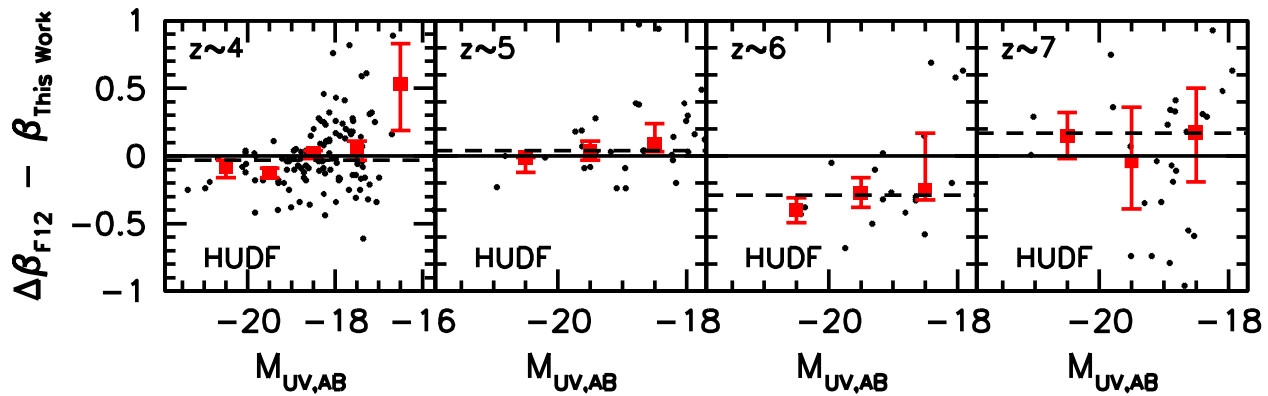}
\caption{Object-by-object differences (\textit{small black points})
  between the $\beta$'s we measure in this work with those measured in
  Finkelstein et al.\ (2012: F12) versus $UV$ luminosity ($M_{UV}$)
  for sources in our $z\sim4$, $z\sim5$, $z\sim6$, and $z\sim7$
  samples from the XDF data set.  When comparing against the
  Finkelstein et al.\ (2012) $\beta$ results, we have corrected their
  raw $\beta$ measurements (Table 3 of their work) for the
  $\Delta\beta\sim0.23$ blueward bias in the faintest $z\sim7$
  galaxies they reported.  The large red squares show the median
  differences in the measured $\beta$ for a bright subsample
  ($-21.5<M_{UV,AB}<-20$), an intermediate-magnitude subsample
  ($-20<M_{UV,AB}<-19$), and a fainter subsample
  ($-19<M_{UV,AB}<-18$).  The dashed line gives the median difference
  in the measured $\beta$ for all sources in common between this work
  and that of Finkelstein et al.\ (2012).  Good agreement is observed
  overall.  Nevertheless, the $\beta$'s that Finkelstein et
  al.\ (2012) derive for the brightest $z\sim4$ galaxies from the HUDF
  are slightly bluer (i.e., $\Delta\beta\sim0.1$) than found here.
  Finkelstein et al.\ (2012) also measure slightly bluer $\beta$'s
  (i.e., $\Delta\beta\sim0.2$) for $z\sim6$ sources than we measure
  here.  A similar offset in the $\beta$'s is apparent in comparing
  the $z\sim6$ $\beta$ results from the ERS+CANDELS
  fields.\label{fig:compfink}}
\end{figure}

\begin{figure}
\epsscale{1.0}
\plotone{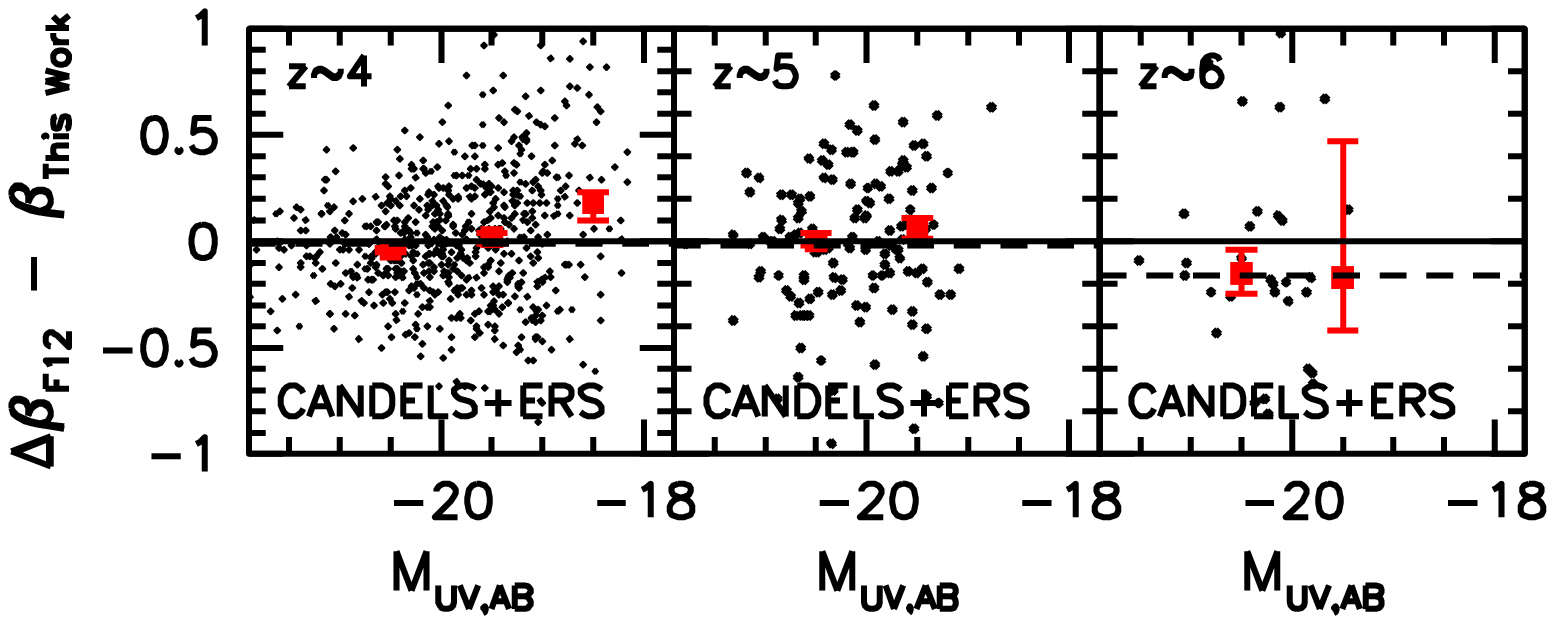}
\caption{Object-by-object differences (\textit{small black points})
  between the $\beta$'s we measure in this work with those measured in
  Finkelstein et al.\ (2012: F12) versus $UV$ luminosity ($M_{UV}$)
  for sources in our $z\sim4$, $z\sim5$, and $z\sim6$ samples from the
  CANDELS-South and ERS data sets.  The large red squares show the
  median differences in the measured $\beta$ for a bright subsample
  ($-21.5<M_{UV,AB}<-20$), an intermediate-magnitude subsample
  ($-20<M_{UV,AB}<-19$), and a fainter subsample
  ($-19<M_{UV,AB}<-18$).  The dashed line gives the median difference
  in the measured $\beta$ for all sources in common between this work
  and that of Finkelstein et al.\ (2012).  While the Finkelstein et
  al.\ (2012) results are in good agreement overall with our results,
  the derived $\beta$'s for the faintest sources in the CANDELS-South
  and ERS field from Finkelstein et al.\ (2012) are offset to redder
  values, by $\Delta\beta\sim0.2$.  As is discussed extensively in
  \S4.7 of Bouwens et al.\ (2012), this likely occurs due to a
  coupling between the derived $\beta$'s for the faintest sources in
  the CANDELS-South field and the derived $UV$ luminosity of these
  sources. \label{fig:compfinkg}}
\end{figure}

\begin{figure}
\epsscale{0.4}
\plotone{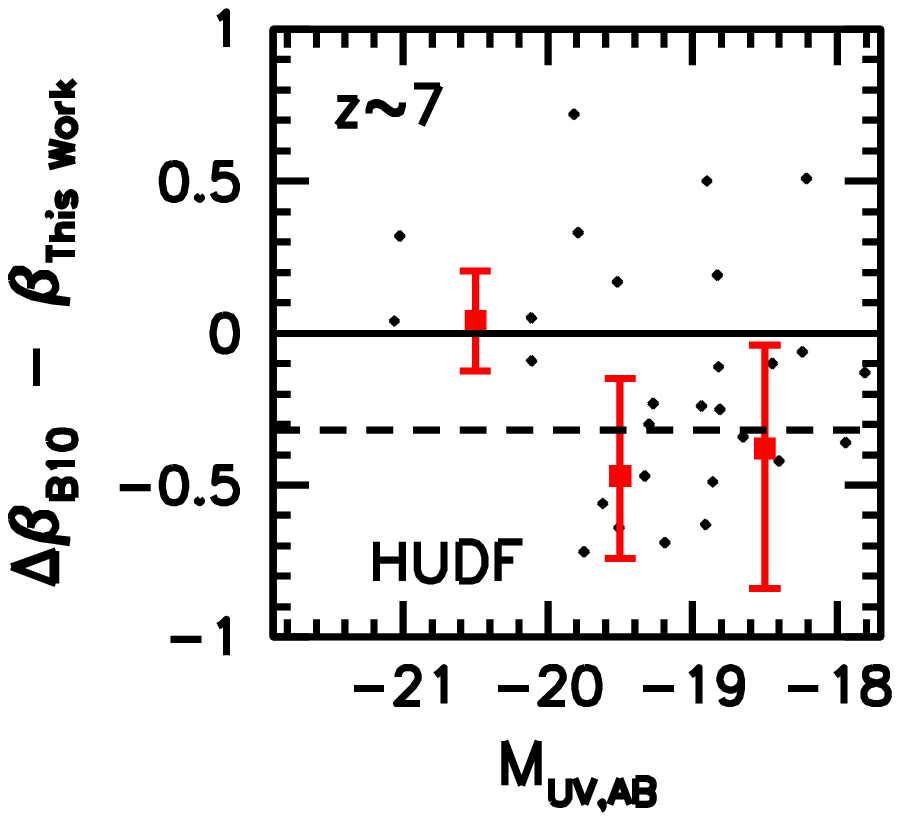}
\caption{Object-by-object differences (\textit{small black points})
  between the $\beta$'s we measure in this work and those measured in
  Bouwens et al.\ (2010: B10) versus $UV$ luminosity ($M_{UV}$) for
  sources in our $z\sim7$ sample from the XDF data set.  The results
  for $z\sim7$ sources from our XDF data set can be shown
  (\textit{represented here as the small points}) as often as twice,
  due to our selecting $z\sim7$ sources on our fields twice (see
  \S4.2).  The large red squares show the median differences in the
  measured $\beta$ for a bright subsample ($-21.5<M_{UV,AB}<-20$), an
  intermediate-magnitude subsample ($-20<M_{UV,AB}<-19$), and a
  fainter subsample ($-19<M_{UV,AB}<-18$).  The dashed line gives the
  median difference in the measured $\beta$ for all sources in common
  between this work and that of Bouwens et al.\ (2010).  Differences
  between the present $\beta$ measurements and those of Bouwens et
  al.\ (2010) can be most easily explained as a result of a
  $\Delta\beta\sim0.2$ systematic bias in the Bouwens et al.\ (2010)
  measurements (see Appendix B.2).\label{fig:compb10}}
\end{figure}

\section{A. Procedure to Obtain Accurate PSF-Matched Observations}

Measurements of the $UV$-continuum slopes $\beta$ of galaxies are
extremely sensitive to the color measurements.  Even $\sim$0.05 mag
errors in the measured colors result in $\Delta\beta \sim 0.12$-0.2
errors in the derived $\beta$'s (depending on how much leverage is
available in wavelength).  Accurate color measurements are therefore
absolutely essential.

Great care has been taken in deriving accurate PSFs for each bandpass
under study and in PSF-matching the multi-band observations.  During
the PSF-matching process, we explicitly verified that the encircled
energy distribution for point sources in our observations were a good
match to those from the $H_{160}$-band, after convolution by our
PSF-matching kernel.  Any residual systematics in the photometry of
our PSF-matched observations should be no larger than 1.2\%.
Figure~\ref{fig:ee} illustrates the general quality of these results,
comparing the encircled energy distribution for our $H_{160}$ PSF and
that for our $J_{125}$ PSF, after PSF-matching the observations to the
$H_{160}$ band.  As is evident from the figure, the results are
accurate to $\lesssim1.2$\%.  Similar quality results are obtained in
the PSF-matching the observations in the other passbands.

The real errors in the color measurements may be even larger than
1.2\%, given that the HST zeropoints themselves are also uncertain
(typically on the $\sim$1\% level) as well as potential uncertainties
in the amount of foreground extinction from our Galaxy.  Considering
all sources of error, we conservatively quote an uncertainty of 3\% on
our absolute color measurements (see \S4.6).  This translates into
systematic uncertainties in our derived $\beta$'s of
$\Delta\beta\lesssim0.06$, $\Delta\beta\lesssim0.08$,
$\Delta\beta\lesssim0.13$, and $\Delta\beta\lesssim0.27$ at
$z\sim4$-5, $z\sim6$, $z\sim7$, and $z\sim8$-8.5, respectively.

\section{B. Comparisons with Previous $\beta$ Measurements}

To provide us with the broadest possible perspective on how our
current $\beta$ measurements compare with previous published
measurements, we conducted a comprehensive set of source-by-source
comparisons with previous work.  Not only did this provide us with an
unparalleled view on how systematics could have affected previous
studies, including our own, but it also provided us with valuable
cross-checks on our results as we were putting together our samples
and measurements.

\subsection{B.1  Comparisons with the $\beta$ Measurements of 
Finkelstein et al.\ 2012}

Independent of our own work on the $UV$-continuum slopes (Bouwens et
al.\ 2009, 2010, 2012), the most comprehensive compilation of $\beta$
measurements in the literature has been provided by Finkelstein et
al.\ (2012) for photometric-redshift-selected galaxy samples at
$z\sim4$, $z\sim5$, $z\sim6$, $z\sim7$, and $z\sim8$.  Finkelstein et
al.\ (2012) derived these $\beta$ measurements for these samples by
fitting their photometry to model SEDs and then marginalizing over the
results.

Figure~\ref{fig:compfink} compares the $\beta$ measurements of
Finkelstein et al.\ (2012) with those from the present work for
sources from the XDF.  For these comparisons, small corrections to the
raw $\beta$ values provided by Finkelstein et al.\ (2012) consistent
with the biases Finkelstein et al.\ (2012) estimate, the most
significant of which is a $\Delta\beta\sim0.23$ redward correction to
the $\beta$ measurements for faint ($M_{UV,AB}\gtrsim-19$) $z\sim7$
galaxies.  A similar comparison is made for $\beta$ measurements from
the CANDELS-South and ERS fields in Figure~\ref{fig:compfinkg}.  The
large solid squares show the median differences between the $\beta$
measurements from Finkelstein et al.\ (2012) and those obtained here.

The median $\beta$'s measured by Finkelstein et al.\ (2012) are
$0.03_{-0.02}^{+0.02}$ bluer, $0.04_{-0.02}^{+0.07}$ redder,
$0.29_{-0.13}^{+0.06}$ bluer, and $0.17_{-0.12}^{+0.08}$ redder for
their $z\sim4$, $z\sim5$, $z\sim6$, and $z\sim7$ HUDF selections,
respectively, than what we measure here.  The median $\beta$'s derived
by Finkelstein et al.\ (2012) for sources from the CANDELS-South and
ERS fields are $0.01_{-0.02}^{+0.01}$ bluer, $0.02_{-0.03}^{+0.03}$
redder, and $0.16_{-0.13}^{+0.04}$ bluer at $z\sim4$, $z\sim5$, and
$z\sim6$, respectively, than the values we derive here.

Overall, the Finkelstein et al.\ (2012) $\beta$ results are in broad
agreement with our own results, with a few noteworthy differences.  At
$z\sim6$, the Finkelstein et al.\ (2012) $\beta$ measurements appear
to be systematically bluer than our measurements on average, by
$\Delta\beta\sim0.2$.  One possible cause for this difference is due
to the effect of Ly$\alpha$ emission in boosting the broadband flux
measurements at the position of the Lyman break.  Since Finkelstein et
al.\ (2012) use the flux information in all HST passbands in deriving
$\beta$, it is possible this could shift their measured $\beta$'s to
bluer values.  Finkelstein et al.\ (2012) explicitly consider this
type of effect in their paper and attempt to determine the impact it
would have on their results.  Finkelstein et al.\ (2012) report
differences as large as $\Delta\beta\sim0.25$ between the two
approaches.  The simulations shown by Rogers et al.\ (2013) also
suggest that similar biases (i.e., $\Delta\beta\sim0.25$-0.5) could be
present in the Finkelstein et al.\ (2013) results due to Ly$\alpha$
emission.

The only other noteworthy difference we find is between the
Finkelstein et al.\ (2012) $\beta$ results at $z\sim4$ and our $\beta$
results.  In particular, Finkelstein et al.\ (2012) measure bluer
$\beta$'s for the brightest $z\sim4$ sources than we measure and
redder $\beta$'s for the faintest $z\sim4$ sources.  The effect is
evident both in the results for the HUDF/XDF data set and from
CANDELS-South and ERS data sets.  The origin of this difference is not
clear.  However, since we find no median difference between the
$\beta$'s we measure for the same $z\sim4$ galaxies based on the XDF
and the CANDELS-South data sets, we believe that it is unlikely to
arise from our own measurements.  By contrast, the median $\beta$'s
that Finkelstein et al.\ (2012) derive for the faintest $z\sim4$
sources from the CANDELS-South field and the HUDF differ by
$\Delta\beta\gtrsim0.2$ (disagreeing at $\sim3\sigma$ significance).

One partial explanation for this offset is discussed in \S4.7 of
Bouwens et al.\ (2012: see also Finkelstein et al.\ 2012).  We might
expect small biases in the Finkelstein et al.\ (2012) $\beta$
vs. $M_{UV}$ results due to the fact that Finkelstein et al.\ (2012)
determine the $UV$ luminosity of sources at approximately the same
rest-frame wavelength as the blue end of the wavelength baseline they
use to derive $\beta$.  This effectively introduces a coupling between
$M_{UV}$ and $\beta$, so that noise scatters sources either (1) to
lower luminosities and redder $\beta$'s or (2) to higher luminosities
and bluer $\beta$'s.

\subsection{B.2  Comparisons with the $\beta$ Measurements of Bouwens et al.\ 2010}

Much of the debate regarding the $UV$-continuum slopes $\beta$ of
faint $z\sim7$ galaxies has centered around the early measurements
made by Bouwens et al.\ (2010), which were very blue, i.e.,
$\beta\sim-3$ (see also Finkelstein et al.\ 2010).  These very blue
measurements contrast with the $\beta$'s of $-2.30\pm0.18$ we measure
in this same general luminosity range $-19.3<M_{UV,AB}<-18.0$ from the
XDF+HUDF09-Ps fields.

\begin{figure}
\epsscale{1.15}
\plotone{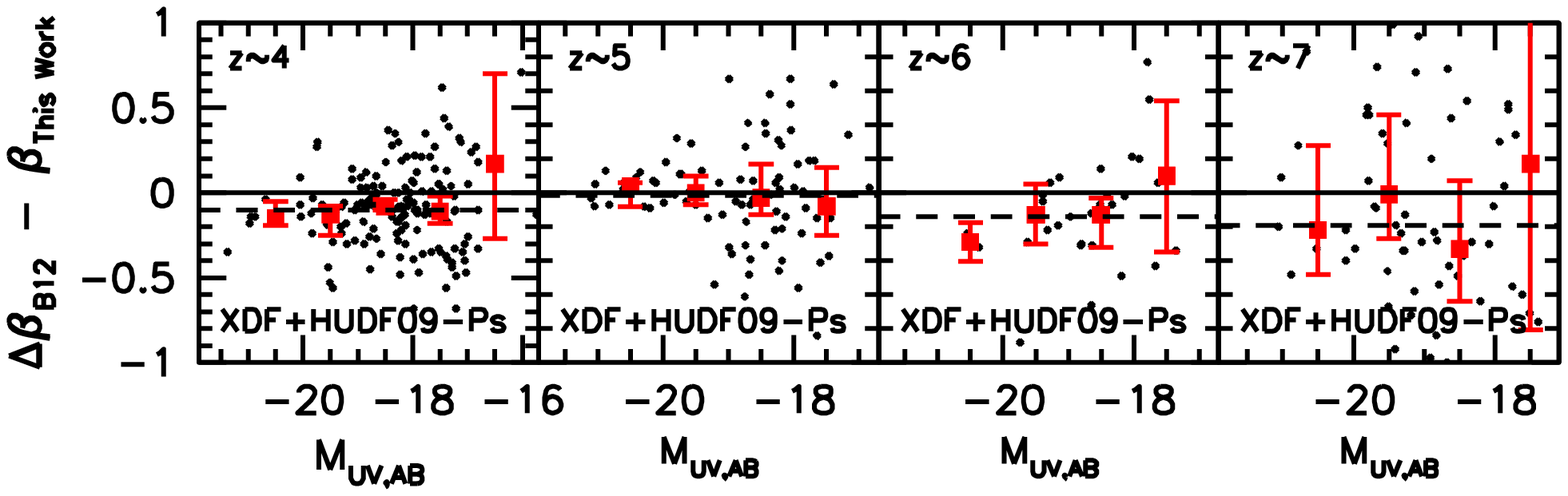}
\caption{Object-by-object differences (\textit{small black points})
  between the $\beta$'s we measure in this work with those measured in
  Bouwens et al.\ (2012: B12) versus $UV$ luminosity ($M_{UV}$) for
  sources in our $z\sim4$, $z\sim5$, $z\sim6$, and $z\sim7$ samples
  from the XDF data set.  The large red squares show the median
  differences in the measured $\beta$ for a bright subsample
  ($-21.5<M_{UV,AB}<-20$), an intermediate-magnitude subsample
  ($-20<M_{UV,AB}<-19$), and a fainter subsample
  ($-19<M_{UV,AB}<-18$).  The dashed line gives the median difference
  in the measured $\beta$ for all sources in common between this work
  and that of Bouwens et al.\ (2012).  The sources considered here
  from Table 8 from Bouwens et al.\ (2012) were corrected for the
  relevant biases estimated in that study.  The most important
  correction was a $\Delta\beta\sim0.2$ redward offset to the
  $\beta$'s measured for the faintest $z\sim7$ sources by Bouwens et
  al.\ (2012).  In general, our current $\beta$ measurements are in
  good agreement with the Bouwens et al.\ (2012) measurements.
  However, we note a systematic bias in the Bouwens et al.\ (2012)
  $\beta$'s to bluer values.  The size of this effect is
  $\Delta\beta\sim0.1$, $\sim$0.0, $\sim$0.14, and $\sim$0.2 for
  $z\sim4$, $z\sim5$, $z\sim6$, $z\sim7$ sources.  For the faintest
  galaxies in the Bouwens et al.\ (2012) $z\sim7$ samples, the median
  $\beta$ may show a slightly larger offset relative to the present
  values.  However, the difference is not statistically
  significant.\label{fig:compb12}}
\end{figure}

Why do the current results differ so significantly from the previous
results?  One factor is that Bouwens et al.\ (2010) required that
sources in their $z\sim7$ selection be detected at 5.5$\sigma$ in the
$J_{125}$ band (Dunlop et al.\ 2012; Rogers et al.\ 2013).  As a
result, Bouwens et al.\ (2010) preferentially selected those $z\sim7$
candidates which were brighter in the $J_{125}$ band (which can arise
both as a result of the intrinsic color variations in galaxies and due
to noise), causing the $J_{125}-H_{160}$ colors of their $z\sim7$
candidates to be biased.

It is possible (at least approximately) to quantify the size of this
bias by requiring that all sources in the $z\sim7$,
$-19<M_{UV,AB}<-18$ selection of Bouwens et al.\ (2010) also be
detected at $5.5\sigma$ significance in the $H_{160}$-band, to mirror
similar criteria applied in the $J_{125}$ band.  As expected, the mean
$\beta$ we measure for this $-19<M_{UV,AB}<-18$ subsample becomes
somewhat redder as a result of imposing this additional criterion.
However the change in the mean $\beta$ is only $\Delta\beta\sim0.1$,
i.e., from $\beta\sim-3$ to $\beta\sim-2.9$.  This is somewhat
surprising, since we might have expected to derive a somewhat larger
bias in $\beta$ drawing on the Rogers et al.\ (2013) simulations
albeit smaller in size than the bias predicted in Figure~6 of Rogers
et al.\ (2013).\footnote{The Rogers et al.\ (2013) simulations likely
  overpredict somewhat (i.e., by a factor of $\gtrsim$1.5) the
  expected biases in the faintest $z\sim7$ sources in the Bouwens et
  al.\ (2010) study.  This is due to Bouwens et al.\ (2010) using
  somewhat smaller ($\sim$1.4$\times$) apertures for their color
  measurements than assumed by Rogers et al.\ (2013) in their
  simulations and due to the mean value for $\beta$ being closer to
  $-2.3$ rather than $-2.0$ (though the latter effect would be
  small).}

If noise-driven systematic biases were not the dominant explanation
for differences with our current $\beta$ measurements, what is the
explanation?  Comparing the current $\beta$ measurements with those
from Bouwens et al.\ (2010) on a source-by-source basis
(Figure~\ref{fig:compb10}), we find some evidence for there being a
systematic offset $\Delta\beta\sim0.2$ offset between the $\beta$
measurements made in Bouwens et al.\ (2010) and those made here.
Sources in both the two lowest-luminosity intervals show exactly the
same offset in the median $\beta$ relative to the current
measurements.  Since we find essentially the same offset between the
measured $\beta$'s from Bouwens et al.\ (2012) and the current study
(Appendix B.2), this suggests the dominant bias in the reported
$\beta$'s from Bouwens et al.\ (2010) may have been a systematic color
measurement bias (see Figure~\ref{fig:illust}) and that the role of
noise-driven biases in producing the blue $\beta$'s may have been
smaller.

Correcting for both effects gives us a mean $\beta$ of $\sim-2.68$,
similar to the mean $\beta$ (i.e., $-2.18\pm0.19$) we measure for
sources in the luminosity interval $-19.3<M_{UV,AB}<-18$ from the
current XDF data set.

\begin{figure}
\epsscale{0.8}
\plotone{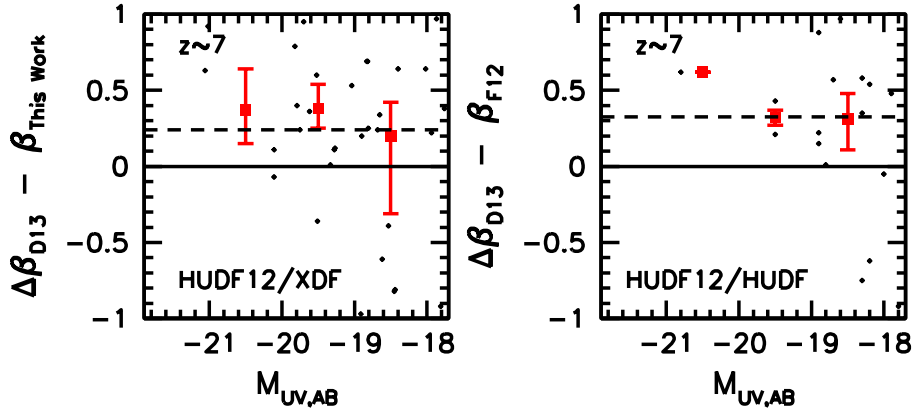}
\caption{(\textit{left}) Object-object differences (\textit{small
    black points}) between the $\beta$'s we measure in this work and
  those measured in Dunlop et al.\ (2013: D13) versus $UV$ luminosity
  ($M_{UV}$) for sources in our $z\sim7$ sample from the HUDF12/XDF
  data set.  The large red squares show the median differences in the
  measured $\beta$ for a bright subsample ($-21.5<M_{UV,AB}<-20$), an
  intermediate-magnitude subsample ($-20<M_{UV,AB}<-19$), and a
  fainter subsample ($-19<M_{UV,AB}<-18$).  The dashed line gives the
  median difference in the measured $\beta$ for all sources in common
  between this work and Dunlop et al.\ (2013).  Interestingly enough,
  the largest offsets between the $\beta$'s we derive and as derived
  by Dunlop et al.\ (2012) are for the brightest sources.  The offsets
  for fainter sources are smaller.  One would expect such a trend, if
  the Dunlop et al.\ (2013) $\beta$'s are biased redward in proportion
  to the size of the sources (see also Figure~\ref{fig:sizebias},
  \S5.4, and Appendix D.4).  (\textit{right}) Similar to the left
  panel, except comparing the $\beta$'s measured by Finkelstein et
  al.\ (2012) with those measured in Dunlop et al.\ (2013) versus the
  $UV$ luminosity as determined by Finkelstein et al.\ (2012).
  $\Delta\beta\sim0.23$ corrections to the individual $\beta$
  measurements for the faintest sources in the Finkelstein et
  al.\ (2012) $z\sim7$ sample were made to account the photometric
  error coupling bias that Finkelstein et al.\ (2012) estimated.
  Comparison of the Dunlop et al.\ (2013) results with those from
  Finkelstein et al.\ (2012) provide independent support for our
  general conclusion that the Dunlop et al.\ (2013) $\beta$
  measurements are systematically biased to redder values (see also
  Figure~\ref{fig:sizebias}).\label{fig:compd12}}
\end{figure}

\subsection{B.3  Comparisons with the $\beta$ Measurements of Bouwens et al.\ 2012}

The most comprehensive study of the $\beta$ distribution for
$z\sim4$-7 galaxies to date is that of Bouwens et al.\ (2012).  Here
we compare the present $\beta$ measurements we obtain for specific
$z\sim4$-7 galaxies in our XDF+HUDF09-Ps and ERS+CANDELS selections
with those from Bouwens et al.\ (2012).  This should allow us to
determine very precisely the extent to which the $\beta$ measurements
from Bouwens et al.\ (2012) exhibit any systematic offsets relative to
those given here.

The object-by-object comparisons are shown in Figure~\ref{fig:compb12}
for sources in our XDF, HUDF09-Ps, ERS, and CANDELS selections.  For
these comparisons, small corrections to the raw $\beta$ values
provided in Table 8 of Bouwens et al.\ (2012) are made.  The most
significant of these is a $\Delta\beta\sim0.2$ redward correction to
the $\beta$ measurements for faint ($M_{UV,AB}>-19$) $z\sim7$ galaxies
that Bouwens et al.\ (2012) found necessary to remove a noise-driven
systematic bias (see \S4.1).

While in general we find broad agreement between the $\beta$
measurements from our two studies, we note the presence of systematic
offsets in the measured $\beta$'s.  Compared to the present $\beta$
measurements for galaxies in our $z\sim4$, $z\sim5$, $z\sim6$, and
$z\sim7$ samples over the XDF and HUDF09-Ps fields, those from Bouwens
et al.\ (2012) are $\Delta\beta\sim0.10_{-0.01}^{+0.02}$,
0.01$_{-0.03}^{+0.03}$, 0.08$_{-0.03}^{+0.08}$, and
0.19$_{-0.14}^{+0.07}$ bluer, respectively, in the median than those
derived in the present study.  Comparing the Bouwens et al.\ (2012)
$\beta$ results from CANDELS-South and the ERS field with those from
the present work, we found that the median $\beta$'s derived by
Bouwens et al.\ (2012) are $0.05\pm0.01$ bluer, $0.10\pm0.01$ bluer,
$0.09_{-0.01}^{+0.03}$ bluer, $0.30_{-0.20}^{+0.06}$ bluer (in the
median) at $z\sim4$, $z\sim5$, $z\sim6$, and $z\sim7$ than what we
derive here.

While the above differences are consistent with the systematic
uncertainties quoted by Bouwens et al.\ (2012) for their $\beta$
measurements, we do note a general shift in our measured $\beta$
values to redder values overall.  The difference is most likely due to
the Bouwens et al.\ (2012) PSFs not containing sufficient light in the
wings due to their basing their PSFs on only a small sample of faint
stars over the HUDF (for simplicity).  If the PSFs for all passbands
contained exactly the same amount of light in their wings, this would
not significantly bias the Bouwens et al.\ (2012) results.  However,
given that the wings of the redder bandpasses contain more light than
for the bluer passbands, this caused Bouwens et al.\ (2012) to measure
colors which were systematically too blue.  As a result, the $\beta$'s
that Bouwens et al.\ (2012) derived are mildly bluer than measured
here.

There was one additional potential bias regarding the Bouwens et
al. (2012) study that was discussed by Dunlop et al. (2013) and
developed further in Rogers et al.\ (2013).  Rogers et al.\ (2013)
show that the Bouwens et al.\ (2012) $\beta$ analysis would be biased
blueward \textit{if} Bouwens et al. (2012) had binned their selections
as a function of the $J_{125}$-band magnitude in evaluating trends in
$\beta$.  However, this was not done by Bouwens et al. (2012), so it
is unclear why Rogers et al.\ (2013) frame this as being potentially
problematic for the Bouwens et al.\ (2012) study.

\subsection{B.4  Comparisons with the $\beta$ Measurements of Dunlop et al.\ 2013}

Finally, we consider the recent $\beta$ measurements obtained by
Dunlop et al.\ (2013) for $z\sim7$ galaxies in the HUDF12/XDF.  The
individual measurements of $\beta$ were obtained from Table A1 of
their study.

\subsubsection{B.4.1  Source-By-Source Comparisons}

A comparison of the Dunlop et al.\ (2013) $\beta$ measurements with
our own $\beta$ determinations is shown in Figure~\ref{fig:compd12}.
Based on the plotted galaxies, we find a median offset of
0.24$_{-0.06}^{+0.18}$ in the value of $\beta$, with the Dunlop et
al.\ (2013) $\beta$'s being redder than those derived here.  Such an
offset in $\beta$ corresponds to a $\sim$0.06$_{-0.02}^{+0.04}$ mag
offset in the median $J_{125}-H_{160}$ color.  A similar conclusion
can be drawn comparing the Dunlop et al.\ (2013) results with those
from Finkelstein et al.\ (2012: \textit{right panel of
  Figure~\ref{fig:compd12}}).

\subsubsection{B.4.2  Effect of the WFC3/IR Reductions on the $\beta$
Measurements}

To investigate the origin of this offset, we first explored whether it
could originate from the WFC3/IR reductions we were utilizing.  We
therefore downloaded the publicly available reductions the HUDF12 team
made available to the community (Koekemoer et al.\ 2013), aligned this
data set with our own, and then resampled these reductions onto the
same grid as our own reductions of the HUDF12/HUDF09 WFC3/IR data set.
We derived the effective PSF for the resampled Koekemoer et
al.\ (2013) reductions, remeasured the fluxes for the sources using
the same apertures, and rederived $\beta$ for all the sources in our
$z\sim7$ and $z\sim8$ samples.  The median $J_{125}-H_{160}$ colors we
derive from the Koekemoer et al.\ (2013) reductions for $z\sim7$
sources from our selections differ by just $\sim$0.01 mag (in the
median) from those derived from our own reductions of the same data
set.  This suggests that the reductions themselves are not the source
of the observed differences.

\begin{figure}
\epsscale{0.5}
\plotone{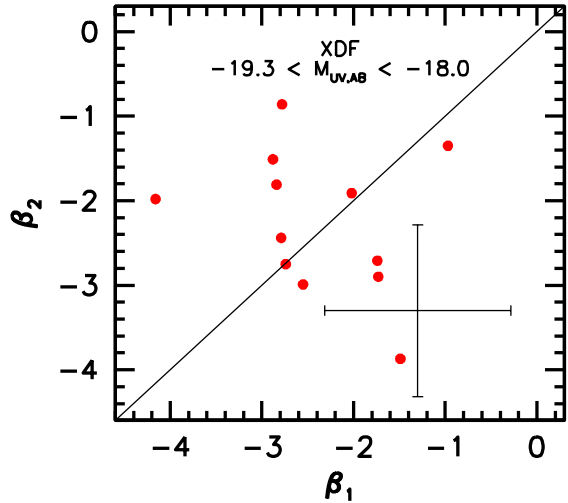}
\caption{A comparison of the two different $UV$-continuum slopes
  $\beta$ estimated for the same $-18.0<M_{UV,AB}<-19.3$ $z\sim7$
  candidates in the XDF (\textit{solid red circles}).  The $\beta$ on
  the horizontal axis ($\beta_1$) is estimated from the second
  $\sim$50\% of the $J_{125}$ and $H_{160}$-band observations.  The
  $\beta$ on the vertical axis ($\beta_2$) is estimated from the first
  50\% of the $J_{125}$ and $H_{160}$-band observations.  Each of the
  two $z\sim 7$ selections contained one source that was not present
  in the other selection.  A $1\sigma$ representative error bar is
  also shown.  The difference between the two $\beta$ measurements are
  consistent with what one would expect from the uncertainties in the
  $J_{125}$ and $H_{160}$-band fluxes.\label{fig:scatplot}}
\end{figure}

\subsubsection{B.4.3  Effect of the Photometric Procedure of the $\beta$ Measurements}

Next, we investigated whether the differences might originate from our
procedures for performing photometry.  We therefore measured
$J_{125}-H_{160}$ colors for sources in our own $z\sim7$-8 selection
using the Dunlop et al.\ (2013) methodology for performing photometry.
Dunlop et al.\ (2013) measure the $J_{125}$, $JH_{140}$, and $H_{160}$
fluxes for sources in 0.44$''$-diameter, 0.47$''$-diameter, and
0.50$''$-diameter circular apertures (with no PSF correction).  Dunlop
et al.\ (2013) have chosen these apertures for their photometry since
these apertures enclose 70\% of the light in the $J_{125}$,
$JH_{140}$, and $H_{160}$-band images for point sources.  [By
  comparison, we measure the fluxes of sources in identical Kron
  apertures, after PSF-correcting the HST observations to match the
  $H_{160}$-band observations.]

To investigate whether differences between the two photometric
procedures could be the source of the differences between the reported
$\beta$'s, we performed photometry on $z\sim7$ sources in our
selection using both procedures.  Overall, we found similar
$J_{125}-H_{160}$ colors using the two photometric procedures, but did
nevertheless note a $\sim$0.03 mag median offset between the measured
$J_{125}-H_{160}$ colors for sources in the luminosity range
$-19<M_{UV,AB}<-18$.\footnote{Utilizing the same PSF corrections as
  were utilized in Bouwens et al.\ (2012), we found the
  $J_{125}-H_{160}$ colors measured by the two procedures to differ by
  0.07-0.12 mag in the median (depending on the $UV$ luminosities of
  the sources considered).}

\subsubsection{B.4.4  Effect of Source Size on the Derived Color}

Why would there be such differences between the measured colors?  One
potential explanation is that such differences could occur if the
candidate $z\sim7$ galaxies were not perfect point sources (as assumed
by Dunlop et al.\ 2013) and in fact showed some spatial extension.

\begin{figure}
\epsscale{0.65} \plotone{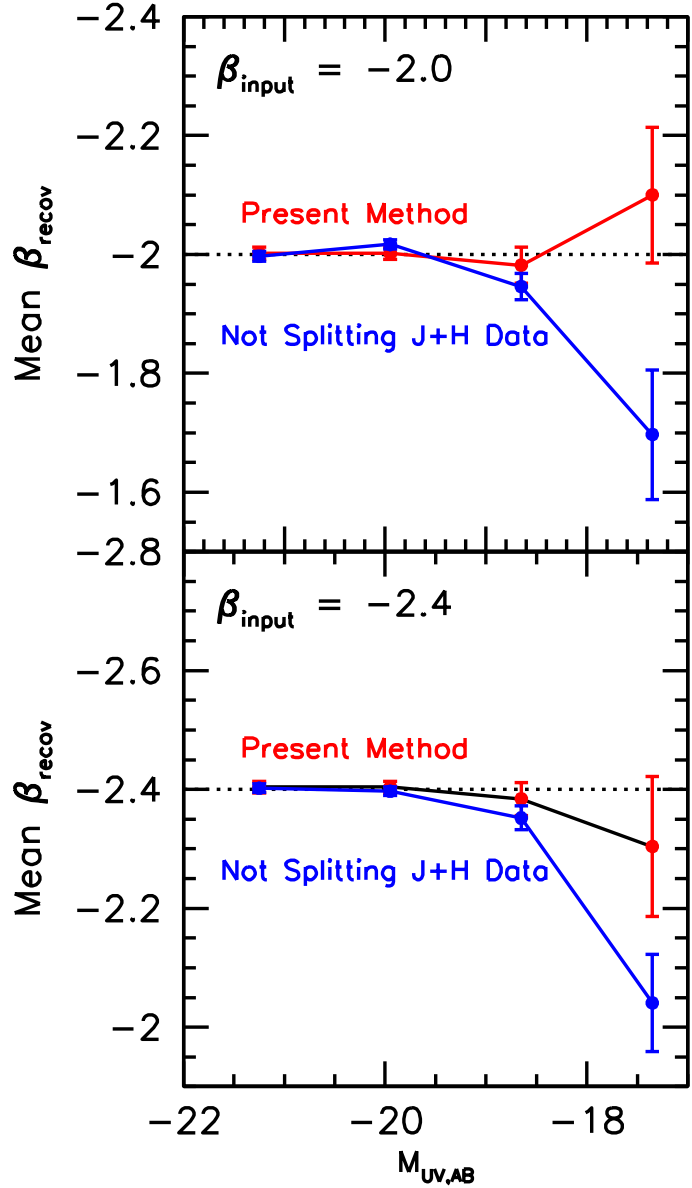}
\caption{The biweight mean $\beta$'s (\textit{red circles} with
  $1\sigma$ errors) we recover versus $UV$ luminosity $M_{UV,AB}$
  using a mock XDF data set and our proposed algorithm of splitting
  the $J_{125}$ and $H_{160}$-band exposures into two equal subsets
  and alternatively using one half of the observations to select
  sources and the other half to measure $\beta$.  The error bars are
  derived from bootstrap resampling our results.  The upper and lower
  panels show results assuming an input $\beta$ of $-2$ and $-2.4$,
  respectively, for all sources.  Also shown (\textit{blue circles})
  are the biweight mean $\beta$'s adopting a procedure where the
  $J_{125}$ and $H_{160}$ data are not split into two equal subsets so
  as to separate the measurement of $\beta$ from the source selection
  process and the determination of $UV$ luminosity.  In deriving these
  results, we used full end-to-end simulations, from the generation of
  images for mock galaxies and the selection of sources to the
  measurement of $\beta$.  All of the mean $\beta$'s we derive from
  the simulations for our preferred algorithm (\textit{red circles})
  are within $1\sigma$ of the input values, strongly suggesting our
  final results should be free of significant systematic errors.  This
  is in contrast to results for a procedure (\textit{blue circles})
  where the $J_{125}$ and $H_{160}$ data are not split into two equal
  subsets to separate the measurement of $\beta$ from source selection
  or the determination of the $UV$ luminosity.\label{fig:sims}}
\end{figure}

To explore the effect the non-zero sizes of $z\sim7$ galaxies have on
the $J_{125}$ and $H_{160}$ band photometry, we stacked the $Y_{105}$,
$J_{125}$, $JH_{140}$, and $H_{160}$-band images for all of the
sources in the lower luminosity ($-19.3<M_{UV,AB}<-18$) $z\sim7$ 
sample.  We then constructed the encircled-energy distributions for
this stacked source in the same way as we constructed this
encircled-energy distribution for the five isolated stars we
identified on the image.  We find that the ratio of light in
$0.50''$-diameter apertures to $0.44''$-diameter apertures is
consistently $\sim$3\% higher for faint ($-19.3<M_{UV,AB}<-18$)
galaxies than it is for stars, independent of whether we consider
stacks of the sources in the $Y_{105}$, $J_{125}$, $JH_{140}$, or
$H_{160}$ bands.

\begin{figure}
\epsscale{0.7} \plotone{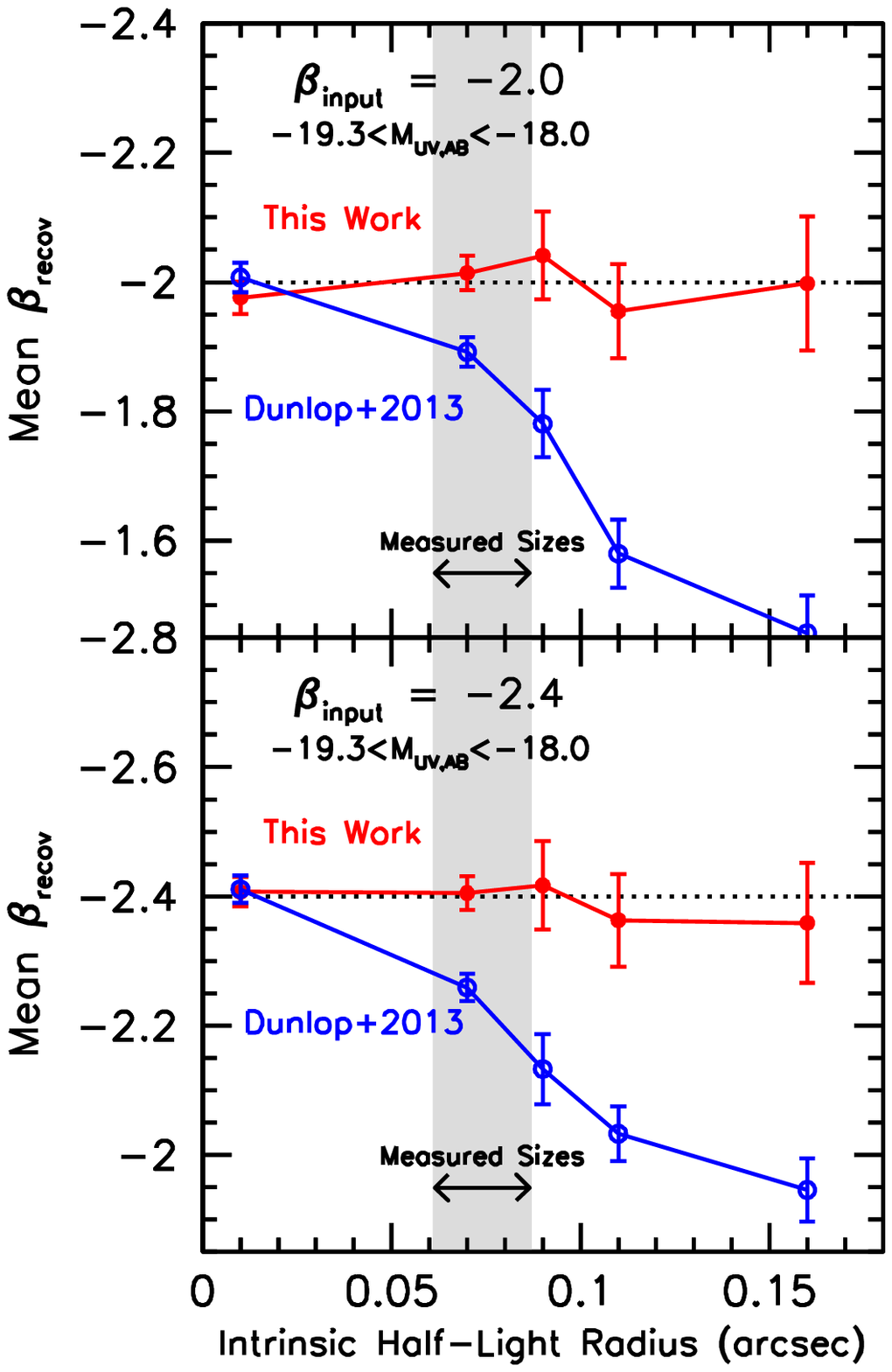}
\caption{The biweight mean $\beta$'s we recover (from full end-to-end
  simulations) versus the half-light radius of sources (\textit{solid
    red circles with $1\sigma$ errors}) using our proposed algorithm
  on a mock XDF data.  Also shown are the mean $\beta$'s we find using
  the photometric procedure of Dunlop et al.\ (2013: \textit{open blue
    circles with $1\sigma$ errors}).  The upper and lower panels show
  results assuming an input $\beta$ of $-2$ and $-2.4$, respectively,
  for all sources.  The error bars are derived from bootstrap
  resampling our simulation results.  While the mean $\beta$'s we
  derive are not significantly biased even in the case of large
  $z\sim7$ galaxies, the mean $\beta$ one would derive using the
  photometric procedure that Dunlop et al.\ (2013) apply is biased
  towards redder $\beta$'s for large $z\sim7$ galaxies (by
  $\Delta\beta\sim0.5$).  Since the faintest sources at $z\sim7$ are
  nevertheless quite small, i.e., $r_{hl}=0.074\pm0.013''$
  (\textit{shaded grey region}: see \S5.4 and Appendix B.4), the
  actual bias in $\beta$ for faint $z\sim7$ galaxies is modest but
  non-zero ($\Delta\beta\sim0.13$ bias for faint $z\sim7$
  galaxies).\label{fig:sims2}}
\end{figure}

Fitting for the half-light radius with \texttt{galfit} (Peng et
al.\ 2002), we find a size of $0.074''$$\pm$0.013$''$, similar to the
half-light radius (i.e., $\sim$0.06$''$ to 0.07$''$) measured by Ono
et al.\ (2013).  We checked that this size roughly reproduces the 3\%
differential excess observed for faint galaxies relative to stars.
Computing the encircled energy distribution expected for sources with
this radial profile and comparing with the results for sources with
zero size, we calculate that sources with this half-light radius would
be subject to a $\sim0.03$ to 0.04 mag bias in their measured
$J_{125}-H_{160}$ colors using the Dunlop et al.\ (2013) photometric
scheme.  Such a color bias would translate into a $\sim0.13$-0.18 bias
in $\beta$.  Different source sizes, of course, translate into
different expected biases in $\beta$.  In Figure~\ref{fig:sizebias}
from \S5.4, we show how the bias in the $J_{125}-H_{160}$ colors and
$\beta$ depends on the average half-light radii of sources under
consideration (see also Figure~\ref{fig:sims2} from Appendix E).

In addition, for $z\sim7$ galaxies in the vicinity of other bright
galaxies, it is also possible that the color measurements could be
biased by light from neighboring sources.  The wider 0.5$''$-diameter
apertures (used for the $H_{160}$-band photometry) would include
proportionally more light from neighboring sources than narrower
$0.44''$-diameter apertures (used for the $J_{125}$-band photometry).
While we would expect this effect to be small for typical sources, the
systematic differences between different studies are on the 3-4\%
level and so even these types of effects could play a role.

\section{C.  Differences between the $\beta$ Measurements We Obtain using Two Independent Splits of the Observations}

In this paper, we have presented an algorithm to obtain $\beta$
measurements for $z\sim7$ samples in a way that is robust against
noise-driven biases.  We accomplish this by splitting our $J_{125}$
and $H_{160}$-band observations into two independent subsets and using
one half of the observations for selecting individual sources and the
other half for measuring $\beta$.  The role of the two 50\% splits of
the data is then reversed.

While we were able to demonstrate that our procedure indeed allows to
measure the mean $\beta$ for faint galaxy samples in a way that is
robust against noise-driven biases (Appendix D), an important question
is how such a procedure performs in practice for faint sources in real
data sets.  An example of this performance is illustrated in
Figure~\ref{fig:scatplot} for faint $-19.3<M_{UV,AB}<-18.0$ galaxies
from the XDF.  Sources which are selected from both halves of the data
set are shown at the horizontal and vertical values corresponding to
their measured $\beta$ determinations in the first and second
selections on each data set, respectively.  Sources not selected in
the first or second halves of the data set are plotted at zero on the
vertical or horizontal axis, respectively.

From this figure, it is clear that for the luminosity interval chosen
noise plays almost no role in moving individual $z\sim7$ galaxies
inside or outside of our $z\sim7$ selection window.  Essentially all
$z\sim7$ candidates are present in both selections.  Moreover, the
measured $\beta$'s from the two selections typically differ by less
than $\Delta\beta\sim0.8$.  While these differences are modest, larger
differences between the two selections are expected near the flux
limit of the XDF data set.

\section{D.  Realistic End-to-End Simulations Demonstrating that Our New Algorithm is Robust Against Noise-Driven Biases}

In the main text of this paper, we have described a new algorithm for
measuring $\beta$ for $z\sim7$ galaxies that is robust against
noise-driven biases.  This approach is valuable in cases where data
are available in only three near-infrared filters, since this would
normally lead to systematic errors from the crosstalk between color
information needed for redshift determination and that needed to
determine $\beta$.  We demonstrate here the effectiveness of this
approach using realistic end-to-end simulations.

We assume that all galaxies have exactly the same intrinsic SED --
which we take to be a perfect power-law spectrum with $\beta=-2$.  We
assume there are an equal number of galaxies at ten discrete
redshifts, $z=6.2$, $z=6.4$, ..., $z=7.6$, and at five apparent
magnitudes in the $H_{160}$ band $H_{160,AB}=27.5$, $H_{160,AB}=28$,
$H_{160,AB}=28.5$, $H_{160,AB}=29.0$, and $H_{160,AB}=29.5$ (before
adding noise).  We include the effect of opacity from the neutral
hydrogen forest following the Madau (1995) prescription, with no
transmittance at redshifts higher than $z=6$.  

Based on this intrinsic SED and opacity model, we calculate the flux
of sources in the $z_{850}$, $Y_{105}$, $J_{125}$, and $H_{160}$
bands.  We then simulated 11 different images of each model galaxy,
including one image each for the
$B_{435}V_{606}i_{775}I_{814}z_{850}Y_{105}JH_{140}$ bands and two
equal-depth images for the $J_{125}$ and $H_{160}$ bands.  We assume
that the half-light radius distribution of galaxies is equally divided
between galaxies with 0.01$''$, 0.06$''$, 0.11$''$, and 0.16$''$
radii.  Axial ratios for sources were assumed to be uniformly
distributed between 1.0 and 1.5.  Simulated sources were then
convolved with our derived PSFs in the different passbands (\S2.2 and
Appendix A) and noise added to match that present in the XDF data set.
100 galaxies were included per arcmin$^2$ on the simulated images.

We then selected and measured $\beta$ for individual sources using
exactly the same procedure as we applied on the XDF data set itself
(and other data sets in this study: see \S4.3).  The biweight mean
$\beta$'s we derive assuming an input $\beta$ of $-2$ and $-2.4$ is
shown in Figure~\ref{fig:sims}, and it is clear that our procedure
works very well in recovering the input $\beta$'s.  The procedure we
use to derive the biweight mean $\beta$ is the same, as what we use on
the observations themselves.  $3\times$10$^4$ sources are used in
these simulations.

We also present our recovered biweight mean $\beta$'s using a simpler
procedure where the $J_{125}$ and $H_{160}$ band data are not split
into two equal pieces.  In this case, the same $J_{125}$ and
$H_{160}$-band flux measurements are used for source selection, the
measurement of $\beta$, and the determination of the $UV$ luminosity.
Whereas our preferred algorithm shows a bias consistent with zero,
this alternate algorithm (i.e. using the same $J_{125}$ and
$H_{160}$-band flux measurements for all three processses which is the
most direct approach) shows a small but clear bias.

\section{E.  The Sensitivity of the Derived $\beta$'s to Source Size 
Using the Present Approach and the Dunlop et al.\ (2013) Photometric Procedure}

As we emphasized in \S5.4, the $\beta$'s one derives for $z\sim7$
galaxies can be biased by the sizes of $z\sim7$ galaxies and the
assumptions one's photometric procedure makes about these sizes.
Since we ensure that our observations are PSF-matched before we
perform our photometry, we would not expect our measured $\beta$'s to
be affected by the size of $z\sim7$ galaxies.  By contrast, Dunlop et
al.\ (2013) derive colors for sources in apertures containing 70\% of
the light for point sources (0.44$''$-diameter apertures in the
$J_{125}$ band, 0.47$''$-diameter apertures in the $JH_{140}$ band,
and 0.50$''$-diameter apertures in the $H_{160}$ band).  Adopting the
latter approach can result in biased measurements of the
$J_{125}-H_{160}$ colors if $z\sim7$ galaxies are not in fact point
sources.

We can use the same set of simulations as we used in the previous
section to test the dependence of the robustness of our $\beta$
measurements on the size of sources.  We will also use these same
simulations to test the dependence of the $\beta$ measurements on
source size, if we adopt the Dunlop et al.\ (2013) photometric
procedure.  We focus on the results in the magnitude interval
$-19.3<M_{UV,AB}<-18.0$, since this has been the focus of the
controversy.

In upper and lower panels of Figure~\ref{fig:sims2}, we show how the
mean $\beta$ we measure depends on the half-light radius of sources,
assuming an input $\beta$ of $-2$ and $-2.4$ in the simulations.  In
both cases, we are able to recover the mean $\beta$'s to which the
quoted uncertainties over the full range of source sizes $0.01''$ to
$0.16''$.  On the same figure, we also plot the mean $\beta$'s we
derive using the Dunlop et al.\ (2013) photometric procedure, again as
a function of the size of $z\sim7$ galaxies input into the
simulations.  Again, while the Dunlop et al.\ (2013) photometric
procedure is successful at recovering $\beta$ quite accurately for
small ($\sim$0.01$''$) sources, we find a redward bias in the measured
$\beta$ as the size of the source increases.  For sources with
half-light radii of $\sim$0.15$''$, we find that that the derived
$\beta$'s are too red by $\Delta\beta \sim 0.5$.  Our results here are
in contrast to those from Dunlop et al.\ (2013), who conclude based on
the end-to-end testing they run that there is no bias in the $\beta$'s
they measure for $z\sim7$ galaxies.  The conclusion of Dunlop et
al.\ (2013) is incorrect because they model faint $z\sim7$ galaxies as
point sources in their simulations, and as shown here and by Ono et
al.\ (2013), this is not an accurate assumption to make.

\begin{figure}
\epsscale{1.17}
\plotone{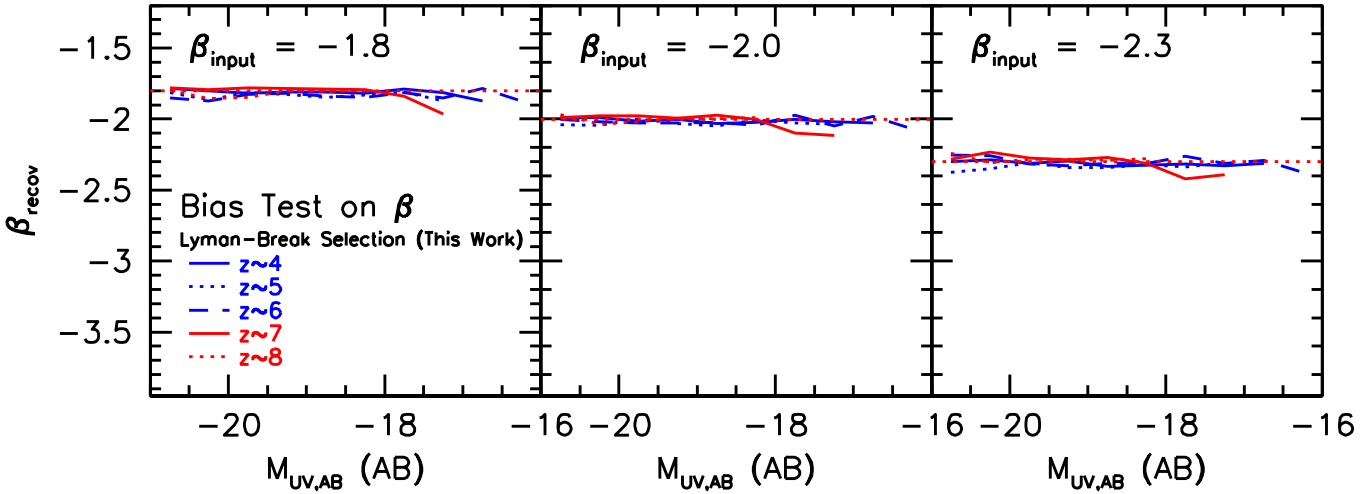}
\caption{Estimated mean $\beta$ we would recover as a function of the
  $UV$ luminosity of sources due to the fact that galaxies with
  intrinsically bluer $\beta$'s are more selectable than galaxies with
  redder $\beta$'s (see Appendix F).  The input $\beta$ distribution
  for these simulations is assumed to have a mean value of
  $\beta=-1.8$ (\textit{left}), $\beta=-2.0$ (\textit{center}), and
  $\beta=-2.3$ (\textit{right}), with a $1\sigma$ scatter of 0.35.
    Similar simulations were performed in Appendix B.1.1 of Bouwens et
    al.\ (2012).  Shown are the results for our $z\sim4$, $z\sim5$,
    $z\sim6$, $z\sim7$, and $z\sim8$ selections over the XDF data set.
    The typical bias in $\beta$ is $\Delta\beta \lesssim 0.05$ for our
    selection criteria.\label{fig:seleff}}
\end{figure}

\begin{figure}
\epsscale{0.5}
\plotone{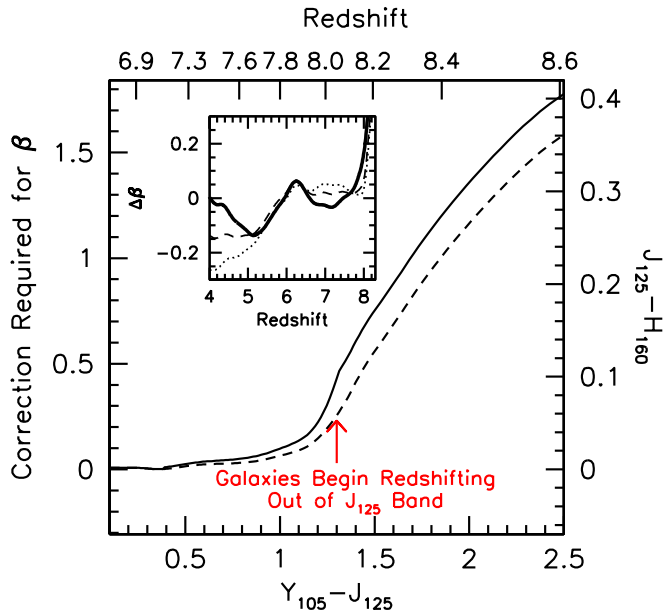}
\caption{Estimated correction (solid line) that must be made to the
  observed spectral slopes for $z\sim8$ galaxies inferred from the
  $J_{125}-H_{160}$ colors.  The correction is relative to the
  intrinsic $\beta$ for the model SED under consideration, which is
  calculated from the 1270\AA$~$to 2600\AA$~$wavelength range
  traditionally used to define $\beta$ (Calzetti et al.\ 1994).  The
  required correction to $\beta$ is a function of the redshift of a
  galaxy, which can be approximately computed based on the
  $Y_{105}-J_{125}$ color.  Corrections are necessary to account for
  two effects: (1) the redshifting of galaxies out of the
  $J_{125}$-band at $z\gtrsim8.1$ (resulting in substantially redder
  $J_{125}-H_{160}$ colors), (2) the impact of the Ly$\alpha$ damping
  wing on the observed colors, and (3) the somewhat redder shape of
  the SED blueward of 1350\AA.  The net results of the three effects
  is that, if uncorrected, galaxies would appear to have redder
  $UV$-continuum slopes $\beta$ at $z\sim8$ than at $z\sim7$, even if
  there is no evolution.  The Ly$\alpha$ damping wing is computed
  assuming a $10^{21}$ cm$^2$ neutral hydrogen column and a $x_e=0.2$
  neutral fraction in the IGM.  The dashed line shows the correction
  not accounting for the Ly$\alpha$ damping wing.  The figure inset
  shows how the correction depends on the precise shape of the SED for
  a star-forming galaxy and is shown over a larger range in redshift.
  The dashed, dotted, and thick solid curves give the corrections
  assuming (1) a $10^8$ year constant star formation model, (2) a
  10-Myr constant star formation model, and (3) an $80$-Myr constant
  star formation model followed by 20-Myr with no star formation.  The
  correction to $\beta$ that we utilize is the average of models (1)
  and (3).  Dunlop et al.\ (2013) do not discuss correcting the
  $\beta$'s they derive at $z\sim8$ for the effect of the IGM on the
  $J_{125}$-band fluxes.  While this effect will not have a huge
  impact on the derived $\beta$'s from Dunlop et al.\ (2013), we do
  estimate a redward bias of $\Delta\beta\sim0.1$ using the redshift
  distribution that Dunlop et al.\ (2013) provide for their $z\sim8$
  sample.\label{fig:beta8}}
\end{figure}

\begin{figure}
\epsscale{0.5}
\plotone{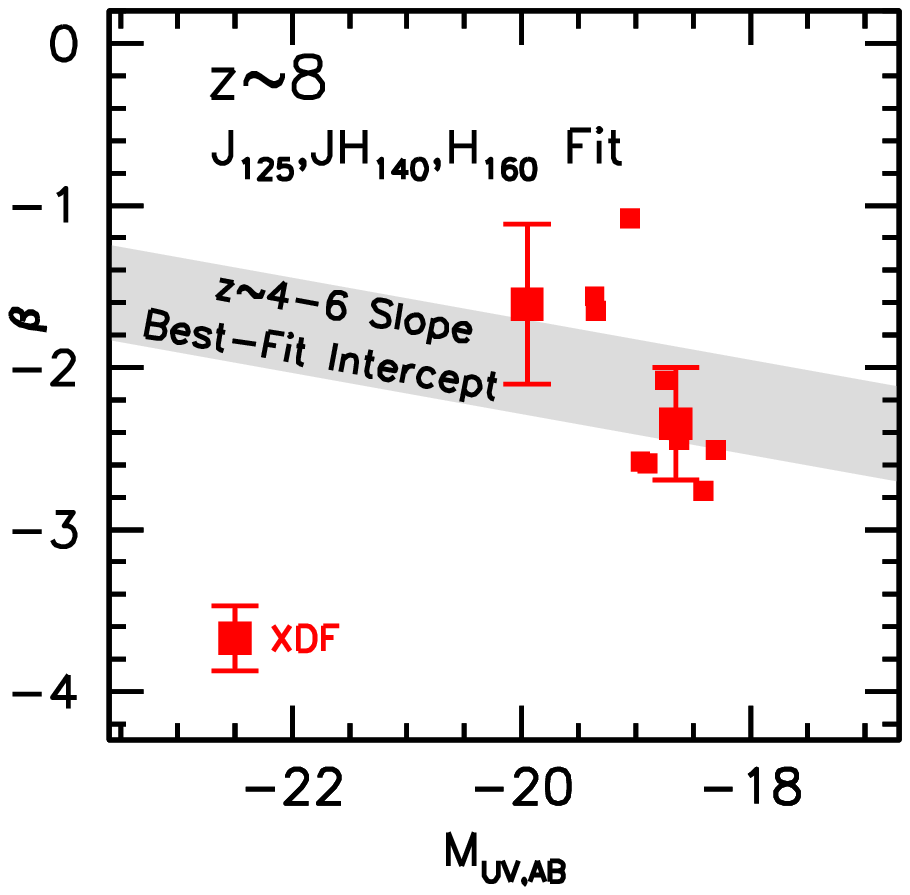}
\caption{Measured $\beta$'s (small red squares) versus the rest-frame
  $UV$ luminosity $M_{UV}$ for galaxies in our XDF sample of $z\sim8$
  galaxy candidates (see Appendix G.2).  The $\beta$'s we derive for
  sources here are based on a power-law fit to the $J_{125}$,
  $JH_{140}$, and $H_{160}$ photometry, in contrast to
  Figure~\ref{fig:colmag8a} where we present our results for our XDF
  sample based on a power-law fit to the $JH_{140}$ and $H_{160}$
  photometry.  The large red squares represent the
  inverse-variance-weighted mean $\beta$'s and $1\sigma$
  uncertainties, in various 1.3-mag bins of $UV$ luminosity.  The
  gray-shaded band shows the $1\sigma$ range allowed for a fit to the
  $\beta$-$M_{UV}$ relation fixing the slope of this relation to the
  average value ($-0.13$) found by Bouwens et al.\ (2012) for
  $z\sim4$-6 galaxies.\label{fig:colmag8b}}
\end{figure}

\section{F.  Computing the ``Selection Volume Bias'': Bias Related to the 
Intrinsic Selectability of Sources}

All photometric criteria used to identify high-redshift galaxies,
whether it be a traditional Lyman-Break selection or a simple
photometric redshift approach, are much more effective in the
selection of sources with bluer spectral slopes than in the selection
of sources with redder slopes.  As a result, the bluest galaxies at
high redshifts will be overrepresented in high-redshift samples
relative to the reddest galaxies (proportionally speaking), resulting
in a biased measurement of the mean $\beta$ in high-redshift samples,
if one does not control for these effects.

In general, the bias in the $\beta$ one derives will be quite large,
if the information one uses to derive $\beta$ is not independent of
the information one uses to select sources.  In this situation, noise
can drive both the selection of sources and the measurement of
$\beta$, resulting in potentially large biases in $\beta$.  Such a
scenario is extensively discussed by Bouwens et al.\ (2012) and Dunlop
et al.\ (2012) and has been called the photometric error coupling bias
by Bouwens et al.\ (2012: see Appendix B.1.2 from that work).

As a result of such concerns, great care has been taken here to ensure
that the $\beta$ measurements reported here are entirely decoupled
from the selection process (and for the definition of the photometric
apertures) for the faintest sources in all of our high redshift
samples.  This ensures, by construction, that the photometric error
coupling bias is identically zero.

However, one can still arrive at a biased measurement of the mean
$\beta$, even if one measures their mean $\beta$ using independent
information from what one uses to select high-redshift sources.  This
is as a result of the fact that sources with intrinsically bluer
colors are selected more efficiently than those with redder colors.

To compute the size of this bias for our $\beta$ results, we added
artificial sources to the real observations over a wide range in
$\beta$ (i.e., $\beta=-4$ to $\beta=2$), $UV$ luminosity, and
redshift.  To simulate the spatial profiles of individual sources we
added to the real data, we made use of similar-luminosity $z\sim4$
galaxies from the XDF data set and artificially redshifted them using
our ``cloning'' machinery (Bouwens et al.\ 1998; Bouwens et al.\ 2003)
to a range of redshifts over the selection window of our high-redshift
samples, scaling their physical sizes as $(1+z)^{-1}$ to match the
observed scalings (e.g., Bouwens et al.\ 2006; Oesch et al.\ 2010;
Mosleh et al.\ 2012).  After adding the sources to the real data, we
then attempted to reselect the sources using the same procedure as we
apply to the real observations, and we computed the selection
efficiency of galaxies as a function of their $UV$ luminosity and
their intrinsic $\beta$.

Then, using the selection efficiency results from the above
simulations and assuming that the inherent $\beta$'s for $z\sim4$-8
galaxies is normally distributed with some mean value and a standard
deviation $\sigma_{\beta}\sim0.35$ (consistent with that derived by
Bouwens et al.\ 2009; Bouwens et al.\ 2012; Castellano et al.\ 2012),
we have calculated the bias in $\beta$.  The results are shown in
Figure~\ref{fig:seleff}.  In general, the bias we expect in $\beta$ is
not especially large, $\Delta\beta \lesssim 0.05$, and should not
appreciably bias the results for all three input $\beta$'s we have
considered, i.e., $\beta\sim-1.8$, $\beta\sim-2.0$, and
$\beta\sim-2.3$.  While the computed bias could be quite a bit larger
if the intrinsic $\beta$ distribution were larger, recent results by
Rogers et al.\ (2014) not only confirm that the intrinsic width of the
$\beta$ distribution is $\sim$0.35, but provide evidence that the
width of this distribution is even narrower for lower luminosity
sources.  This strongly suggests that any biases resulting from
differences in the intrinsic selectability of sources should be small,
i.e., $\Delta\beta \lesssim 0.05$.  Similar simulations were
previously presented in Appendix B.1.1 of Bouwens et al. (2012).

\section{G.  $\beta$ Determinations for $z\sim8$ Galaxies by Fitting to the Measured $J_{125}+JH_{140}+H_{160}$ Fluxes}

\subsection{G.1  General Procedure}

Our primary method for deriving $\beta$ at $z\sim8$ only makes use of
the flux measurements in the $JH_{140}$ and $H_{160}$ bands.  The
advantage of this approach is that for both bands, the full wavelength
coverage lies firmly in the $UV$-continuum and redward of the Lyman
break, and therefore the $JH_{140}-H_{160}$ color provides a reliable
measurement of the $UV$ slope $\beta$.  One significant drawback,
however, in restricting ourselves to the use of these bands in
deriving $\beta$ is the limited leverage in wavelength the two bands
provide for determining $\beta$, resulting in much larger statistical
uncertainties in the derived $\beta$'s.

Fortunately, one can consider making use of the observed
$J_{125}$-band flux to gain additional leverage for constraining the
$UV$-continuum slope $\beta$.  The challenge with including
$J_{125}$-band flux measurements in the fits is that both Ly$\alpha$
and the Lyman break begin to enter the $J_{125}$ band at $z\gtrsim8.1$
(see Figure~\ref{fig:beta8}).  This can significantly bias the
measured $\beta$'s for individual $z\sim8$ sources if this effect is
not corrected for.  The Lyman break would make $z\sim8$ galaxies too
faint in the $J_{125}$-band relative to the baseline continuum values,
while Ly$\alpha$ emission would make the $J_{125}$-band fluxes too
bright.  While one might imagine that this issue would be isolated to
only those sources with redshift estimates in excess of 8.0, the
redshift of each source in our $z\sim8$ sample is uncertain and so
could potentially be in excess of 8 (and therefore be biased by this
issue, if not corrected).  Corrections for IGM absorption can also be
somewhat model dependent.  It is because of these challenges that we
opted not to use this as our primary approach.

In order to overcome these issues and still make use of the observed
$J_{125}$-band fluxes to derive $\beta$, it is clear that a correction
to the $J_{125}$-band fluxes of $z\sim8$ sources is required.  Here we
estimate the approximate effect the Lyman-break has on the
$J_{125}$-band flux based on the redshifts we infer for sources using
their measured $Y_{105}-J_{125}$ colors.  Three different SED
templates are considered in correcting the $J_{125}$-band fluxes for
$z\sim8$ galaxies in our samples: the first assuming constant star
formation for 100 Myr, the second assuming constant star formation for
10 Myr, and the third assuming constant star formation for $80$ Myr
followed by 20 Myr with no star formation.  Different star formation
histories were explored to see what effect the shape of the
$UV$-continuum has on the derived $UV$-continuum slopes.  SED
templates were derived for each of these star-formation histories
using the Bruzual \& Charlot (2003) spectral synthesis code and 0.2
$Z_{\odot}$ metallicities.  From the inset of Figure~\ref{fig:beta8},
it is clear that the unknown star-formation histories for individual
sources could have a modest effect on the results.  The $\beta$
corrections we adopt are taken to be an average of the results for the
first and third model star-formation histories.

The observed $J_{125}$-band fluxes for $z\sim8$ galaxies are also
attenuated somewhat due to damped absorption from neutral hydrogen gas
in the galaxy and likely an increasingly neutral IGM.  Assuming a
neutral hydrogen column of $\sim$1$\times$10$^{21}$ cm$^{-2}$ in
individual galaxies, we estimate that the Ly$\alpha$ damping wing
could attenuate the $J_{125}$-band flux seen from individual sources
by $\sim$2-3\%.  We also account for some absorption from neutral
hydrogen in the IGM (Miralda-Escude 1998), assuming a neutral fraction
$x_e$ of 0.2 at $z\sim8$ (Bouwens et al.\ 2012; Kuhlen \&
Faucher-Giguere 2012; Robertson et al.\ 2013).  The observed
$J_{125}$-band fluxes are corrected for this effect based on the
redshifts we infer for sources from their $Y_{105}-J_{125}$ colors.

Accounting for the effect of Ly$\alpha$ emission on the observed
$J_{125}$-band fluxes would also be useful.  However, spectroscopic
follow-up of current $z\sim7$-8 samples (e.g., Ono et al.\ 2012;
Schenker et al.\ 2012; Pentericci et al.\ 2011; Caruana et al.\ 2012)
seems to suggest that Ly$\alpha$ emission is quite rare in these
sources (likely due to an increasingly neutral medium).  Therefore, we
would not expect Ly$\alpha$ emission to have a large effect on the
observed colors.  If this is not the case for all sources, this could
bias the measurements.

This second approach to deriving $\beta$ for $z\sim8$ galaxies is very
similar to one recently utilized by Dunlop et al.\ (2013) in examining
the spectral slopes of $z\sim8$ galaxies over the HUDF12.  However,
Dunlop et al.\ (2013) appear to have made no correction to the
$J_{125}$-band fluxes used in their fits, since this issue is not
discussed in their paper.  Using the redshift distribution Dunlop et
al.\ (2013) estimate for their $z\sim8$ sample (Appendix A of Dunlop
et al.\ 2013), we estimate that the bias in $\beta$ will be $\sim 0.1$
to the red.  While this bias is likely not huge (and here we assume
that the contribution from Ly$\alpha$ emission is not substantial), it
will cause the $\beta$'s inferred by Dunlop et al.\ (2013) to be
slightly redder than what we find in \S6.2 (at least for the $z\sim8$
$\beta$'s they derive from the $J_{125}-H_{160}$ colors).

\subsection{G.2  Results for $z\sim8$ Sample}

In this section, we apply the approach we describe in the previous
section to the three deepest HST fields to derive the mean $\beta$ for
faint $z\sim8$ galaxies.  For consistency with our primary $\beta$
determinations for $z\sim8$ galaxies in \S6, we use the same criteria
for selecting $z\sim8$ galaxies as given in \S6.1.  There we derived
the mean $\beta$ based on a small $z\sim8$ sample from the XDF data
set using a power-law fit to the $JH_{140}$ and $H_{160}$-band fluxes.

Similar to the strategy we employ on our $z\sim7$ samples, we split
the $J_{125}$-band observations into two equal subsets and
alternatively select and measure $\beta$ on each half of the data.
This is to ensure that our results are not biased due to a coupling
between the $J_{125}$-band flux used to derive $\beta$ and that used
to select the sources.  Given the sensitivity of our results to having
accurate $Y_{105}-J_{125}$ colors for candidates in our $z\sim8$ (see
Figure~\ref{fig:beta8}), we restrict our analysis to only those
sources from the XDF data set in determining the mean $UV$-continuuum
slopes $\beta$.

We plot the observed $\beta$'s for individual sources in
Figure~\ref{fig:colmag8b} as a function of their measured
luminosities.  Subdividing our $z\sim8$ sample by $UV$ luminosity, we
derive the inverse-variance-weighted mean $\beta$'s for sources in
each luminosity bin.  The results are then presented in
Figure~\ref{fig:colmag8b} as the large blue squares.  The
inverse-variance-weighted mean $\beta$ in the luminosity interval
$-19.3<M_{UV,AB}<-18.0$ is $-2.39\pm0.35\pm0.13$.  This determination
is quite consistent with the $\beta$ for derive for $z\sim8$ in a
similar luminosity interval in \S6.2, i.e., $-$1.88$\pm$0.74$\pm$0.27.
While it might seem surprising that our determination using the full
$J_{125}+JH_{140}+H_{160}$ flux information is bluer than the
determination from \S6.2 given what was argued about the impact of the
IGM on the $J_{125}$ flux for $z\gtrsim8$ sources, we emphasize that
the $z\sim8$ $\beta$'s we derive in this section are based on
$J_{125}$-band fluxes \textit{corrected} for the impact of the IGM.

Finally, we fit the $\beta$ vs. $M_{UV}$ relationship to a line,
fixing the slope of this relationship to the average slope found for
$z\sim4$-6 galaxies by Bouwens et al.\ (2012), i.e., $-0.15$.  The
best-fit intercept for this relationship at $M_{UV}$ luminosity of
$-19.5$ mag is $-2.01\pm0.28\pm0.13$.  This value is in excellent
agreement with that given in \S6.2.

\section{H.  Complete Set of $\beta$ and $M_{UV}$ Measurements for Our $z\sim4$-8
Samples}

To facilitate comparisons with the present results (and in the
interests of transparency), we provide a complete list of the sources
in our $z\sim4$, $z\sim5$, $\sim6$, $z\sim7$, and $z\sim8$ samples in
Table~\ref{tab:mrtable}.  This table includes the coordinates of the
sources as well as their derived $\beta$'s and $UV$ luminosities
($M_{UV}$).  A complete set of the flux measurements used to derive
$\beta$ for these sources is provided in Table~\ref{tab:phottable}.

\begin{deluxetable}{ccccccc}
\tablecaption{A complete list of the UV-continuum slopes $\beta$ we 
       measure for sources in our $z\sim4$, $z\sim5$, $z\sim6$, $z\sim7$, and $z\sim8$
       samples\tablenotemark{a}\label{tab:mrtable}}
\tablehead{\colhead{Source ID} & \colhead{R.A.} & \colhead{Declination} & \colhead{$M_{UV,AB}$} & \colhead{$\beta$} & \colhead{$<z>$} & \colhead{Data Set\tablenotemark{b}}}
\startdata
XDF-23876748271 & 03:32:38.76 & $-$27:48:27.11 & $-$19.54 &  0.00$\pm$0.14 & 4 & 1  \\
XDF-23848748246 & 03:32:38.49 & $-$27:48:24.64 & $-$16.54 & $-$3.00$\pm$1.42 & 4 & 1  \\
XDF-23848748214 & 03:32:38.49 & $-$27:48:21.44 & $-$17.92 & $-$1.82$\pm$0.22 & 4 & 1  \\
XDF-23842748186 & 03:32:38.42 & $-$27:48:18.69 & $-$16.45 & $-$2.40$\pm$0.55 & 4 & 1  \\
XDF-23796748182 & 03:32:37.96 & $-$27:48:18.22 & $-$16.00 &  0.36$\pm$1.32 & 4 & 1  \\
XDF-23766748168 & 03:32:37.66 & $-$27:48:16.88 & $-$17.24 & $-$2.08$\pm$0.35 & 4 & 1  \\
XDF-23800748165 & 03:32:38.00 & $-$27:48:16.51 & $-$19.14 & $-$1.16$\pm$0.13 & 4 & 1  \\
XDF-23884748189 & 03:32:38.84 & $-$27:48:18.97 & $-$20.24 & $-$0.95$\pm$0.12 & 4 & 1  \\
XDF-23859748162 & 03:32:38.60 & $-$27:48:16.23 & $-$18.06 & $-$2.01$\pm$0.23 & 4 & 1  \\
XDF-23836748145 & 03:32:38.37 & $-$27:48:14.51 & $-$15.84 & $-$3.07$\pm$1.96 & 4 & 1  
\enddata
\tablenotetext{a}{Table~\ref{tab:mrtable} is published in its entirety in the electronic edition of the Astrophysical Journal.  A portion is shown here for guidance regarding its form and content.}
\tablenotetext{b}{The data set from which the source was selected and in
          which its UV-continuum slope beta derived (1 = XDF,
	  2 = HUDF09-1, 3 = HUDF09-2, 4 = CANDELS/ERS)}
\end{deluxetable}

\begin{deluxetable}{ccccccccc}
\tablecaption{Complete set of flux measurements used to derive $\beta$ for our $z\sim4$,
  $z\sim5$, $z\sim6$, $z\sim7$, and $z\sim8$ samples\tablenotemark{a}\label{tab:phottable}}
\tablehead{
\colhead{} & \multicolumn{8}{c}{Measured Flux (100 pJy)} \\
\colhead{Source ID} & \colhead{$i_{775}$} & \colhead{$I_{814}$} & \colhead{$z_{850}$} & \colhead{$Y_{098}$} & \colhead{$Y_{105}$} & \colhead{$J_{125}$} & \colhead{$JH_{140}$} & \colhead{$H_{160}$}}
\startdata
XDF-23876748271     & 760$\pm$24 & 827$\pm$66 & 871$\pm$40 & --- & --- & 1862$\pm$42 & --- & --- \\
XDF-23848748246     & 62$\pm$15 & $-$28$\pm$42 & 13$\pm$25 & --- & --- & 40$\pm$26 & --- & --- \\
XDF-23848748214     & 211$\pm$11 & 198$\pm$32 & 235$\pm$19 & --- & --- & 216$\pm$18 & --- & --- \\
XDF-23842748186     & 60$\pm$9 & 29$\pm$24 & 61$\pm$15 & --- & --- & 48$\pm$13 & --- & --- \\
XDF-23796748182     & 11$\pm$10 & 26$\pm$26 & 27$\pm$16 & --- & --- & 52$\pm$17 & --- & --- \\
XDF-23766748168     & 112$\pm$11 & 133$\pm$30 & 99$\pm$19 & --- & --- & 111$\pm$18 & --- & --- \\
XDF-23800748165     & 578$\pm$11 & 594$\pm$32 & 698$\pm$19 & --- & --- & 823$\pm$18 & --- & --- \\
XDF-23884748189     & 1545$\pm$12 & 1550$\pm$34 & 1866$\pm$21 & --- & --- & 2430$\pm$19 & --- & --- \\
XDF-23859748162     & 247$\pm$15 & 213$\pm$42 & 286$\pm$26 & --- & --- & 259$\pm$23 & --- & --- \\
XDF-23836748145     & 25$\pm$9 & 17$\pm$24 & 24$\pm$14 & --- & --- & 7$\pm$13 & --- & --- 
\enddata
\tablenotetext{a}{Table~\ref{tab:phottable} is published in its entirety
  in the electronic edition of the Astrophysical Journal.  A portion
  is shown here for guidance regarding its form and content.}
\end{deluxetable}


\begin{thebibliography}{} 
\bibitem[Alavi et al.(2014)]{2014ApJ...780..143A} Alavi, A., Siana, B., 
Richard, J., et al.\ 2014, \apj, 780, 143
\bibitem[Beckwith et al.(2006)]{2006AJ....132.1729B} Beckwith, S.~V.~W., et 
al.\ 2006, \aj, 132, 1729
\bibitem[Beers et al.(1990)]{1990AJ....100...32B} Beers, T.~C., Flynn, K., 
\& Gebhardt, K.\ 1990, \aj, 100, 32 
\bibitem[Bouwens, Broadhurst and Silk (1998)]{1998ApJ...506..557B} Bouwens,
R., Broadhurst, T.\ and Silk, J.\ 1998, \apj, 506, 557
\bibitem[Bouwens et al.(2003)]{2003ApJ...593..640B} Bouwens, R., 
Broadhurst, T., \& Illingworth, G.\ 2003, \apj, 593, 640 
\bibitem[Bouwens et al. (2006)]{2006Bouwens} Bouwens, R.J., Illingworth,
G.D., Blakeslee, J.P., \& Franx, M.  2006, \apj, 653, 53 
\bibitem[Bouwens et al.(2007)]{2007ApJ...670..928B} Bouwens, R.~J., 
Illingworth, G.~D., Franx, M., \& Ford, H.\ 2007, \apj, 670, 928
\bibitem[Bouwens et al.\ (2009)]{2009ApJ...705..936B} Bouwens, R.~J., et
  al.\ 2009, \apj, 705, 936
\bibitem[Bouwens et al.(2010)]{2010ApJ...709L.133B} Bouwens, R.~J., 
Illingworth, G.~D., Oesch, P.~A., et al.\ 2010, \apjl, 709, L133
\bibitem[Bouwens et al.(2011)]{2011ApJ...737...90B} Bouwens, R.~J., 
Illingworth, G.~D., Oesch, P.~A., et al.\ 2011, \apj, 737, 90 
\bibitem[Bouwens et al.(2012)]{2012ApJ...754...83B} Bouwens, R.~J., 
Illingworth, G.~D., Oesch, P.~A., et al.\ 2012, \apj, 754, 83
\bibitem[Bouwens et al.(2012)]{2012ApJ...752L...5B} Bouwens, R.~J., 
Illingworth, G.~D., Oesch, P.~A., et al.\ 2012a, \apjl, 752, L5 
\bibitem[Bouwens et al.(2013)]{2012arXiv1211.2230B} Bouwens, R., Bradley, 
L., Zitrin, A., et al.\ 2013, \apj, submitted, arXiv:1211.2230
\bibitem[Bouwens et al.(2014)]{2014arXiv1403.4295B} Bouwens, R.~J., 
Illingworth, G.~D., Oesch, P.~A., et al.\ 2014, arXiv:1403.4295
\bibitem[Bromm 
\& Larson(2004)]{2004ARA&A..42...79B} Bromm, V., \& Larson, R.~B.\ 2004, \araa, 42, 79 
\bibitem[Bruzual 
\& Charlot(2003)]{2003MNRAS.344.1000B} Bruzual, G., \& Charlot, S.\ 2003, \mnras, 344, 1000 
\bibitem[Bunker et al.(2010)]{2010MNRAS.409..855B} Bunker, A.~J., Wilkins, 
S., Ellis, R.~S., et al.\ 2010, \mnras, 409, 855
\bibitem[Calzetti et al.(1994)]{1994ApJ...429..582C} Calzetti, D., Kinney, 
A.~L., \& Storchi-Bergmann, T.\ 1994, \apj, 429, 582
\bibitem[Calzetti et al.(2000)]{2000ApJ...533..682C} Calzetti, D., Armus, 
L., Bohlin, R.~C., Kinney, A.~L., Koornneef, J., 
\& Storchi-Bergmann, T.\ 2000, \apj, 533, 682 
\bibitem[Caruana et al.(2012)]{2012MNRAS.427.3055C} Caruana, J., Bunker, 
A.~J., Wilkins, S.~M., et al.\ 2012, \mnras, 427, 3055 
\bibitem[Castellano et al.(2012)]{2012A&A...540A..39C} Castellano, M.,
  Fontana, A., Grazian, A., et al.\ 2012, \aap, 540, A39
\bibitem[Dav{\'e} et al.(2006)]{2006MNRAS.370..273D} Dav{\'e}, R., 
Finlator, K., \& Oppenheimer, B.~D.\ 2006, \mnras, 370, 273 
\bibitem[Dayal \& Ferrara(2012)]{2012MNRAS.tmp.2461D} Dayal, P., \&
  Ferrara, A.\ 2012, \mnras, 2461
\bibitem[Dayal et al.(2013)]{2013MNRAS.tmp.1089D} Dayal, P., Libeskind, 
N.~I., \& Dunlop, J.~S.\ 2013, \mnras, 1089

\bibitem[de Barros et al.(2014)]{2014A&A...563A..81D} de Barros, S.,
  Schaerer, D., \& Stark, D.~P.\ 2014, \aap, 563, A81

\bibitem[Dressel et al. (2012)]{dressel} Dressel, L., et al.\ 2012. 
“Wide Field Camera 3 Instrument Handbook, Version 5.0” (Baltimore: STScI)
\bibitem[Dunlop et al.(2012)]{2012MNRAS.420..901D} Dunlop, J.~S., McLure, 
R.~J., Robertson, B.~E., et al.\ 2012a, \mnras, 420, 901 
\bibitem[Dunlop et al.(2013)]{2013MNRAS.432.3520D} Dunlop, J.~S., Rogers, 
A.~B., McLure, R.~J., et al.\ 2013, \mnras, 432, 3520 
\bibitem[Ellis et al.(2013)]{2013ApJ...763L...7E} Ellis, R.~S., McLure, 
R.~J., Dunlop, J.~S., et al.\ 2013, \apjl, 763, L7 
\bibitem[Erb et al.(2006)]{2006ApJ...644..813E} Erb, D.~K., Shapley, A.~E., 
Pettini, M., Steidel, C.~C., Reddy, N.~A., 
\& Adelberger, K.~L.\ 2006a, \apj, 644, 813 
\bibitem[Finkelstein et al.(2010)]{2010ApJ...719.1250F} Finkelstein, S.~L., 
Papovich, C., Giavalisco, M., Reddy, N.~A., Ferguson, H.~C., Koekemoer, 
A.~M., \& Dickinson, M.\ 2010, \apj, 719, 1250
\bibitem[Finkelstein et al.(2012)]{2012ApJ...756..164F} Finkelstein, S.~L., 
Papovich, C., Salmon, B., et al.\ 2012, \apj, 756, 164 
\bibitem[Finlator et al.(2011)]{2011MNRAS.410.1703F} Finlator, K., 
Oppenheimer, B.~D., \& Dav{\'e}, R.\ 2011, \mnras, 410, 1703 
\bibitem[Giavalisco et al.(2004)]{2004ApJ...600L..93G} Giavalisco, M., 
Ferguson, H.~C., Koekemoer, A.~M., et al.\ 2004, \apjl, 600, L93
\bibitem[Gonz{\'a}lez et al.(2011)]{2011ApJ...735L..34G} Gonz{\'a}lez, V., 
Labb{\'e}, I., Bouwens, R.~J., et al.\ 2011, \apjl, 735, L34
\bibitem[Gonz{\'a}lez et al.(2012)]{2012ApJ...755..148G} Gonz{\'a}lez, V., 
Bouwens, R.~J., Labb{\'e}, I., et al.\ 2012, \apj, 755, 148 
\bibitem[Gonz{\'a}lez et al.(2014)]{2014ApJ...781...34G} Gonz{\'a}lez, V., 
Bouwens, R., Illingworth, G., et al.\ 2014, \apj, 781, 34
\bibitem[Grogin et al.(2011)]{2011ApJS..197...35G} Grogin, N.~A., Kocevski, 
D.~D., Faber, S.~M., et al.\ 2011, \apjs, 197, 35 
\bibitem[Illingworth et al.(2013)]{2013ApJS..209....6I} Illingworth, G.~D., 
Magee, D., Oesch, P.~A., et al.\ 2013, \apjs, 209, 6
\bibitem[Jaacks et al.(2012)]{2012MNRAS.427..403J} Jaacks, J., Nagamine, 
K., \& Choi, J.~H.\ 2012, \mnras, 427, 403
\bibitem[Jiang et al.(2013)]{2013ApJ...772...99J} Jiang, L., Egami, E., 
Mechtley, M., et al.\ 2013, \apj, 772, 99
\bibitem[Kimble et al.(2008)]{2008SPIE.7010E..43K} Kimble, R.~A., MacKenty, 
J.~W., O'Connell, R.~W., \& Townsend, J.~A.\ 2008, \procspie, 7010
\bibitem[Koekemoer et al.(2003)]{2003hstc.conf..337K} Koekemoer, A.~M., 
Fruchter, A.~S., Hook, R.~N., 
\& Hack, W.\ 2003, HST Calibration Workshop : Hubble after the Installation of the ACS and the NICMOS Cooling System, 337 
\bibitem[Koekemoer et al.(2011)]{2011ApJS..197...36K} Koekemoer, A.~M., 
Faber, S.~M., Ferguson, H.~C., et al.\ 2011, \apjs, 197, 36 
\bibitem[Koekemoer et al.(2013)]{2013ApJS..209....3K} Koekemoer, A.~M., 
Ellis, R.~S., McLure, R.~J., et al.\ 2013, \apjs, 209, 3
\bibitem[Kron (1980)]{kron} Kron, R. G. 1980, \apjs, 43, 305
\bibitem[Kuhlen 
\& Faucher-Gigu{\`e}re(2012)]{2012MNRAS.423..862K} Kuhlen, M., \& Faucher-Gigu{\`e}re, C.-A.\ 2012, \mnras, 423, 862 
\bibitem[Labb{\'e} et al.(2007)]{2007ApJ...665..944L} Labb{\'e}, I., et 
al.\ 2007, \apj, 665, 944 
\bibitem[Labb{\'e} et al.(2010)]{2009arXiv0911.1356L} Labb{\'e}, I., et
  al.\ 2010, \apjl, 716, L103
\bibitem[Labb{\'e} et al.(2013)]{2013ApJ...777L..19L} Labb{\'e}, I., Oesch, 
P.~A., Bouwens, R.~J., et al.\ 2013, \apjl, 777, L19
\bibitem[Lee et al.(2012)]{2012ApJ...752...66L} Lee, K.-S., Ferguson, 
H.~C., Wiklind, T., et al.\ 2012, \apj, 752, 66
\bibitem[Lotz et al.(2014)]{2014AAS...22325401L} Lotz, J., Mountain, M., 
Grogin, N.~A., et al.\ 2014, American Astronomical Society Meeting 
Abstracts, 223, \#254.01
\bibitem[Madau(1995)]{1995ApJ...441...18M} Madau, P.\ 1995, \apj, 441, 18 
\bibitem[Maiolino et al.(2004)]{2004Natur.431..533M} Maiolino, R., 
Schneider, R., Oliva, E., Bianchi, S., Ferrara, A., Mannucci, F., Pedani, 
M., \& Roca Sogorb, M.\ 2004, \nat, 431, 533 
\bibitem[Maiolino et 
al.(2008)]{2008A&A...488..463M} Maiolino, R., et al.\ 2008, \aap, 488, 463 
\bibitem[Meurer et al.(1999)]{1999ApJ...521...64M} Meurer, G.~R., Heckman, 
T.~M., \& Calzetti, D.\ 1999, \apj, 521, 64 
\bibitem[Miralda-Escude(1998)]{1998ApJ...501...15M} Miralda-Escude, J.\ 
1998, \apj, 501, 15
\bibitem[Monna et al.(2014)]{2014MNRAS.438.1417M} Monna, A., Seitz, S., 
Greisel, N., et al.\ 2014, \mnras, 438, 1417
\bibitem[Mosleh et al.(2012)]{2012ApJ...756L..12M} Mosleh, M., Williams, 
R.~J., Franx, M., et al.\ 2012, \apjl, 756, L12 
\bibitem[Oesch et al.(2012)]{2012ApJ...759..135O} Oesch, P.~A., Bouwens, 
R.~J., Illingworth, G.~D., et al.\ 2012, \apj, 759, 135
\bibitem[Oesch et al.(2013)]{2013ApJ...772..136O} Oesch, P.~A., Labb{\'e}, 
I., Bouwens, R.~J., et al.\ 2013a, \apj, 772, 136 
\bibitem[Oesch et al.(2013)]{2013ApJ...773...75O} Oesch, P.~A., Bouwens, 
R.~J., Illingworth, G.~D., et al.\ 2013b, \apj, 773, 75 
\bibitem[Oke \& Gunn(1983)]{1983ApJ...266..713O} Oke, J.~B., \& Gunn, 
J.~E.\ 1983, \apj, 266, 713 
\bibitem[Ono et al.(2012)]{2012ApJ...744...83O} Ono, Y., Ouchi, M., 
Mobasher, B., et al.\ 2012, \apj, 744, 83
\bibitem[Ono et al.(2013)]{2013ApJ...777..155O} Ono, Y., Ouchi, M., 
Curtis-Lake, E., et al.\ 2013, \apj, 777, 155
\bibitem[Pannella et al.(2009)]{2009ApJ...698L.116P} Pannella, M., Carilli, 
C.~L., Daddi, E., et al.\ 2009, \apjl, 698, L116 
\bibitem[Papovich et al.(2004)]{2004ApJ...600L.111P} Papovich, C., et al.\ 
2004, \apjl, 600, L111
\bibitem[Peng et al.(2002)]{2002AJ....124..266P} Peng, C.~Y., Ho, L.~C., 
Impey, C.~D., \& Rix, H.-W.\ 2002, \aj, 124, 266
\bibitem[Pentericci et al.(2011)]{2011ApJ...743..132P} Pentericci, L., 
Fontana, A., Vanzella, E., et al.\ 2011, \apj, 743, 132
\bibitem[Postman et al.(2012)]{2012ApJS..199...25P} Postman, M., Coe, D., 
Ben{\'{\i}}tez, N., et al.\ 2012, \apjs, 199, 25
\bibitem[Reddy et al.(2010)]{2010ApJ...712.1070R} Reddy, N.~A., Erb, D.~K., 
Pettini, M., Steidel, C.~C., \& Shapley, A.~E.\ 2010, \apj, 712, 1070
\bibitem[Robertson et al.(2010)]{2010Natur.468...49R} Robertson, B.~E., 
Ellis, R.~S., Dunlop, J.~S., McLure, R.~J., 
\& Stark, D.~P.\ 2010, \nat, 468, 49
\bibitem[Rogers et al.(2013)]{2013MNRAS.429.2456R} Rogers, A.~B., McLure, 
R.~J., \& Dunlop, J.~S.\ 2013, \mnras, 429, 2456
\bibitem[Rogers et al.(2014)]{2014MNRAS.440.3714R} Rogers, A.~B., McLure, 
R.~J., Dunlop, J.~S., et al.\ 2014, \mnras, 440, 3714
\bibitem[Schlafly \& Finkbeiner(2011)]{2011ApJ...737..103S} Schlafly,
  E.~F., \& Finkbeiner, D.~P.\ 2011, \apj, 737, 103
\bibitem[Schaerer 
\& de Barros(2009)]{2009A&A...502..423S} Schaerer, D., \& de Barros, S.\ 2009, \aap, 502, 423
\bibitem[Schenker et al.(2012)]{2012ApJ...744..179S} Schenker, M.~A., 
Stark, D.~P., Ellis, R.~S., et al.\ 2012, \apj, 744, 179
\bibitem[Skelton et al.(2014)]{2014arXiv1403.3689S} Skelton, R.~E., 
Whitaker, K.~E., Momcheva, I.~G., et al.\ 2014, arXiv:1403.3689
\bibitem[Smit et al.(2014)]{2014ApJ...784...58S} Smit, R., Bouwens, R.~J., 
Labb{\'e}, I., et al.\ 2014, \apj, 784, 58 
\bibitem[Stanway et al.(2005)]{2005MNRAS.359.1184S} Stanway, E.~R., 
McMahon, R.~G., \& Bunker, A.~J.\ 2005, \mnras, 359, 1184 
\bibitem[Stark et al.(2009)]{2009ApJ...697.1493S} Stark, D.~P., Ellis, 
R.~S., Bunker, A., et al.\ 2009, \apj, 697, 1493 
\bibitem[Stark et al.(2010)]{2010MNRAS.408.1628S} Stark, D.~P., Ellis, 
R.~S., Chiu, K., Ouchi, M., \& Bunker, A.\ 2010, \mnras, 408, 1628 
\bibitem[Stark et al.(2013)]{2013ApJ...763..129S} Stark, D.~P., Schenker, 
M.~A., Ellis, R., et al.\ 2013, \apj, 763, 129
\bibitem[Steidel et al.\ (1999)]{1999ApJ...519....1S} Steidel, C.\ C.,
Adelberger, K.\ L., Giavalisco, M., Dickinson, M.\ and Pettini, M.\ 1999,
\apj, 519, 1
\bibitem[Szalay et al.(1999)]{1999AJ....117...68S} Szalay, A.~S.,
Connolly, A.~J., \& Szokoly, G.~P.\ 1999, \aj, 117, 68
\bibitem[Tremonti et al.(2004)]{2004ApJ...613..898T} Tremonti, C.~A., et 
al.\ 2004, \apj, 613, 898
\bibitem[Tokunaga 
\& Vacca(2005)]{2005PASP..117.1459T} Tokunaga, A.~T., \& Vacca, W.~D.\ 2005, \pasp, 117, 1459
\bibitem[van Dokkum et al.(2013)]{2013arXiv1305.2140V} van Dokkum, P., 
Brammer, G., Momcheva, I., et al.\ 2013, arXiv:1305.2140 
\bibitem[Vanzella et al.(2014)]{2014ApJ...783L..12V} Vanzella, E., Fontana, 
A., Zitrin, A., et al.\ 2014, \apjl, 783, L12
\bibitem[Wilkins et al.(2011)]{2011MNRAS.417..717W} Wilkins, S.~M., Bunker, 
A.~J., Stanway, E., Lorenzoni, S., \& Caruana, J.\ 2011, \mnras, 417, 717 
\bibitem[Wilkins et al.(2013)]{2013MNRAS.430.2885W} Wilkins, S.~M., Bunker, 
A., Coulton, W., et al.\ 2013, \mnras, 430, 2885
\bibitem[Windhorst et al.(2011)]{2011ApJS..193...27W} Windhorst, R.~A., 
Cohen, S.~H., Hathi, N.~P., et al.\ 2011, \apjs, 193, 27 
\bibitem[Wise et al.(2012)]{2012ApJ...745...50W} Wise, J.~H., Turk, M.~J., 
Norman, M.~L., \& Abel, T.\ 2012, \apj, 745, 50 
\bibitem[Zheng et al.(2012)]{2012Natur.489..406Z} Zheng, W., Postman, M., 
Zitrin, A., et al.\ 2012, \nat, 489, 406 (Z12) 
\bibitem[Zitrin et al.(2012)]{2012ApJ...747L...9Z} Zitrin, A., Moustakas, 
J., Bradley, L., et al.\ 2012, \apjl, 747, L9
\end{thebibliography}
\end{document}